\documentclass[journal, twocolumn, 10pt]{IEEEtran}

\usepackage[OT1]{fontenc} 
\usepackage{cite}
\usepackage[cmex10]{amsmath}
\usepackage{amssymb}
\usepackage{graphicx}
\usepackage{color}
\usepackage{subfigure}
\usepackage{tabularx}
\usepackage{arydshln}
\usepackage{mathtools}
\usepackage{nomencl}
\usepackage{url}
\makenomenclature

\newcommand*{\abbr}[3]{\nomenclature[z-#3]{#3}{#2}#1 (#3)\index{#3}}

\def\a{\mathbf{a}}
\def\A{\mathbf{A}}
\def\b{\mathbf{b}}

\def\C{\mathbf{C}}
\def\diag{\mathrm{diag}}
\def\D{\mathbf{D}}
\def\DH{D_\mathrm{H}}
\def\DT{D_\mathrm{T}}
\def\DR{D_\mathrm{R}}

\def\etal{\textit{et al.}~}

\def\Frm{\mathrm{F}}

\def\H{\mathbf{H}}

\def\Hrm{\mathrm{H}}

\def\I{\mathbf{I}}

\def\L{\mathcal{L}}

\def\Mb{\bar{M}}

\def\Na{N_{\rm{a}}}

\def\P{\mathbf{P}}

\def\prm{\mathrm{p}}
\def\pb{\mathbf{p}}
\def\Q{\mathbf{Q}}

\def\s{\mathbf{s}}
\def\S{\mathbf{S}}
\def\Sh{\hat{\mathbf{S}}}

\def\Trm{\mathrm{T}}

\def\v{\mathbf{v}}
\def\vec{\mathrm{vec}}
\def\V{\mathbf{V}}

\def\wb{\mathbf{w}}
\def\W{\mathbf{W}}
\def\x{\mathbf{x}}
\def\X{\mathbf{X}}
\def\Xt{\tilde{\mathbf{X}}}
\def\Xh{\hat{\mathbf{X}}}

\def\Y{\mathbf{Y}}

\def\0{\mathbf{0}}
\def\1{\mathbf{1}}

\def\({\left(}
  \def\){\right)}
\def\[{\left[}
    \def\]{\right]}

\begin{document}
%
%
\title{50 Years of Permutation, Spatial and Index Modulation: From Classic RF to Visible Light Communications and Data Storage}
%
\author{{Naoki~Ishikawa,~\IEEEmembership{Member,~IEEE}},~{Shinya~Sugiura,~\IEEEmembership{Senior~Member,~IEEE}, and Lajos~Hanzo,~\IEEEmembership{Fellow,~IEEE}}
  
  \thanks{Manuscript received August 5, 2017; accepted March 10, 2018.}
  \thanks{N.~Ishikawa is with the Graduate School of Information Sciences, Hiroshima City University, Hiroshima 731-3194, Japan (e-mail: naoki@ishikawa.cc).}
 \thanks{S.~Sugiura is with the Department of Computer and Information Sciences, Tokyo University of Agriculture and Technology, Koganei, Tokyo 184-8588, Japan. (sugiura@ieee.org)}
  \thanks{L.~Hanzo is with the School of Electronics and Computer Science, University of Southampton, Southampton SO17 1BJ, UK (e-mail: lh@ecs.soton.ac.uk).}
  \thanks{The work of N.~Ishikawa was partially supported by the Japan Society for the Promotion of Science (JSPS) KAKENHI (Grant Numbers 16J05344, 17H07036).
    The work of S.~Sugiura was supported in part by the JSPS KAKENHI (Grant Numbers 26709028, 16KK0120). The work of L.~Hanzo was supported by the EPSRC projects EP/N004558/1 and EP/L018659/1, as well as of the European Research Council’s Advanced Fellow Grant under the Beam-Me-Up project and of the Royal Society’s Wolfson Research Merit Award.
  }
}
%
\markboth{Accepted for publication in IEEE Communications Surveys \& Tutorials}
{Shell \MakeLowercase{\textit{et al.}}: Bare Demo of IEEEtran.cls for Journals}
\maketitle
%
\begin{abstract}
  In this treatise, we provide an interdisciplinary survey on spatial modulation (SM), where multiple-input multiple-output microwave and visible light, as well as single and multicarrier communications are considered. Specifically, we first review the permutation modulation (PM) concept, which was originally proposed by Slepian in 1965. The PM concept has been applied to a wide range of applications, including wired and wireless communications and data storage. By introducing a three-dimensional signal representation, which consists of spatial, temporal and frequency axes, the hybrid PM concept is shown to be equivalent to the recently proposed SM family. In contrast to other survey papers, this treatise aims for celebrating the hitherto overlooked studies, including papers and patents that date back to the 1960s, before the invention of SM. We also provide simulation results that demonstrate the pros and cons of PM-aided low-complexity schemes over conventional multiplexing schemes.
\end{abstract}

\IEEEpeerreviewmaketitle
\section{Introduction \label{sec:intro}}
\abbr{\IEEEPARstart{S}{patial} modulation}{Spatial Modulation}{SM} has attracted tremendous attention in the \abbr{multiple-input multiple-output}{Multiple-Input Multiple-Output}{MIMO} literature due to its reduced-complexity structure both at the transmitter and the receiver \cite{ishikawa2017thesis,renzo2011commag,renzo2014spatial,yang2015smdesign,yang2016sm,kadir2015stsk,renzo2016smbook,basar2017im}.
Specifically, the SM scheme allocates additional information bits for selecting a single antenna out of multiple transmit antennas.
Because the SM architecture reduces the number of data streams to be transmitted, attractive reduced-complexity detectors have been proposed for the SM scheme \cite{rajashekar2014reduced,sugiura2011rcc,wen2015dsm,zhang2016dsm,lee2015smsphere,xu2017survdec}.

The wide range of SM studies has demonstrated the performance or hardware complexity advantages of SM over conventional MIMO schemes in specific scenarios.
For example, the performance advantages have been observed and verified in a range of fields: \abbr{space-time block codes}{Space-Time Block Codes}{STBCs} \cite{basar2011stsm,sugiura2012universal,helmy2016stm,le2014ostbc}, differential MIMO communications \cite{bian2013dsm,wen2014berdsm,ishikawa2014udsm,bian2015dsm,wen2015dsm,rajashekar2017dsm}, \abbr{millimeter-wave communications}{Millimeter-Wave Communications}{MWCs} \cite{ishikawa2017mmgsm,liu2015ssk,liu2016mmsm}, \abbr{visible light communications}{Visible Light Communications}{VLCs} \cite{mesleh2011osm,fath2011osm,fath2013comparison,ishikawa2015osm}, and classic multicarrier communications \cite{abu2009sim,tsonev2011sim,basar2016im,basar2016simmimo,basar2013sim,wen2015mar,ishikawa2016sim}.

\begin{figure*}[!t] 
  \centering
  \includegraphics[clip,scale=0.54]{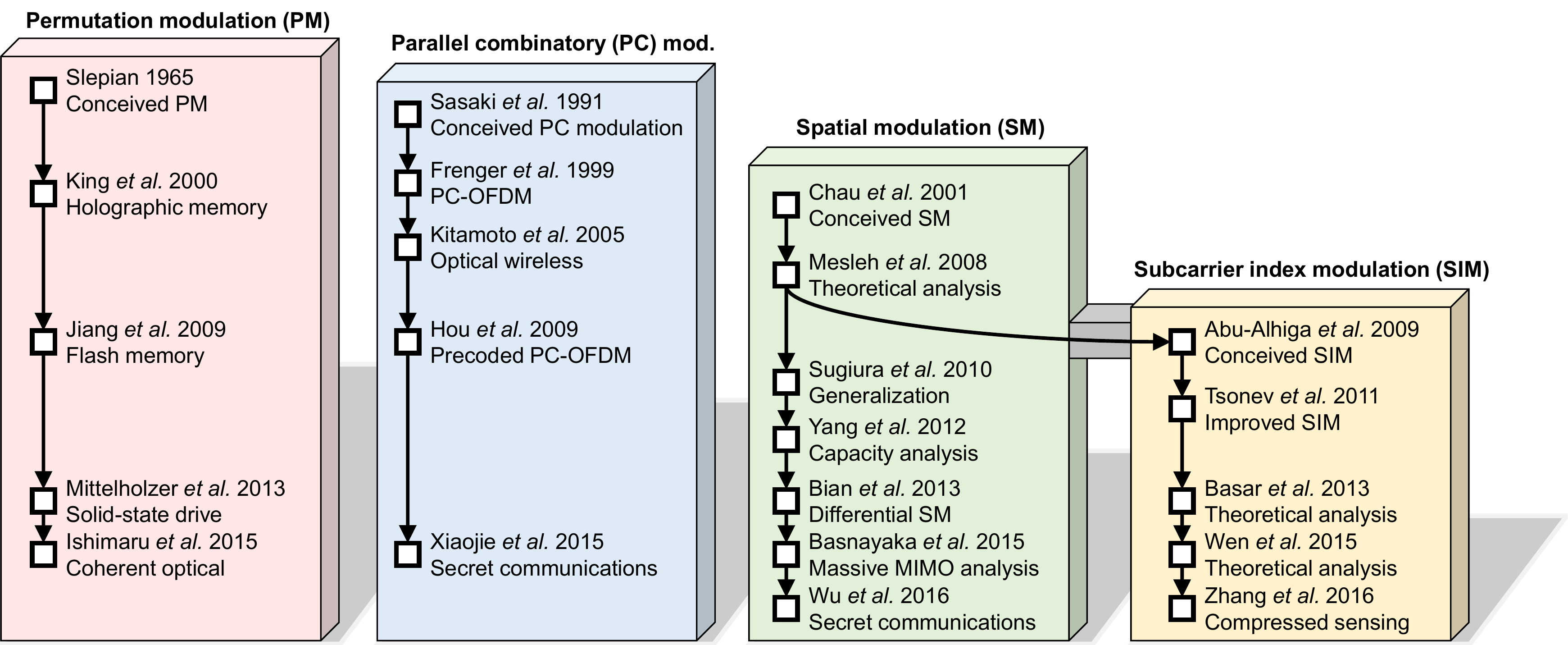}
  \caption{Milestones of the permutation modulation family including parallel combinatory, spatial modulation, and subcarrier index modulation.}
  \label{fig:pm-milestones}
\end{figure*}
Fig.~\ref{fig:pm-milestones} shows milestones of the SM-related schemes.
The SM concept was first proposed in 2001 \cite{chau2001sm}, and its theoretical analysis by Mesleh \etal \cite{mesleh2006sm,mesleh2008spatial} sparked off a paradigm-shift both in the coherent and non-coherent MIMO literature.
In addition, the SM concept was exported to \abbr{orthogonal frequency-division multiplexing}{Orthogonal Frequency-Division Multiplexing}{OFDM} \cite{abu2009sim}, which was later termed as \abbr{subcarrier index modulation}{Subcarrier Index Modulation}{SIM}.
Before the invention of SM \cite{chau2001sm}, \abbr{permutation modulation}{Permutation Modulation}{PM} \cite{slepian1965pm} and \abbr{parallel combinatory}{Parallel Combinatory}{PC} modulation \cite{sasaki1991pcss} were independently developed in 1965 and 1991, respectively.
In contrast to the SM studies, PM research has flourished in the data storage research area, which includes steganography \cite{mittelholzer1999pm}, holographic memories \cite{king2000sparse}, flash memories \cite{jiang2009flash} and solid-state storage \cite{mittelholzer2013patent}, due to the inherent sparsity in data symbols.
More specifically, the PM scheme compresses the input bits by selecting a permutation of a set of sequences, where the sequences consist of ``on'' and ``off'' states for example.
By reducing the number of ``on'' states recorded in a physical material, the PM scheme succeeded in increasing the storage capacity, while maintaining low-latency low-complexity reading and writing \cite{king2000sparse}.
Most recently, the time-domain IM counterpart of SIM was proposed in \cite{nakao2017ftn,sugiura2017access}, which is capable of attaining benefits of SIM, while maintaining a low \abbr{peak-to-average power ratio}{Peak-to-Average Power Ratio}{PAPR}. Furthermore, the time-domain IM scheme was extended to the scenarios of faster-than-Nyquist signaling~\cite{ishihara2017im} and of dual-mode IM~\cite{nakao2017im}.

All of the conventional PM, PC, SM, and SIM schemes rely on the same permutation philosophy.
The SM and SIM schemes have been termed as {\it \abbr{index modulation}{Index Modulation}{IM}} \cite{basar2016im,wen2016underwater,wen2017imbook,shamasundar2017im} since 2016.
For example, in \cite{wen2017imbook}, the SM and SIM schemes were referred to as {\it space domain IM} and {\it frequency domain IM}, respectively.
In this treatise, we use the term {\it permutation modulation}, because the PM concept can be regarded as the origin of the current SM, PC, and SIM schemes. Hence, the novel contributions of this treatise interpreted in the spirit of a survey paper are as follows:
\begin{table*}[t]
	\centering
	\small
	\caption{Comparisons of this survey with other valuable surveys.\label{table:pmsurveys}}
	\begin{tabular}{|l|c|c|c:c:c:c:c|c|c|}
		\hline
    	&Published & Dates & \multicolumn{5}{c|}{SM concept} & PC & PM \\
		& in & back to & \multicolumn{1}{c}{Coherent} & \multicolumn{1}{c}{Differential} & \multicolumn{1}{c}{MWC} & \multicolumn{1}{c}{VLC} & \multicolumn{1}{c|}{OFDM} & concept & concept \\
		\hline \hline
		Sugiura \etal \cite{sugiura2012universal} & 2012 & 2006 & \checkmark & \checkmark & & & & & \\
		Renzo \etal \cite{renzo2014spatial} & 2013 & 2001 & \checkmark &  & & \checkmark & & & \\
		Yang \etal \cite{yang2015smdesign} & 2014 & 2001 & \checkmark  & \checkmark  & &\checkmark &\checkmark & & \\
		Kadir \etal \cite{kadir2015stsk} & 2014 & 2001 & \checkmark & \checkmark & & & \checkmark & & \\
		Yang \etal \cite{yang2016sm} & 2016 & 1980 & \checkmark &  & & & & & \\
	    Ishikawa \etal \cite{ishikawa2016sim} & 2016 & 1991 &  &  & & &\checkmark &\checkmark & \\
		Wen \etal \cite{wen2017imbook} & 2017 & 1986 & \checkmark & \checkmark & & & \checkmark & & \\
		Shamasundar \etal \cite{shamasundar2017im} & 2017 & 2001 & \checkmark &  & & &\checkmark & & \\
		This survey & & 1965 & \checkmark & \checkmark &\checkmark &\checkmark &\checkmark &\checkmark &\checkmark \\
		\hline
	\end{tabular}
\end{table*}
\begin{itemize}
\item Against the backcloth of the existing valuable surveys on the
  popular PM-derivatives of SM and IM
  schemes~\cite{basar2016im,wen2016underwater,wen2017imbook,shamasundar2017im},
  we survey the broad spectrum of historic contributions on the
  general PM and PC concepts, which have been hitherto
  somewhat overlooked in the SM and SIM literature.  Thus, this
  treatise has been conceived for celebrating the tremendous
  contributions of the past five decades since 1965, when Slepian
  coined the term of \textit{permutation modulation}~\cite{slepian1965pm,slepian1965patent}. These historic
  contributions have inspired a spate of sophisticated recent
  developments in SM and SIM. The novel contributions of this survey over other surveys are summarized in Table~\ref{table:pmsurveys}.\footnote{Magazine papers were excluded from this table for fair comparisons, such as \cite{renzo2011commag,basar2016im,wen2016underwater}. In addition, the survey papers focused on the state-of-the-art technologies were also excluded, such as \cite{basar2017im,sugiura2017access}.}

\item Explicitly, we adopt a broad interdisciplinary perspective on
  PM-related schemes by including both microwave and visible light as
  well as single and multicarrier communications.

\item In more technical terms, the intricate interplay between the
  classic modulation constellation and the spatial antenna-domain as
  well as frequency index-domain is detailed. Several metrics are
  considered in the context of the coherent vs. non-coherent as well
  as single-versus multiple-RF design-dilemma, including the mutual
  information, the Eucledian distance and the error probability.
  
 \item This treatise is designed to enable readers to reproduce the simulation results, since the associated channel models are defined in a unified manner for coherent MIMO, differential MIMO, MIMO-MWC, MIMO-VLC and multicarrier communications. This would help readers to understand the state-of-the-art in the IM concept.
\end{itemize}
%

The remainder of this treatise is organized as follows.
Section~\ref{sec:philo} reviews the original PM philosophy.
Section~\ref{sec:sys} defines our system model, while Section~\ref{sec:app} introduces the family of PM schemes proposed for single and multicarrier microwave as well as visible light communications.
Section~\ref{sec:metrics} describes our performance metrics, while Section~\ref{sec:comp} provides performance comparisons between the PM-based schemes and conventional schemes in terms of the metrics described in Section~\ref{sec:metrics}.
Section~\ref{sec:conc} concludes this treatise.
\begin{figure}[tbp]
  \centering
  \includegraphics[clip,scale=0.54]{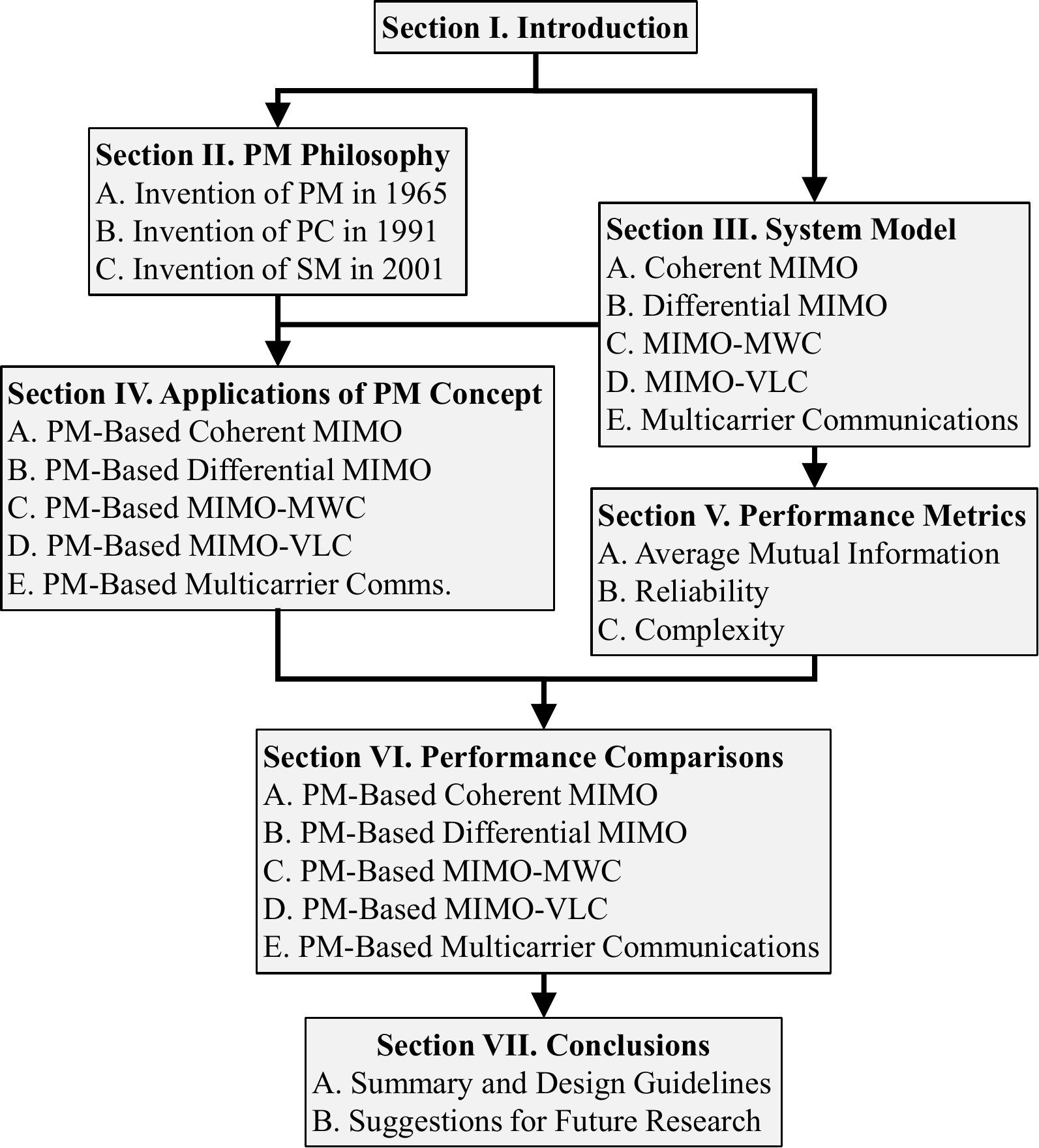}
  \caption{The structure of this treatise.\label{fig:structure}}
\end{figure}
The structure of this contribution is detailed in Fig.~\ref{fig:structure}.
\begin{figure}[tbp]
  \centering
  \subfigure[A data symbol]{ 
    \includegraphics[clip, scale=0.54]{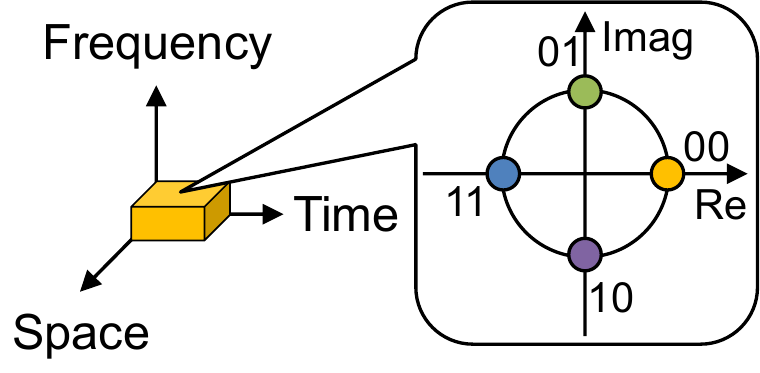} 
    \label{fig:signal-representation-siso}
  }
  \subfigure[Spatial domain]{
    \includegraphics[clip, scale=0.54]{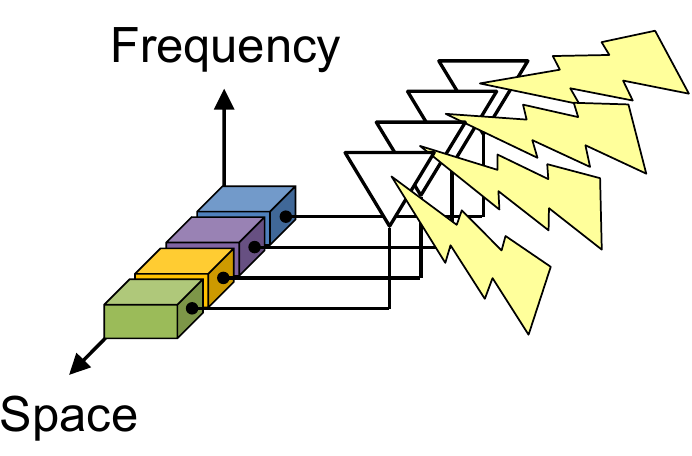} 
    \label{fig:signal-representation-space}
  }
  \subfigure[Temporal domain]{
    \includegraphics[clip, scale=0.54]{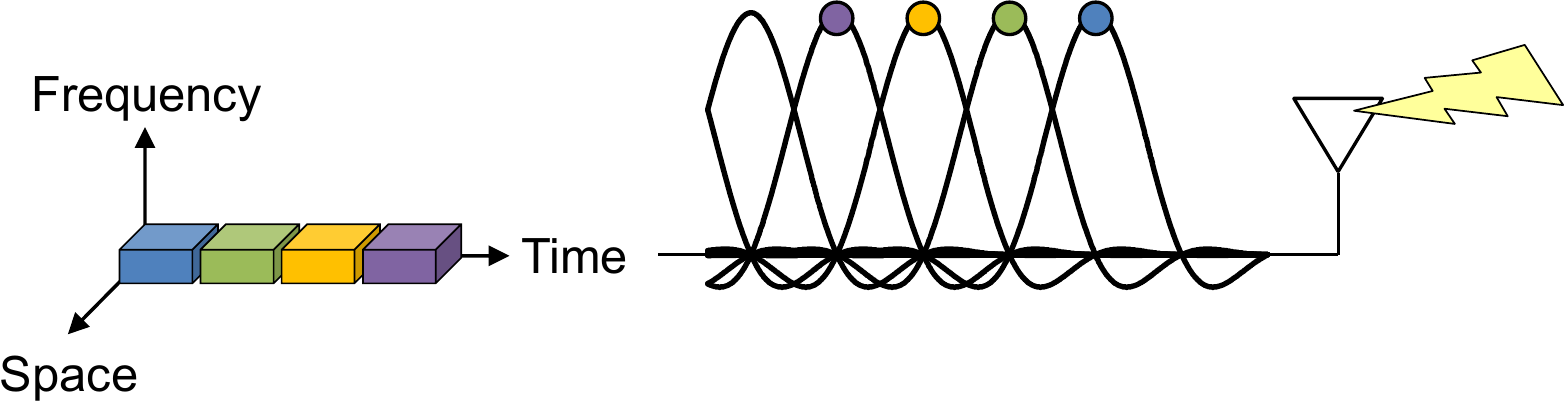} 
    \label{fig:signal-representation-time}
  }
  \subfigure[Frequency domain]{
    \includegraphics[clip, scale=0.54]{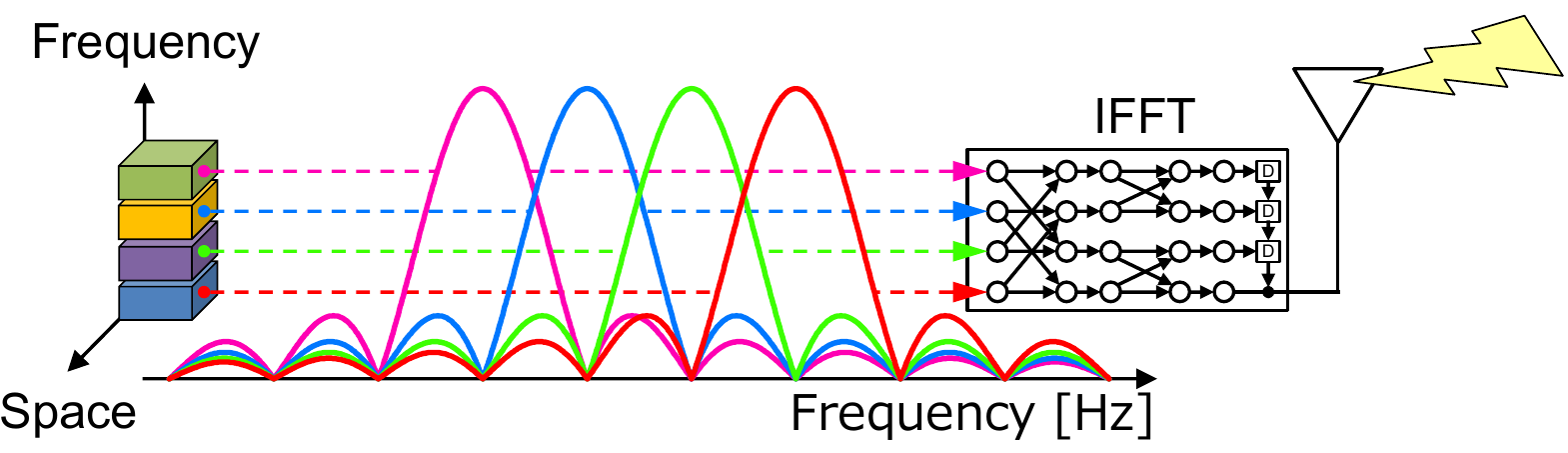} 
    \label{fig:signal-representation-frequency}
  }
  \caption{Three-dimensional signal representation.}
  \label{fig:signal-representation}
\end{figure}
Fig.~\ref{fig:signal-representation} shows the three-dimensional signal representation used in this treatise.
In Fig.~\ref{fig:signal-representation-siso}, a complex-valued data symbol is represented as a colored cube with space, time, and frequency axes.
As shown in Figs.~\ref{fig:signal-representation}(b) and (c), the space axis corresponds to the independent transmit antennas, and the time axis represents the discrete transmission index.
The frequency axis represents the subcarrier index of the OFDM signal, where the frequency domain symbols are transformed to the time domain by the \abbr{inverse fast Fourier transform}{Inverse Fast Fourier Transform}{IFFT}.

We use the following notations throughout this treatise.
Italicized symbols represent scalar values, and bold symbols represent vectors/matrices.
$( \cdot )^{\mathrm{T}}$ denotes the transpose of a matrix 
and $( \cdot )^{\mathrm{H}}$ denotes the Hermitian transpose of a matrix.
Furthermore, ${\cdot \choose \cdot}$ denotes the binomial coefficient.
$\mathcal{CN}(\mu,\sigma^2)$ denotes the complex normal distribution of a random variable having mean $\mu$ and variance $\sigma^2$.
$\mathbb{C}^{m \times n}$ represents a set of complex-valued matrices with $m$ rows and $n$ columns.
$\mathbb{R}$ denotes the field of real numbers, while $\mathbb{Z}$ and $\mathbb{B}$ represent the ring of integers and binary numbers $[0,1]$, respectively.
$\I_m$ represents the $(m \times m)$-sized identity matrix.
$\lfloor \cdot \rfloor$ denotes the floor function.

\section{PM Philosophy and Its Related Family\label{sec:philo}}

\subsection{Invention of PM in 1965\label{sec:philo:pm}}
In 1965, Slepian proposed the PM concept~\cite{slepian1965pm}, which was published in the {\it Proceedings of the IEEE}\footnote{Note that the PM concept was patented in \cite{slepian1965patent}.}.
The transmission codewords of the PM scheme are generated by permuting the order of a set of numbers.
In the original PM, the initial codeword $\s^{(1)} \in \mathbb{R}^{M \times 1}$ is defined by \cite{slepian1965pm}:
\begin{align}
  \s^{(1)} = [
  \overbrace{
    \underbrace{\mu_1\cdots\mu_1}_{M_1~\mathrm{rows}}~~
    \underbrace{\mu_2\cdots\mu_2}_{M_2~\mathrm{rows}}~~\cdots~~
    \underbrace{\mu_k\cdots\mu_k}_{M_k~\mathrm{rows}}
  }^{M~\mathrm{rows}}
  ]^\Trm,
  \label{eq:intr:pm-definiton}
\end{align}
where we have an integer $M = M_1 + M_2 + \cdots + M_k$ and real values $\mu_1<\mu_2<\cdots < \mu_k$.
In Eq.~\eqref{eq:intr:pm-definiton}, $\mu_1$ is repeated $M_1$ times.
Then, the other codewords are generated by permuting the order of $\s^{(1)}$.
The cardinality of possible codewords $N_c$ is calculated by \cite{slepian1965pm}:
\begin{align}
  N_c = \frac{M!}{M_1!M_2! \cdots M_k!},
  \label{eq:intr:pm-ncodewords}
\end{align}
which increases with the factorial order.
The \abbr{pulse position modulation}{Pulse Position Modulation}{PPM} and \abbr{pulse code modulation}{Pulse Code Modulation}{PCM} are subsumed by the PM scheme \cite{slepian1965pm}.
Specifically, the PPM codewords are generated by the following initial codeword:
\begin{align*}
  \s^{(1)} = [\underbrace{~~~~0~~~~~0~~~~~0~~~~}_{M_1 = 3}~\underbrace{1}_{M_2=1}]^\Trm \in \mathbb{R}^4,
\end{align*}
where $M = M_1 + M_2 = 4$ and $(\mu_1,\mu_2) = (0,1)$.
The total $N_c = M! / \(M_1! \cdot M_2! \) = 4! / \(3! \cdot 1!\) = 4$ number of codewords are generated by the permutation of four numbers as follows:
\begin{align}
  \s^{(1)} = [0~0~0~1]^\Trm,~ \s^{(2)} = [0~0~1~0]^\Trm, \nonumber \\
  \s^{(3)} = [0~1~0~0]^\Trm,~ \s^{(4)} = [1~0~0~0]^\Trm.
  \label{eq:intr:pm-example}
\end{align}
Observe that the codewords in Eq.~\eqref{eq:intr:pm-example} are the same as those of the \abbr{space shift keying}{Space Shift Keying}{SSK} scheme, which uses only a single antenna at any transmission time instant.
Similarly, the well-known PPM scheme conveys the input bits by selecting a single time index.
Thus, the SSK scheme is the spatial domain counterpart of the PPM scheme, which maps the information bits to the temporal domain.
Let us examine another example.
If we have the initial codeword of $\a_1 = [1,2] \in \mathbb{Z}^2$, all the $N_c = 2!/(1!\cdot 1!)=2$ number of PM codewords are given by $\a_1 = [1,2]$ and $\a_2 = [2,1]$.
Suppose that we map the vectors $\a_1$ and $\a_2$ onto space-time matrices, then an example set of codewords is given by
\begin{align}
  \mathbf{A}_1 =
  \begin{bmatrix}
    1	&	0	\\
    0	&	1	\\
  \end{bmatrix},~
  \mathbf{A}_2 =
  \begin{bmatrix}
    0	&	1	\\
    1	&	0	\\
  \end{bmatrix},
  \label{eq:intr:pm-matrices}
\end{align}
each of which is known as a {\it permutation matrix} in linear algebra\footnote{The relationship between PM codewords and permutation matrices is also detailed in \cite{slepian1965pm}.}.
We will show that the permutation matrices of Eq.~\eqref{eq:intr:pm-matrices} are the basis of the SM-aided STBC \cite{sugiura2012universal} and its differential counterpart \cite{bian2013dsm}.

\subsubsection*{Hybrid PM Signaling}
\begin{figure}
  \centering
  \subfigure[Conventional signaling]{
    \includegraphics[clip, scale=0.54]{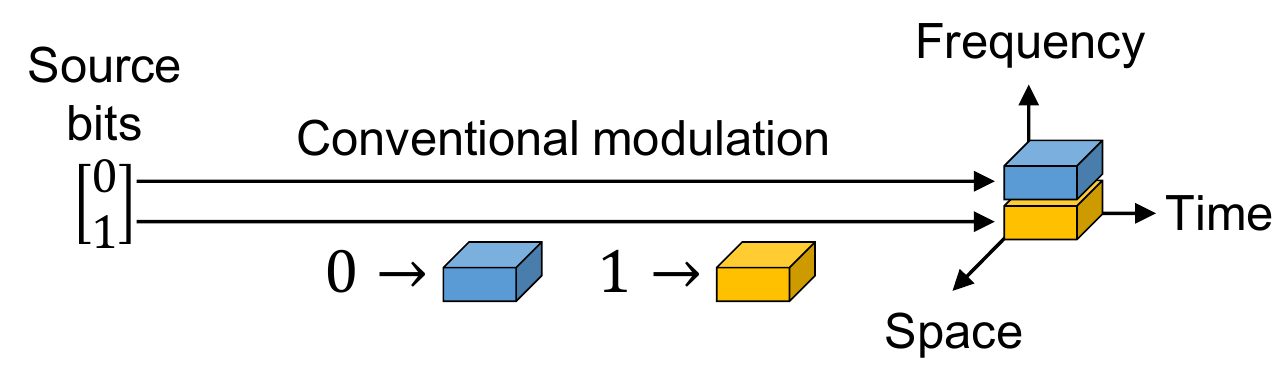} \label{fig:concept-conv}
  }
  \subfigure[Hybrid PM-based signaling \cite{li1996permutation,yongacoglu1997sim}]{
    \includegraphics[clip, scale=0.54]{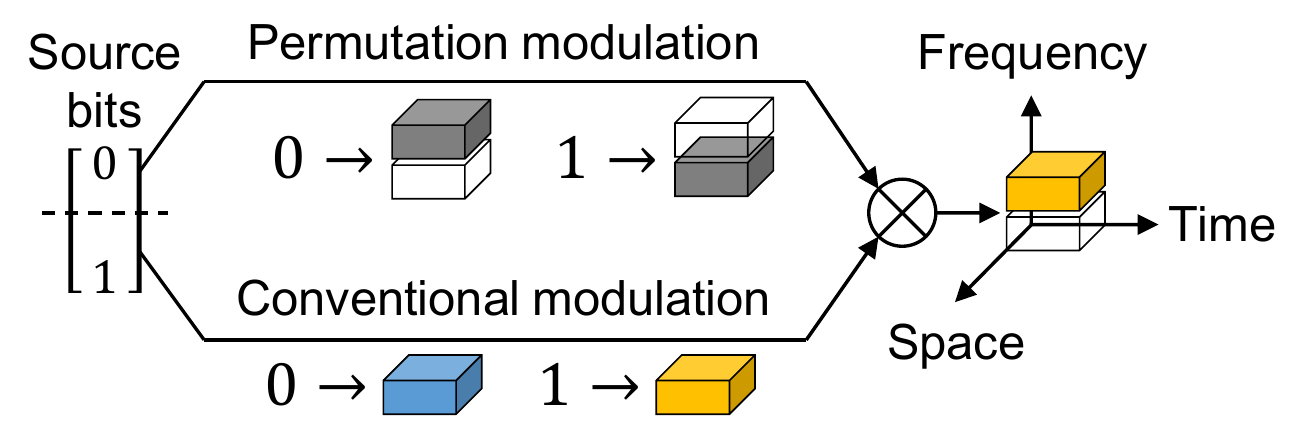} \label{fig:concept-pm}
  }
  \caption{Comparison of conventional and PM signaling. \label{fig:concept}}
\end{figure}
The original PM scheme \cite{slepian1965pm} processes the input bits by outputting a permuted sequence, whereas the hybrid PM scheme \cite{li1996permutation,yongacoglu1997sim} divides the input bits into two parts.
Here, the first bits are used for selecting a set of frequencies, while the remaining bits are used for generating the conventional \abbr{amplitude and phase-shift keying}{Amplitude and Phase-Shift Keying}{APSK} symbols that are carried by the selected frequencies.
Fig.~\ref{fig:concept} exemplifies the concept of hybrid PM.
In contrast to conventional signaling, the hybrid PM scheme maps the permuted sequence to the frequency domain \cite{li1996permutation}, which is based on the original PM concept of \cite{slepian1965pm}.
Let us assume that we map the sequence to the spatial domain. Then, the hybrid PM scheme becomes equivalent to the \abbr{generalized spatial modulation}{Generalized Spatial Modulation}{GSM} of \cite{jeganathan2008generalized}, which is a coherent MIMO scheme.

In this treatise, we interpret the above PM concept in a generalized context.
The diverse PC-, SM-, or SIM-aided schemes convey additional bits by selecting
a spreading sequence \cite{sasaki1991pcss},
a subcarrier-index \cite{frenger1999pcofdm},
a transmit antenna \cite{chau2001sm},
a transmit \abbr{light emitting diode}{Light Emitting Diode}{LED} \cite{mesleh2011osm},
a dispersion matrix \cite{sugiura2012universal},
a permutation matrix \cite{bian2013dsm},
and/or a transmission subarray \cite{ishikawa2017mmgsm}.
These schemes commonly rely on the basic permutation structure of \cite{slepian1965pm,li1996permutation}.
Hence, we regard these schemes as families of the PM concept.
Specifically, we regard an arbitrary modulation scheme as a PM-aided scheme, if it relies on the following combination matrix \cite{frenger1999pcofdm}:
\begin{align}
  \C_{M,P} =
  \begin{bmatrix}
    \mathbf{1} & \C_{M-1,P-1} \\
    \mathbf{0} & \C_{M-1,P} \\
  \end{bmatrix}
  \in \mathbb{B}^{ {M \choose P} \times M },
  \label{eq:intr:combination-matrix}
\end{align}
where $P$ arbitrary elements are selected out of $M$ candidates.
In Eq.~\eqref{eq:intr:combination-matrix}, $\mathbf{0}$ denotes a zero vector of length ${M-1 \choose P}$.
Similarly, $\mathbf{1}$ denotes a one vector of length ${M-1 \choose P-1}$.
Thus, the combination matrix of Eq.~\eqref{eq:intr:combination-matrix} represents the on-off state of arbitrary elements, such as transmit antennas and subcarrier indices.
For example, if we consider the $(M,P)=(4,1)$ case, the combination matrix is given by
\begin{align}
  \C_{4,1} =
  \begin{bmatrix}
    1&0&0&0\\
    0&1&0&0\\
    0&0&1&0\\
    0&0&0&1\\
  \end{bmatrix}
  \in \mathbb{B}^{ 4 \times 1 },
\end{align}
where each row corresponds to the SSK codewords given in Eq.~\eqref{eq:intr:pm-example}.
Furthermore, for the $(M,P)=(4,2)$ case, the combination matrix is given by
\begin{align}
  \C_{4,2} =
  \begin{bmatrix}
	1&1&0&0\\
    1&0&1&0\\
    1&0&0&1\\
    0&1&1&0\\
    0&1&0&1\\
    0&0&1&1\\
  \end{bmatrix}
  \in \mathbb{B}^{ 6 \times 4 },
  \label{eq:combination-matrix-42}
\end{align}
where each row is the same as the PM codewords generated from $\s^{(1)} = [0~0~1~1]^\Trm$ having $(M_1,M_2)=(2,2)$.

Because the number of PM codewords increases with the factorial order, as given in Eq.~\eqref{eq:intr:pm-ncodewords}, PM has been applied to physical data storage systems to increase the storage capacity.
For example, PM and its relatives have been proposed for steganography \cite{mittelholzer1999pm}, volume holographic storage \cite{king2000sparse}, flash memory \cite{jiang2009flash} and solid-state storage \cite{mittelholzer2013patent}.
\begin{table*}[t]
  \centering
  \small
  \caption{Contributions to PM as well as its background and applications.\label{table:intr-pm}}
  \begin{tabularx}{\linewidth}{|c|l|X|}
    \hline
    Year & Authors & Contribution \\
    \hline \hline
    1960 & Lehmer \cite{lehmer1960code} & Conceived an algorithm that generates a permutation of a sequence, which was later called {\it Lehmer code}.\\
    \hline
    1965 & Slepian \cite{slepian1965pm} & Proposed a PM concept, which generates codewords by permuting an initial sequence.\\
    \hline
    1972 & Berger \etal \cite{berger1972permutation} & Proposed a channel coding scheme based on the PM concept. \\
    \hline
    1989 & Atkin and Corrales \cite{atkin1989modulation} & Proposed a PM-based \abbr{frequency shift keying}{Frequency Shift Keying}{FSK} scheme that selects a group of elements, with the aim to achieve higher bandwidth efficiency.\\
    \hline
    1996 & Li \cite{li1996permutation} & Proposed a hybrid permutation scheme that combines the PSK-aided FSK with the PM concept.\\
    \hline
    1997 & Savage \cite{savage1997combinatorial} & Surveyed the combinatorial Gray codes consisting of bits or integers.\\
    \hline
    1999 & Mittelholzer \cite{mittelholzer1999pm} & Applied the PM concept to steganography, which has robustness against attacks.\\
    \hline
    2000 & King and Neifeld \cite{king2000sparse} & Applied the PM concept to volume holographic memories to decrease the number of ``on'' states. This decrease resulted in mitigating interpixel crosstalk and improving capacity. Later, the research of \cite{king2000sparse} was generalized to support multi-levels \cite{king2001permutation}.\\
    \hline
    2005 & Silva and Finamore \cite{silva2005permutation} & Generalized the PM concept to support vectors.\\
    \hline
    2009 & Jiang \etal \cite{jiang2009flash} & Applied the PM concept to the flash memory system to decrease the number of charged cells.\\
    \hline
    2010 & Shi \etal \cite{shi2010permutation} & Proposed a PM-based MIMO-CDMA system.\\
    \hline
    2013 & Mittelholzer \etal \cite{mittelholzer2013patent} & Patented a PM concept applied to a solid-state storage device to mitigate the effects of drift noise.\\
    \hline
    2015 & Ishimura and Kikuchi \cite{ishimura2015permutation} & Applied the PM concept to coherent optical communications.\\
    \hline
  \end{tabularx}
\end{table*}
The contributions to the development of the PM concept detailed in this section are summarized in Table~\ref{table:intr-pm}.

\subsection{Invention of PC in 1991\label{sec:philo:pc}}
Apart from the PM concept \cite{slepian1965pm}, in 1991 another PM concept was independently proposed for spread spectrum communications \cite{sasaki1991pcss}.
The PM-based spread spectrum scheme of \cite{sasaki1991pcss} conveys the additional bits by selecting $P$ out of $M$ spread sequences, which is referred to as {\it parallel combinatory (PC)} concept.
This concept subsumes the conventional $M$-ary spread spectrum communications \cite{tachikawa1992ss,dillard2003ccsk}, where only $P=1$ is selected out of $M$ sequences.
Based on the PC concept of \cite{sasaki1991pcss}, the PC-aided OFDM scheme was proposed in \cite{frenger1999pcofdm}, where a set of subcarriers were selected out of all the subcarrier-activation patterns.
Because the subcarrier-activation process of \cite{frenger1999pcofdm} is based on the combination matrix of Eq.~\eqref{eq:intr:combination-matrix}, the PC-aided scheme is considered as one of the PM families in this treatise.

The PC concept has been researched in the wideband wireless context.
For example, a phase-rotation based PC-OFDM system is proposed in \cite{hou2009papr}, and its secret communication method is proposed in \cite{xiaojie2015sec}.
Note that the modulation principle of the conventional PC-aided OFDM scheme \cite{frenger1999pcofdm} is the same as that of the newly proposed GSM scheme \cite{jeganathan2008generalized}.
Since 2009, the PM-aided OFDM scheme has been gaining increased attention because it improves both the frequency diversity and the coding gain in comparison to the conventional OFDM scheme \cite{abu2009sim,basar2013sim,wen2015mar,ishikawa2016sim}.
In 2015, the code index modulation was proposed \cite{kaddoum2015cim}, which can be subsumed by the PC concept.
The BER and complexity of the generalized code index modulation were analyzed in \cite{kaddoum2016cim}.

\subsection{Invention of SM in 2001\label{sec:philo:sm}}
The hardware complexity of MIMO systems is typically high due to the multiple \abbr{radio frequency}{Radio Frequency}{RF} chains, which process high-frequency signals.
For massive MIMO systems, a huge number of transmit antennas are used for achieving a competitive performance gain, which leads to a high energy consumption.
To address this limitation, the SM concept was proposed for reducing the complexity both at the transmitter and the receiver, without decreasing the spectrum efficiency of conventional systems.
Note that the invention of SM \cite{chau2001sm,mesleh2006sm} was independent of the classic PM and PC concepts.

Again, SM-based research has captured the imagination of scientists on a benefit of its reduced number of transmit RF chains.
A transmit RF chain per antenna is typically composed of digital-to-analog converters, low-pass filters, bandpass filters, synchronizers, and an amplifier.
Together, these lead to a high-complexity and high-cost implementation.
The SM scheme has been shown to be capable of operating single-RF-aided transmissions \cite{renzo2011commag,renzo2014spatial,yang2015smdesign,mesleh2017imp} with the aid of antenna switching.
However, antenna switching at high frequencies is a challenging task \cite{yang2016sm}.
It was shown in \cite{ishibashi2014sm} that the single-RF SM transmitter has to transmit each time-domain symbol relying on symbol-wise antenna switching. Hence, the bandwidth-efficient raised cosine filter is unsuitable for the single-RF SM transmitter. To this end, increasing the number of transmit RF chains at the SM transmitter was proposed for solving this problem \cite{ishibashi2014sm}, while maintaining a low transmitter cost.
It is worth noting that the number of required receive RF chains is identical for both the classic MIMO and the SM schemes, where a receive RF chain per antenna is composed of sophisticated filters and amplifiers.

The full-RF-aided SM transmitter, which is equipped with $M$ transmit RF chains for $M$ antennas, still has advantages over the \abbr{spatial multiplexing}{Spatial MultipleXing}{SMX} scheme in terms of both its higher \abbr{minimum Euclidean distance}{Minimum Euclidean Distance}{MED} \cite{mesleh2008spatial} and its lower computational complexity \cite{xu2013reduced,younis2013spsm,rajashekar2014reduced,lee2015smsphere}.
Hence, it is suitable for open-loop large-scale MIMO scenarios \cite{basnayaka2015massive}.
Similar advantages were also observed in MWC and VLC channels \cite{mesleh2010osm,liu2015ssk},
where the associated channel matrices contain strong \abbr{line-of-sight}{line-of-Sight}{LoS} elements due to their specific propagation properties.
In such channels, the rank of channel matrices tends to be low, and the performance gain of MIMO systems is eroded.
The SM scheme circumvented this issue \cite{mesleh2010osm,liu2015ssk} as a benefit of its reduced number of data streams.
Hence, the SM scheme is capable of operation in low-rank channels.

\section{System Model\label{sec:sys}}
In this treatise, we assume narrowband statistical channel models, such as the Rayleigh, the Rician, and the Jakes channels, where the delay spread is much lower than the reciprocal of the bandwidth.
The numbers of transmit and receive antennas are denoted by $M$ and $N$, respectively.
At the transmission index $i~(i \geq 0)$, based on an input bit segment of length $B$, a specific space-time codeword $\S(i) \in \mathbb{C}^{M \times T}$ is generated out of the $N_c = 2^B$ number of legitimate codewords.
Basically, the codeword $\S(i)$ contains the complex-valued APSK symbols, such as BPSK, QPSK and \abbr{quadrature amplitude modulation}{Quadrature Amplitude Modulation}{QAM}.
Then, codeword $\S(i)$ is transmitted through the $M$ antennas.
The discrete-time and baseband representation of the received block is given by
\begin{align}
  \Y(i) = \H(i)\S(i) + \V(i),
  \label{COH:eq:blockmodel}
\end{align}
where
\begin{description}
\item[$\Y(i) \in \mathbb{C}^{N \times T}$] \ \ \ \ \ \ \ \ \ \ \ is the $i$th received block,
\item[$\H(i) \in \mathbb{C}^{N \times M}$] \ \ \ \ \ \ \ \ \ \ \ is the $i$th channel matrix,
\item[$\S(i) \in \mathbb{C}^{M \times T}$] \ \ \ \ \ \ \ \ \ \ \ is the $i$th space-time codeword, and
\item[$\V(i) \in \mathbb{C}^{N \times T}$] \ \ \ \ \ \ \ \ \ \ \ is the $i$th additive noise.
\end{description}   
In Eq.~\eqref{COH:eq:blockmodel}, the channel coefficient $\H(i)$ denotes the amplitude and phase fluctuation between the $m$th transmit and the $n$th receive antennas, where $1 \leq m \leq M$ and $1 \leq n \leq N$.
Each symbol in $\S(i)$ is transmitted through the $m$th antenna at the time index $t ~ (1 \leq t \leq T)$, which ranges from $\( i \cdot T \cdot T_s + (t-1)T_s \)$ to $\(i \cdot T \cdot T_s + tT_s \)$ [sec],\footnote{[$\cdot$] denotes a unit.}  where $T_s$ represents a symbol duration.
We assume that the noise element in $\V(i)$ obeys the \abbr{independent and identically distributed}{independent and identically distributed}{i.i.d.}\footnote{The i.i.d. assumption implies that each random variable is mutually independent and follows an identical distribution.} \abbr{additive white Gaussian noise}{Additive White Gaussian Noise}{AWGN} with the variance of $\sigma_v^2$, i.e., $\mathcal{CN}(0,\sigma_v^2)$.
Note that the variance-covariance matrix of $\V(i)$ is calculated by $\mathrm{E}\[\vec(\V(i)) \cdot \vec(\V(i))^\Hrm\]$, which converges to $\sigma_v^2 \cdot \I_{NT}$ on average.
We omit the transmission index $i$ if it is not needed.
The received \abbr{signal-to-noise ratio}{Signal-to-Noise Ratio}{SNR} $\gamma$ is defined by
\begin{align}
  \gamma = \frac{\sum_{k=1}^{N_c} \left\| \S^{(k)} \right\|^2_\mathrm{F}}{M \cdot T \cdot \sigma_v^2},
  \label{eq:COH:snr}
\end{align}
where $\S^{(k)} ~ (1 \leq k \leq N_c = 2^B)$ denotes the space-time codeword associated with the $B$ input bits.
Throughout our simulations, we adjust the mean power $\sum_{k=1}^{N_c} \left\| \S^{(k)} \right\|^2_\mathrm{F}$ to $M \cdot T$ for all the schemes.
The random channel matrix $\H(i)$ depends on the channel setup, such as the uncorrelated/correlated Rayleigh, the Rician and the Jakes fading channels.

\subsection{Coherent MIMO\label{sec:sys:coh}}
In 1942, Peterson patented a diversity receiver concept, which exploits the diversity of the channel coefficients \cite{peterson1942patent}.
In 1973, Schmidt \etal patented the space-division multiple access concept, where the received signals are spatially separable \cite{clark1966patent,schmidt1973patent}.
In 1987, Winters derived the ergodic capacity of MIMO channels \cite{winters1987capacity}.
This analysis was inspired by the dually polarized \abbr{single-input single-output}{Single-Input Single-Output}{SISO} channel \cite{amitay1984channel}, which is equivalent to a $2 \times 2$ MIMO channel.
With the aid of virtual independent paths, the SMX scheme of \cite{foschini1996layered,foschini1998limits,wolniansky1998v} performs well in rich-scattering scenarios.
The SMX scheme is also known as \abbr{Bell Laboratories layered space-time}{Bell Laboratories Layered Space-Time}{BLAST} architecture.
The $M$ independent symbols are transmitted through the $M$ antennas and then received by the $M$ antennas.
The key contribution of \cite{foschini1996layered} was the successive nulling concept, where the transmitted symbols are copied or spread over $M$ time slots.
This redundancy mitigates the inter-channel interference at the receiver and improves the communications reliability.
The SMX scheme maximizes the multiplexing gain, whereas the \abbr{orthogonal space-time block code}{Orthogonal Space-Time Block Code}{OSTBC} \cite{alamouti1998std} maximizes the diversity gain.
The simple OSTBC scheme of \cite{alamouti1998std} embeds two APSK symbols in a $2 \times 2$ space-time codeword.
The embedded symbols are spread over the two time slots.
As proved in \cite{zheng2003dam}, all systems have to obey the diversity-multiplexing tradeoff due to the limited number of independent channel paths.
The OSTBC scheme is also capable of avoiding inter-channel interference at the receiver with the aid of the unitary nature of OSTBC codewords.
Note that the conventional BLAST and OSTBC schemes have been subsumed by the general MIMO schemes of \cite{hassibi2002hrc,sugiura2012universal}, hence we can analyze the pros and cons in a comprehensive manner.

Many transmit antennas are also capable of realizing \abbr{beamforming}{BeamForming}{BF}.
The BF scheme improves the received SNR and the spectrum efficiency, as well as the inter-user interference, which is known as BF gain \cite{spencer2004introduction,xu2017bf}.
One of the simplest schemes is the conjugate BF, where the codewords are multiplied by the Hermitian transpose of the estimated channel matrix \cite{marzetta2010noncooperative}.
Specifically, when assuming a large number of transmit antennas at the base station, $\H \H^\Hrm$ converges to a diagonal form on average, and this leads to interference-free detections at the user terminal.
Thus, this structure facilitates low-complexity transmission and reception, even though a large number of antennas are employed.

\subsubsection*{Channel Model\label{sec:sys:coh:channel}}
Radio waves are propagated at the speed of light, attenuated by distance, and reflected by clusters of scatterers.
The scatterers create independent paths and delay the radio waves due to the difference in propagation distances of each path.
The resultant {\it delay spread} $T_t$ [sec] is an important metric, which is defined by the duration between the first and the last arrivals of the radio propagation.
If the delay spread is larger than the reciprocal of the bandwidth $B_w^{-1}$, then the received signals are significantly distorted.
The independent multi-path components may cause amplitude and phase fluctuations destructively, which is called fading.
In addition, when the mobile terminal moves faster, the received radio waves experience Doppler shift, which is typically severe in high-speed trains and airplanes.
The random time-varying behavior of radio waves makes the wireless channel unreliable.

Again, in this treatise, we assume narrowband statistical channel models, such as the Rayleigh and the Rician channels.
The Rayleigh fading channel model is a basic statistical model that assumes a large number of scatterers.
If the scatterers are uniformly distributed, the channel coefficients are approximated by a Gaussian random process \cite{goldsmith2005wireless} on the basis of the central limit theorem.
Furthermore, if the transmit and receive antennas are sufficiently separated, for example, if the spacing is over ten times as large as the wavelength, the correlation between the adjacent channel coefficients can be ignored.
Then, each coefficient of the channel matrix $\H$ can be approximated by the i.i.d. complex-valued Gaussian symbol having a mean of zero and a variance of 1, i.e., $\mathcal{CN}(0,1)$.
Other MIMO channels models were reported in \cite{goldsmith2005wireless,proakis2008}.

\subsubsection*{Detection}
In this contribution, we assume \abbr{maximum-likelihood}{Maximum-Likelihood}{ML} detection at the receiver, which achieves the lowest possible error rate at the cost of a high system complexity \cite{yang2015detect}. 
Here, we review a hard detector for the general MIMO scheme.
The \abbr{maximum {\it a posteriori}}{Maximum {\it A Posteriori}}{MAP} detector searches the best $\Sh$ that maximizes the {\it a posteriori} probability of $\prm(\S|\Y)$, where the received symbol of $\Y$ is given in advance.
Based on Bayes' theorem\footnote{Bayes' theorem \cite{bayes1763} is given by $\prm(X|Y) = \prm(Y|X) \cdot \prm(X) / \prm(Y)$, where $X$ and $Y$ are random variables.}, the relationship between the {\it a priori} and the {\it a posteriori} probabilities is given by
\begin{align}
  \prm(\S|\Y)=\frac{\prm(\Y|\S)\prm(\S)}{\prm(\Y)}.
  \label{eq:GSTSK:bayse}
\end{align}
We assume that $\prm(\S)$ is constant because the input bits are randomly generated and that the associated codeword $\S$ is transmitted at the equal probability of $1 / 2^{N_c}$.
In addition, the {\it a priori} probability $\prm(\Y)$ is unknown in the hard decision process.
Hence, maximizing the {\it a posteriori} probability $\prm(\S|\Y)$ is equivalent to maximizing the likelihood $\prm(\Y|\S)$, which is defined as follows:
\begin{align}
  \prm(\Y|\S) = \frac{1}{\(\pi \sigma_v^2\)^{NT}} \mathrm{exp} \left( -\frac{\|\Y - \H \S \|_{\Frm}^2}{\sigma_v^2} \right).
  \label{GSTSK:eq:cond}
\end{align}
According to Eq.~\eqref{GSTSK:eq:cond}, the decision metric is given by
\begin{align}
  \Sh = \arg \max_{\S} \prm(\Y|\S)
      = \arg\min_{\S} {\| \Y - \H \S \|_{\Frm}^2}. \label{eq:GSTSK:mld-2}
\end{align}
Note that the Frobenius norm calculation of Eq.~\eqref{eq:GSTSK:mld-2} is carried out over $N_c=2^B$ number of trials, and its detection complexity exponentially grows with the input bit segment of length $B$.
The estimated bit sequence might contain errors.
The number of errors between the original bits from the transmitter and the estimated bits at the receiver is referred to as \abbr{bit error ratio}{Bit Error Ratio}{BER}, which is detailed in Section~\ref{sec:metrics:rdc}.

\subsection{Differential MIMO\label{sec:sys:diff}}
The family of coherent MIMO schemes \cite{foschini1996layered,foschini1998limits,wolniansky1998v,alamouti1998std,hassibi2002hrc,sugiura2012universal} relies on estimating the channel matrix $\H$ at the receiver.
Here, the pilot symbols are transmitted in order to estimate the channel coefficients, which are also known as \abbr{channel state information}{Channel State Information}{CSI}.
For example, the simplest scheme transmits the pilot symbols of $\I_M$ through $M$ antennas over $M$ time slots.
At the receiver, based on the received symbols of $\Y = \H \I_M + \V = \H + \V$, the channel matrix is estimated to be $\hat{\H} = \H + \V$.
Because the estimated channel matrix $\hat{\H}$ contains the AWGN of $\V$, the accuracy of the channel estimation is degraded.
The inaccuracy of $\hat{\H}$ degrades the reliability of the coherent MIMO scheme, which typically exhibits an error floor in uncoded scenarios \cite{sugiura2010cad}.
In addition, the inserted pilot symbols reduce the effective transmission rate.
For example, the pilot symbol of $\I_M$ may occupy $M$ time slots, and thus it increases linearly with the number of transmit antennas.
If we consider fast-fading scenarios having a large normalized Doppler frequency $F_dT_s$, it is a challenging task to accurately track the channel coefficients at the receiver, because they change rapidly.
Furthermore, the number of channel coefficients that have to be estimated is calculated by $N \cdot M$, which increases with the number of transmit and receive antennas.
Hence, the channel estimation problem is especially severe for large-scale MIMO systems in fast-moving scenarios.

To circumvent the channel estimation problem, \abbr{differential space-time block code}{Differential Space-Time Block Code}{DSTBC} was proposed in 2000 \cite{hochwald2000ray,tarokh2000dostbc,hughes2000dstm,wang2014diff}.
The DSTBC scheme circumvents the pilot insertion and the channel estimation process with the aid of unitary matrices.
The successive space-time codewords $\S(i-1)$ and $\S(i)$ have a certain relationship, i.e., $\S(i)=\S(i-1) \X(i)$, which is termed as differential encoding.
Here, $\X(i)$ represents a data-carrying matrix.
At the receiver, the previously received symbol $\Y(i-1)$ acts as the {\it pilot symbol} of the coherent MIMO scenario.
Hence, the major benefit of the DSTBC scheme is its capability of operating without the estimated channel matrix $\hat{\H}(i)$.
Basically, most DSTBC schemes rely on the unitary matrix \cite{hochwald2000ray,tarokh2000dostbc,hochwald2000unitary,hughes2000dstm,hochwald2000dstm,hassibi2002cdu}.
By contrast, some DSTBC schemes use the non-unitary matrix to increase the transmission rate \cite{xia2002differential,zhu2005differential,bhatnagar2009differential}.
However, the differential MIMO scheme cannot be readily combined with a large number of transmit antennas due to the unitary constraint; the only exception is the solution found in \cite{ishikawa2017rdsm}.

\subsubsection*{Channel Model}
The channel model of differential MIMO communications is the same as that of its coherent MIMO counterpart. The narrowband Rayleigh fading channel having no delayed taps is typically assumed \cite{hochwald2000ray,tarokh2000dostbc,hochwald2000unitary,hughes2000dstm,hochwald2000dstm,hassibi2002cdu,xia2002differential,zhu2005differential,bhatnagar2009differential,ishikawa2017rdsm}.

\subsubsection*{Detection}
Let us now introduce the hard ML detector for general DSTBC schemes.
The following detector is suitable for any DSTBC scheme, if and only if the data matrix $\X(i)$ is unitary.
Here, we assume that the successive channel matrices $\H(i)$ and $\H(i-1)$ are the same, i.e., $\H(i) = \H(i-1)$.
We define this assumption as the quasi-static channel.
Because we have the relationships of $\S(i)=\S(i-1)\X(i)$ and $\Y(i-1)=\H(i-1)\S(i-1)+\V(i-1)$,
the ML detector of general DSTBC schemes is given by
\begin{align}
  \Xh(i) = \arg\min_{\X} \| \Y(i) - \Y(i-1)\X  \|_\mathrm{F}^2.
  \label{DIFF:eq:mld}
\end{align}
We observe that Eq.~\eqref{DIFF:eq:mld} does not contain the channel matrix $\H(i)$, which implies that the receiver can dispense with the high-complexity channel estimation process.
However, the total noise variance is doubled compared to the coherent case given in Eq.~\eqref{eq:GSTSK:mld-2}, because both the received symbols $\Y(i)$ and $\Y(i-1)$ contain AWGN.
This limitation imposes the well-known 3 [dB] SNR loss\footnote{$10 \log_{10}2 = 3.01029 \cdots \approx 3.0$ [dB]} in the differential scheme.
Hence, the BER curve of the differential scheme is shifted by 3 [dB] as compared to that of the idealized coherent scheme that has perfect estimates of the channel matrix.

\subsection{MIMO-Aided Millimeter-Wave Communications \label{sec:sys:mwc}}
The capacity of wireless communications linearly increases with the bandwidth \cite{shannon1948mathematical,proakis2008}.
In MWCs \cite{rappaport2013mm,busari2017massive,hemadeh2018mm}, relatively large bandwidths are available, as compared to current mobile networks operated within the 2 to 5 [GHz] spectrum.
Millimeter waves have wavelengths ranging from 1 to 10 mm, and the associated frequency ranges from 30 to 300 [GHz].
Hence, in MWCs, the resultant capacity is higher than the current networks due to the wider bandwidth of MWCs.

Typically, MWCs suffer from high propagation losses imposed by the nature of the short wavelength.
For example, if we consider the free-space path loss model, the path loss increases with the square of the wavelength $\lambda$, i.e., $10 \log_{10}\(\lambda^2\)$ [dB] \cite{goldsmith2005wireless}.
To circumvent the path loss problem \cite{rappaport2013mm,rangan2014mm}, millimeter wave transmitters and receivers have to obtain BF gain with the aid of a large number of antenna elements \cite{atta1959array}.
It is unrealistic for commercial devices to use a large number of RF circuits connected to each antenna element, because the RF circuits of MWC are complex, expensive and power-thirsty \cite{celik2008exp}.
In the microwave MIMO context, the hybrid BF scheme that combines \abbr{analog beamforming}{Analog BeamForming}{ABF} and \abbr{digital beamforming}{Digital BeamForming}{DBF} has been proposed \cite{zhang2003hybrid,sugiura2008iet}.
Specifically, the hybrid scheme divides large antenna array into subarrays, where each subarray is connected to a single RF circuit. 
This structure reduces the number of RF chains both at the transmitter and the receiver.
It was demonstrated in \cite{celik2008exp,torkildson2011mmmimo,guo2012hybrid,alkhateeb2013hybrid,ayach2014sparse,han2015mm5g} that this hybrid BF approach is efficient for MIMO-MWC.

\subsubsection*{Channel Model\label{sec:sys:mwc:channel}}
The channel models of indoor and outdoor MWCs have been extensively studied \cite{shoji2009sv,rappaport2013mm,rangan2014mm}.
Shoji \etal \cite{shoji2009sv} proposed the indoor MWC channel model based on the Saleh-Valenzuela model \cite{saleh1987sv}, where the LoS components have a dominant effect on the channel coefficients.
B{\o}hagen \etal \cite{bohagen2007los} proposed the optimal antenna alignment technique for the uniform linear array. This technique combats the detrimental effects of LoS MWC channels.
Cluster-based ray-tracing channel models were investigated in \cite{smulders1997prop,xu2002mmch,ayach2014sparse}, although some studies \cite{nix2007mmk,torkildson2011mmmimo,liu2015ssk,ishikawa2017mmgsm} assumed having Rician fading for MWC channels.
Rappaport \etal \cite{rappaport2013mm} investigated the potential of cellular MWCs in the 5G context, where the basic propagation characteristics were measured in urban areas.
Sridhar \etal \cite{sridhar2016bf5g} proposed a parametric channel model for the 5G MWCs, which is applicable to general RF communications.
The parametric model of \cite{sridhar2016bf5g} enables us to estimate the channel coefficients accurately and to obtain a massive array gain with the aid of superresolution BF.
We consider indoor and LoS MWCs \cite{nix2007mmk,shoji2009sv,torkildson2011mmmimo,zhou2015fast,liu2015ssk} instead of outdoor or \abbr{non-line-of-sight}{Non-Line-of-Sight}{NLoS} channel environments. 
Throughout our simulations, we employ the frequency-flat Rician channel model.
\begin{figure}[tbp]
  \centering
  \includegraphics[clip,scale=0.43]{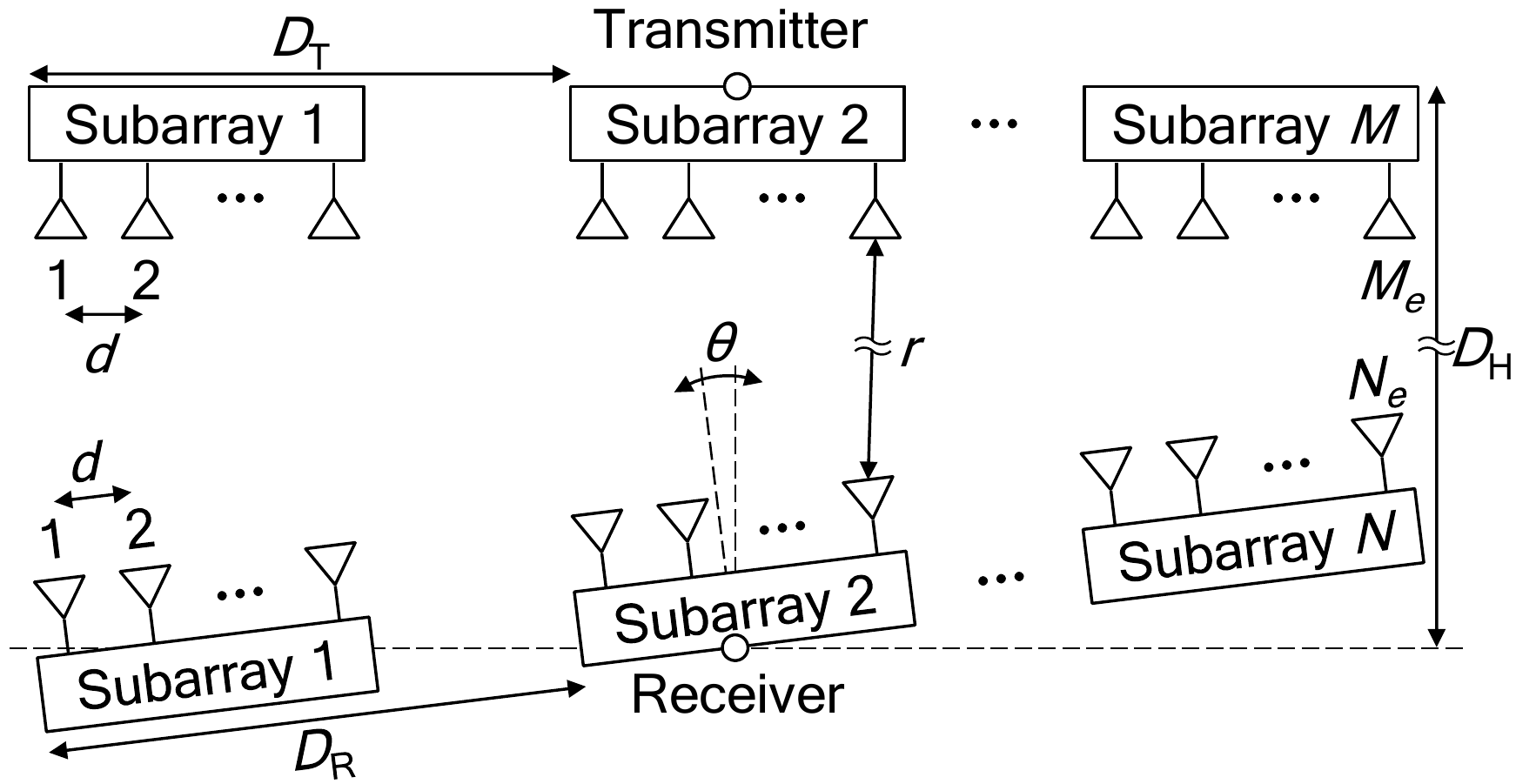}
  \caption{Physical arrangement of the transmitter and the receiver. \copyright IEEE \cite{ishikawa2017mmgsm} \label{mm:fig:channel}}
\end{figure}
Fig.~\ref{mm:fig:channel} shows the arrangement of the transmitter and the receiver, including $M_e$ antenna elements at the transmitter and $N_e$ antenna elements at the receiver.
Each ABF array is separated by a spacing of $\DT$ [m] at the transmitter and $\DR$ [m] at the receiver.
The spacing of each antenna element embedded in an ABF array is $d$ [m].
The transmitter and the receiver are separated by a length of $\DH$ [m], where the receiver is tilted at angle $\theta$.
The channel matrices that follow the Rician fading channels are given by \cite{bohagen2007los,torkildson2011mmmimo,liu2015ssk}:
\begin{align}
	\H_{\mathrm{MWC}} = \sqrt{\frac{K}{K+1}} \H_\mathrm{LoS} + \sqrt{\frac{1}{K+1}} \H_\mathrm{NLoS} \in \mathbb{C}^{N_e \times M_e},
	\label{mm:eq:H}
\end{align}
where $K$ is the Rician factor, which represents the power ratio of LoS elements over NLoS elements.
It was reported in \cite{nix2007mmk} that the Rician $K$ factor was in the range between 8.34 and 12.04 [dB] in 60 [GHz] indoor communications scenarios.
In Eq.~\eqref{mm:eq:H}, the LoS elements are defined by $\H_\mathrm{LoS}[n,m] = \exp\left(-j \cdot \left(2 \pi / \lambda\right) \cdot r[n,m]\right)$, where $r[n,m]$ is the distance between the $m$th transmit antenna element and the $n$th receive antenna element.\footnote{The element at row $n$ and column $m$ of a matrix $\A$ is denoted by $\A[n,m]$, which are row-column indices.}
Here, $\lambda$ represents the wavelength of the transmitted signal.
Furthermore, the NLoS element $\H_\mathrm{NLoS}[n,m]$ obeys the complex-valued Gaussian distribution of $\mathcal{CN}(0,1)$.

At the transmitter, the codeword $\S(i)$ is precoded by an ABF $\P = \mathrm{bdiag}(\pb_1, \cdots, \pb_M) \in \mathbb{C}^{M_e \times M}$ \cite{torkildson2011mmmimo,zhou2015fast}, where $\mathrm{bdiag}(\cdot)$ represents the block diagonalization.
A weight vector $\pb_k \in \mathbb{C}^{(M_e / M) \times 1} ~ (1 \leq i \leq M)$ corresponds to the $k$th ABF array at the transmitter, which has the constraint of $\left\|\pb_k \right\|^2=1$.
Similarly, at the receiver, the ABF weights are represented by $\W = \mathrm{diag}(\wb_1, \cdots, \wb_N) \in \mathbb{C}^{N_e \times N}$ \cite{torkildson2011mmmimo,zhou2015fast}, where $\wb_k \in \mathbb{C}^{(N_e / N) \times1} ~ (1 \leq k \leq N)$ represents a weight vector of the $k$th ABF at the receiver, and each weight $\wb_k$  has the constraint of $\left\|\wb_k\right\|^2=1$.
Based on the general model of Eq.~\eqref{COH:eq:blockmodel}, the channel matrix $\H$ for MIMO-MWC is represented as 
\begin{align}
	\H = \W^{\Hrm} \H_{\mathrm{MWC}} \P \in \mathbb{C}^{N \times M}.
\end{align}

\begin{figure}[tbp]
  \centering
  \includegraphics[clip,scale=0.58]{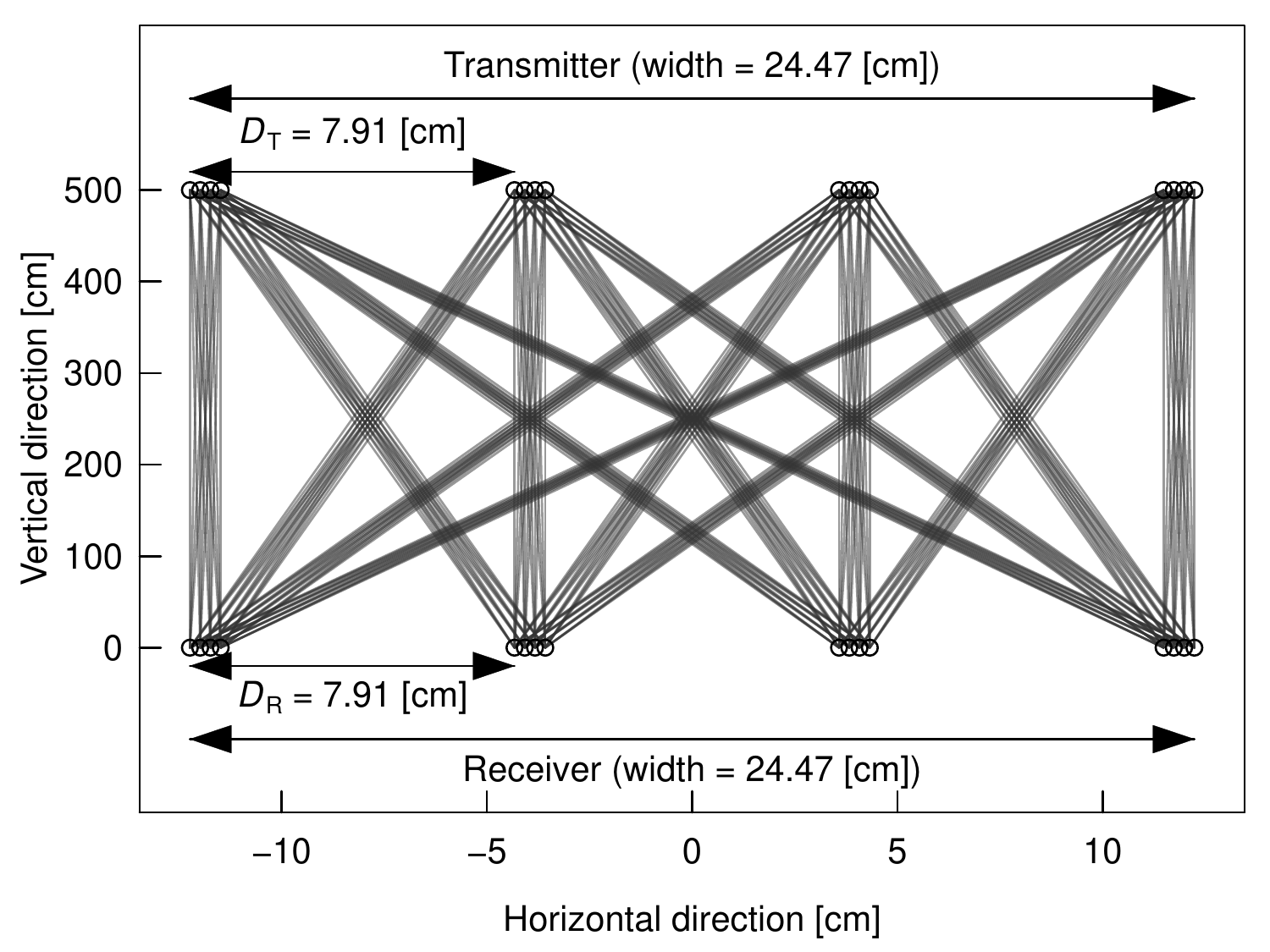}
  \caption{Examples of ABF arrangements, where $(M_e,M,N_e,N)=(16,4,16,4)$ and  $(\DH, \lambda, d) = (500, 0.50, 0.25)$ [cm]. \label{mm:fig:ex-arrangement}}
\end{figure}

In MWCs, a large number of antenna elements are packed in a small space in order to achieve a BF gain.
Typically, the rank of the channel matrix of indoor MWCs is low due to the similarity between adjacent channel coefficients.
In such a low-rank scenario, the performance gains offered by the MIMO techniques are typically reduced.
The optimum antenna alignment scheme that mitigates the above low-rank problem was proposed \cite{bohagen2007los}.
The alignment criterion of \cite{bohagen2007los} recovers the rank of the channel matrix in MIMO-MWCs.
To attain the optimum performance that maximizes the channel rank, the separations of $\DT$ and $\DR$ of ABFs have to satisfy the following relationship \cite{bohagen2007los,torkildson2011mmmimo} :
\begin{align}
  \DT \DR = \frac{\lambda R}{\max(M,N) \cos(\theta)}.
  \label{mm:eq:spacing}
\end{align}
With the aid of Eq.~\eqref{mm:eq:spacing}, the channel rank is increased to $\mathrm{rank}(\H) = \min(M, N)$ for pure LoS scenarios.
For example, if we have a transmitter height of $\DH=5$ [m], receiver tilt of $\theta=0^\circ$, and carrier frequency of 60 [GHz], its wavelength becomes $\lambda=0.5$ [cm].
Here, the spacing between antenna elements embedded in each subarray is $d=\lambda/2=0.25$ [cm].
Furthermore, we consider $M_e=N_e=16$, $M=N=4$, and $\DT = \DR$.
Then, based on Eq.~\eqref{mm:eq:spacing}, the spacing between the subarrays is calculated by
\begin{align}
  \DT = \DR &= \sqrt{\frac{\lambda \DH}{\max(M,N) \cos(\theta)}}\\
  &= \sqrt{\frac{0.005 \times 5}{\max(4,4) \cos(0)}} \approx 7.91~\mathrm{[cm]}.
  \nonumber
\end{align}
Fig.~\ref{mm:fig:ex-arrangement} illustrates the above $\DT=\DR=7.91$ [cm] case, where $(M_e,M,N_e,N)=(16,4,16,4)$.
In this case, the mean rank of the channel matrices is equal to $\mathrm{E}[\mathrm{rank}(\H)]=\min(M,N)=4$.

\subsection{MIMO-Aided Visible Light Communications\label{sec:sys:vlc}}
In 1880, Alexander Graham Bell invented a phone system that conveyed audio signals by means of sunlight \cite{bell1880photophone}.
With the rapid development of LEDs, the brightness, production cost and response time have all been improved by the invention of semiconductor materials such as indium gallium nitride. 
In 2002, Kamiya, who worked at a chemical engineering company, patented the indoor VLC system that uses energy-efficient white LEDs \cite{kamiya2002patent}.
In contrast to conventional infrared communications, the proposed system \cite{kamiya2002patent} used LED illumination as a data-conveying optical wireless channel.
Based on the high brightness and energy efficiency of LEDs, LED-aided VLC research was initiated in \cite{tanaka2003vlc,komine2004vlc,komine2005shadowing}, where the LEDs were used as an illumination bulb as well as a wireless data transmitter.
In VLC, the transmitter modulates the intensity of an LED by the information bits, and the receiver directly detects its change in intensity.

Visible light waves, which are electromagnetic waves that humans can perceive, have a frequency ranging from 430 to 770 [THz].
The available VLC bandwidth is on the THz order, but the practically attainable VLC bandwidth is determined by the modulator and the LED specifications, which are typically limited to the MHz range \cite{minh2008vlc,karunatilaka2015vlc}.
However, license-free and security-aware bandwidths of several dozen MHz are still attractive for our daily lives from the viewpoint of the spectrum shortage in the current networks.

In contrast to microwave communications, the VLC constellation symbols have to be positive and real-valued.
If we consider using OFDM, it has to satisfy the Hermitian symmetry in the frequency-domain \cite{wang2015vlc}, which results in having only positive signals.
LED nonlinearity is a major problem in VLCs \cite{elgala2011vlc} because it may distort the OFDM signals.
The output signal has to satisfy the amplitude constraint in order to avoid clipping distortions \cite{mostafa2015vlcsec}.

\subsubsection*{Channel Model\label{sec:sys:vlc:channel}}
The VLC channel model, which is basically the same as the LED-aided infrared wireless communication model, has been investigated since 1952 \cite{hamme1955infrared,gfeller1980infrared,kahn1997ow}.
The VLC channel coefficients are positive, real-valued and quasi-static for both the outdoor and indoor scenarios \cite{komine2005shadowing,chan2006fso}.
The white LED- and \abbr{photodetector}{PhotoDetector}{PD}-aided VLC channel was analyzed in \cite{tanaka2003vlc,komine2004vlc,komine2005shadowing}, and the employment of the \abbr{complementary metal oxide semiconductor}{Complementary Metal Oxide Semiconductor}{CMOS} imaging sensor was considered in \cite{zeng2009highrate}.
The rank of the channel matrix in the CMOS-aided VLC is generally high due to the additional receiver complexity.

In this treatise, we assume the simplified path loss channel model of \cite{kahn1997ow,fath2013comparison} in our simulations, where intensity modulation and direct detection are employed.
We use the general MIMO system model of Eq.~\eqref{COH:eq:blockmodel},
but the channel matrix $\H$ is replaced by \cite{fath2013comparison,ishikawa2015osm}
\begin{align}
  \H = R_{\mathrm{PD}} \H_{\mathrm{VLC}} \in \mathbb{R}^{N \times M},
\end{align}
where $R_{\mathrm{PD}} \in \mathbb{R}$ [A/W] denotes the response of the PD.
Each element of the matrix $\H_{\mathrm{VLC}}$ is given by \cite{fath2013comparison}
{\footnotesize
\begin{align}
  \H_{\mathrm{VLC}}[n,m] =
  \begin{dcases}
    \frac{(\xi+1)A_{\mathrm{PD}}}{2\pi d[n,m]^2} \cos^{\xi+1}\phi[n,m]  & \left(0 \leq \phi[n,m] \leq \Psi_\frac{1}{2}\right) \\
    0 & \left(\phi[n,m] > \Psi_\frac{1}{2}\right) \\
  \end{dcases},
  \label{vlc:eq:pathloss}
\end{align}
}%
where $\xi = - \ln(2) / \ln\left(\cos\Phi_\frac{1}{2}\right)$.
In Eq.~\eqref{vlc:eq:pathloss}, $A_{\mathrm{PD}}$ denotes the physical area of the PD at the receiver,
$d[n,m]$ represents the distance between the $m$th light source and the $n$th PD,
while $\phi[n,m]$ denotes the angle of incidence from the $m$th light source to the $n$th PD.
Still referring to Eq.~\eqref{vlc:eq:pathloss}, $\Phi_\frac{1}{2}$ represents the transmitter semi-angle, and $\Psi_\frac{1}{2}$ represents the field-of-view semi-angle of the receiver.
The received SNR is defined as follows: \cite{mesleh2011osm}
\begin{align}
  \( \frac{1}{N}\sum_{n=1}^{N} \sigma_r^{(n)} \)^2 / \sigma_v^2,
  \label{survvlc:eq:defsnr}
\end{align}
where $\sigma_r^{(n)}$ is the received optical power at the $n$th PD.

\subsection{Multicarrier Communications\label{sec:sys:ofdm}}
OFDM \cite{hanzo2003ofdm,goldsmith2005wireless,hanzo2010mimoofdm} is an established multicarrier communication technology that has played a key role in numerous communication standards, such as wireless local area network, cellular network, and digital television broadcasting.
The OFDM scheme is capable of exploiting the limited bandwidth, where a number of symbols are simultaneously transmitted via orthogonal subcarriers.
Specifically, the OFDM transmitter multiplexes symbols in the frequency domain, and these symbols are projected onto the time domain by the efficient butterfly-algorithm-aided IDFT.
Then, a redundant signal, which is called guard interval, is added in the time domain.
The received signals are decoded in the frequency domain, which mitigates the inter-carrier interference caused by delayed taps.

Concepts similar to OFDM have been proposed since the 1950s.
In 1958, Mosier and Clabaugh developed a bandwidth-efficient and high-capacity communication system that multiplexes a number of symbols in the frequency domain \cite{mosier1958ofdm}.
In 1966, Chang proposed the basic principle of orthogonal multiplexing, where a number of data symbols are transmitted through a band-limited channel without inter-carrier interference \cite{chang1966ofdm}.\footnote{At the same time the proposed principle was patented in \cite{chang1970patent}.}
Then, Weinstein and Ebert introduced the use of IDFT and ``guard space'' \cite{weinstein1971ofdm}, or guard interval, OFDM was shown to be effective in mobile wireless communications \cite{cimini1985ofdm}.

However, the OFDM scheme still has some open issues.
In the frequency domain, OFDM suffers from out-of-band radiation, which may be suppressed by adding null symbols at the spectrum edge.
In the time domain, the OFDM signal has a high PAPR \cite{ochiai2001papr,ochiai2015moment}, which requires a high dynamic range amplifier per transmit antenna.
Hence, single-carrier transmission combined with frequency-domain equalization may be used for uplink channels \cite{sari1995ofdm,pancaldi2008fde}.
About 60 years have passed since 1958, OFDM still inspires academic attention and many attractive alternatives have been proposed \cite{banelli2014ofdm}.

\subsubsection*{Channel Model\label{sec:sys:ofdm:channel}}
The wideband multipath channel results in delayed paths.
The received symbols are represented by a linear convolution of the transmitted symbols and the channel impulse response. This convolution leads to interference between independent symbols.
The OFDM scheme mitigates this interference problem by concatenating a guard interval, which converts the linear convolution into a circular convolution.
In this treatise, we assume that the guard interval is longer than the maximum delay spread.
Also, carrier-frequency offsets are assumed to be perfectly estimated at the receiver.
Based on the general model of Eq.~\eqref{COH:eq:blockmodel}, the channel matrix $\H$ is represented as $\H = \diag(h_1,\cdots, h_M) \in \mathbb{C}^{M \times M}$ \cite{basar2013sim}, where the coefficients $h_1,\cdots, h_M$ obey the complex-valued Gaussian distribution of $\mathcal{CN}(0,1)$.

\section{Applications of the PM Concept\label{sec:app}}

\subsection{PM-Based Coherent MIMO\label{sec:app:coh}}
In this section, we introduce the PM concept proposed in the coherent MIMO literature \cite{chau2001sm}.
The first PM-based coherent MIMO scheme known as SSK was proposed by Chau and Yu in 2001 \cite{chau2001sm}.
\begin{table*}[t]
  \centering
  \small
  \caption{Contributions to PM-based coherent MIMO schemes.\label{table:survey-cohmimo}}
  \begin{tabularx}{\linewidth}{|c|l|X|}
    \hline
    Year & Authors & Contribution \\
    \hline \hline
    1990 & Baghdady \cite{baghdady1990directional} & Proposed a modulation system based on antenna hopping, where antenna switching results in a phase shift that conveys data.  \\
    \hline
    2001&Chau and Yu \cite{chau2001sm}&Invented an SSK scheme for coherent MIMO communications.\\
    \hline
    2006& Mesleh \etal \cite{mesleh2006sm}& Proposed an SM scheme that activates a single antenna out of multiple transmit antennas.\\
    \hline
    2008 &Yang \etal \cite{yang2008capacity}&Proposed a channel-hopping-based MIMO scheme for high-rate communications and derived its ergodic capacity.\\
    \cline{2-3}
    &Jeganathan \etal \cite{jeganathan2008spatial}&Proposed an optimum detector for the SM scheme of \cite{mesleh2008spatial}.\\
    \cline{2-3}
    &Jeganathan \etal \cite{jeganathan2008generalized}&Extended the SSK concept to one using multiple transmit antennas at the same time.\\%
	\hline
    2010&Sugiura \etal \cite{sugiura2010cad}&Generalized the SM concept and proposed its differential counterpart.\\
    \hline
    2011&Ngo \etal \cite{ngo2011stfsk}&Applied the SM concept to the space-time-frequency domain.\\
    \cline{2-3}
    &Sugiura \etal \cite{sugiura2011gstsk}&Proposed a generalization concept for the SM scheme that subsumes conventional SM-related schemes within the STBC context.\\
    \hline
    2012&Yang and A{\'{i}}ssa \cite{yang2012smcapacity}&Derived the ergodic capacity for the GSM system.\\
    \hline
    2013&Rajashekar \etal \cite{rajashekar2013select}&Conceived a transmit antenna selection scheme for SM systems that maximized its MED of codewords or its capacity.\\
	\cline{2-3}
    &Rajashekar \etal \cite{rajashekar2014reduced}&Proposed a reduced-complexity detector for the SM scheme, where its complexity is free from the constellation size.\\
    \hline
    2014&Ishibashi and Sugiura \cite{ishibashi2014sm}&Clarified that the single-RF SM transmitter has to transmit each symbol during each symbol interval due to symbol-wise antenna switching. Hence, the bandwidth-efficient raised cosine filter is unavailable for the single-RF SM transmitter.\\
    \hline
    2015& Wu \etal \cite{wu2016sec}&Proposed a precoding-aided spatial modulation system for secret communications, which reduced the detection complexity at the receiver.\\
	\cline{2-3}
    &Basnayaka \etal \cite{basnayaka2015massive}&Proved that the SM scheme is effective for large-scale MIMO scenarios in terms of its ergodic capacity.\\
    \hline
    2017& Wang and Zhang \cite{wang2017huff}&Proposed a Huffman coding based adaptive spatial modulation, where the transmitter was assumed to have perfect channel estimates. The antenna activation probability was determined so as to maximize its capacity. \\
    \hline
  \end{tabularx}
\end{table*}
The contributions to the PM-based coherent MIMO schemes are summarized in Table~\ref{table:survey-cohmimo}.
\begin{figure}
  \centering
  \subfigure[SM \cite{chau2001sm}]{
    \includegraphics[clip, scale=0.53]{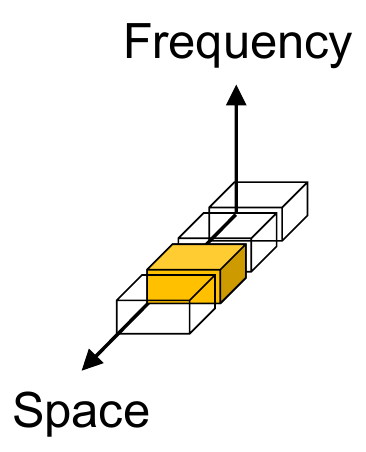} \label{fig:pmcoh-sm}
  }
  \subfigure[GSM \cite{jeganathan2008generalized}]{
    \includegraphics[clip, scale=0.53]{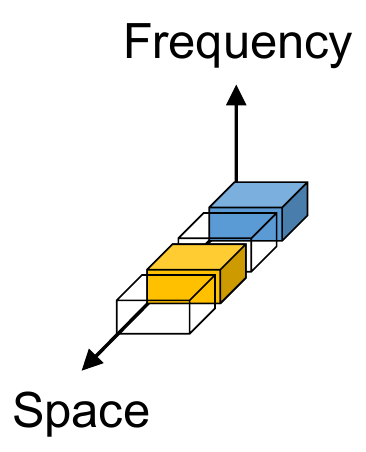} \label{fig:pmcoh-gsm}
  }
  \subfigure[ASTSK \cite{sugiura2012universal}]{ 
    \includegraphics[clip, scale=0.53]{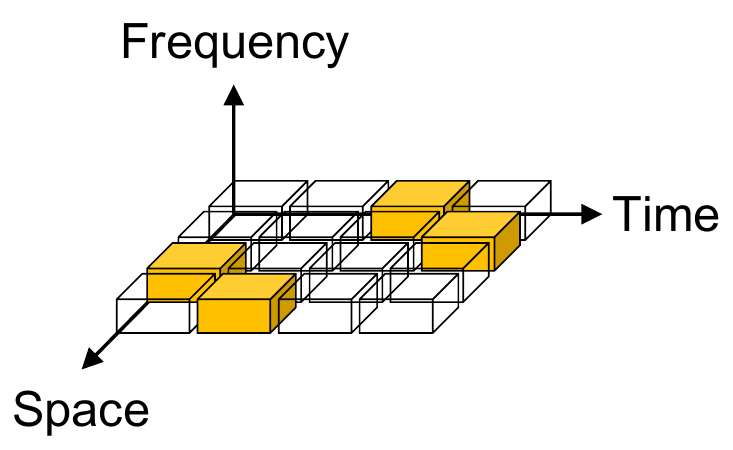} \label{fig:pmcoh-astsk}
  }
  \caption{PM-based coherent MIMO schemes, where we have $M=4$ transmit antennas. \label{fig:pm-coherent-schemes}}
\end{figure}
Fig.~\ref{fig:pm-coherent-schemes} shows the codewords of the SM-related schemes, which are discussed in this subsection.
As shown in Fig.~\ref{fig:pm-coherent-schemes}, each of the codewords has zero and non-zero symbols based on the on-off structure of Eq.~\eqref{eq:intr:combination-matrix}.

In the following, we introduce the \abbr{generalized space-time shift keying}{Generalized Space-Time Shift Keying}{GSTSK} scheme, which is capable of representing conventional MIMO schemes, such as the SM, SSK, GSM, and BLAST schemes, with the aid of its flexible \abbr{dispersion matrix}{Dispersion Matrix}{DM} architecture \cite{sugiura2011gstsk}.
The GSTSK framework allows us to analyze STBC-based MIMO encoding schemes in a comprehensive manner.
In advance of signal transmissions, the GSTSK scheme requires carefully designed DMs $\A_q \in \mathbb{C}^{M \times T}~(1\leq q \leq Q)$, which are obtained off-line.
The DMs are designed to maximize the specific criterion considered, such as the constrained \abbr{average mutual information}{Average Mutual Information}{AMI} of Section~\ref{sec:metrics:ami} as well as the rank and determinant criterion of Section~\ref{sec:metrics:rdc}.
A systematic DM construction method was proposed in \cite{rajashekar2013structured}.
Each DM $\A_q$ has the following energy constraint:
\begin{align}
  \mathrm{tr}\left[ \A_q \A_q^H \right] = \frac{T}{P}~(1 \leq q \leq Q).
  \label{eq:GSTSK:powerc}
\end{align}
The additional information bits are allocated by selecting $P$ DMs out of the set of $Q$ DMs ${\A_1,\cdots,\A_Q}$.
We represent the number of DM selection patterns as $\Na=2^{\left\lfloor \log_2 {Q \choose P} \right\rfloor}$.
The selected DM indices are denoted by $\a_k \in \mathrm{Z}^{P}$ for $1\leq k \leq \Na$, as determined by the on-off combination matrix of $\C_{Q,P}$.
The vector $\a_k$ consists of the $P$ number of sorted integers ranging from 1 to $Q$; these integers represent the activated DM indices.
For example, $\a_1 = [1,2]$ implies that the first and second DMs $\A_1,\A_2$ are activated.
The \abbr{natural binary code}{Natural Binary Code}{NBC} \cite{frenger1999pcofdm} maps $\a_k$ to the $k$th row of $\C_{Q,P}$.
By contrast, the \abbr{look-up table}{Look-Up Table}{LUT} method \cite{frenger1999pcofdm,basar2013sim} maps $\a_k$ to the manually selected row of $\C_{Q,P}$.

The $B=B_1+B_2$ input bits are partitioned into two sequences: $B_1= \log_2(\Na)$ bits and $B_2 = P \log_2(\L)$ bits, where $\L$ denotes a constellation size.
Based on the first $B_1$ bits, the $k$th index vector of $\a_k$ is selected out of the $\Na$ combinations, i.e., $1 \leq k \leq \Na$.
Then, based on the second $B_2$ bits, the $P$ number of APSK symbols $s_1,\cdots,s_P \in \mathbb{C}$ are generated.
Finally, the space-time codeword of the GSTSK scheme is generated by
\begin{align}
  \S = \sum^P_{p=1}s_{p}\A_{\a_k(p)}.
  \label{GSTSK:eq:generateS}
\end{align}
The bit per channel-use throughput of the GSTSK scheme is given by
\begin{align}
  R = \frac{B}{T} = \frac{B_1 + B_2}{T} = \frac{\left\lfloor \log_2 {Q \choose P} \right\rfloor + P \log_2 \L }{T} \ \ \mathrm{[bits/symbol]}.
  \label{GSTSK:eq:rate}
\end{align}
In this treatise, we use the notation of GSTSK($M,N,T,Q,P$) for simplicity.

\begin{figure}[tbp]
  \centering
  \includegraphics*[scale=0.43]{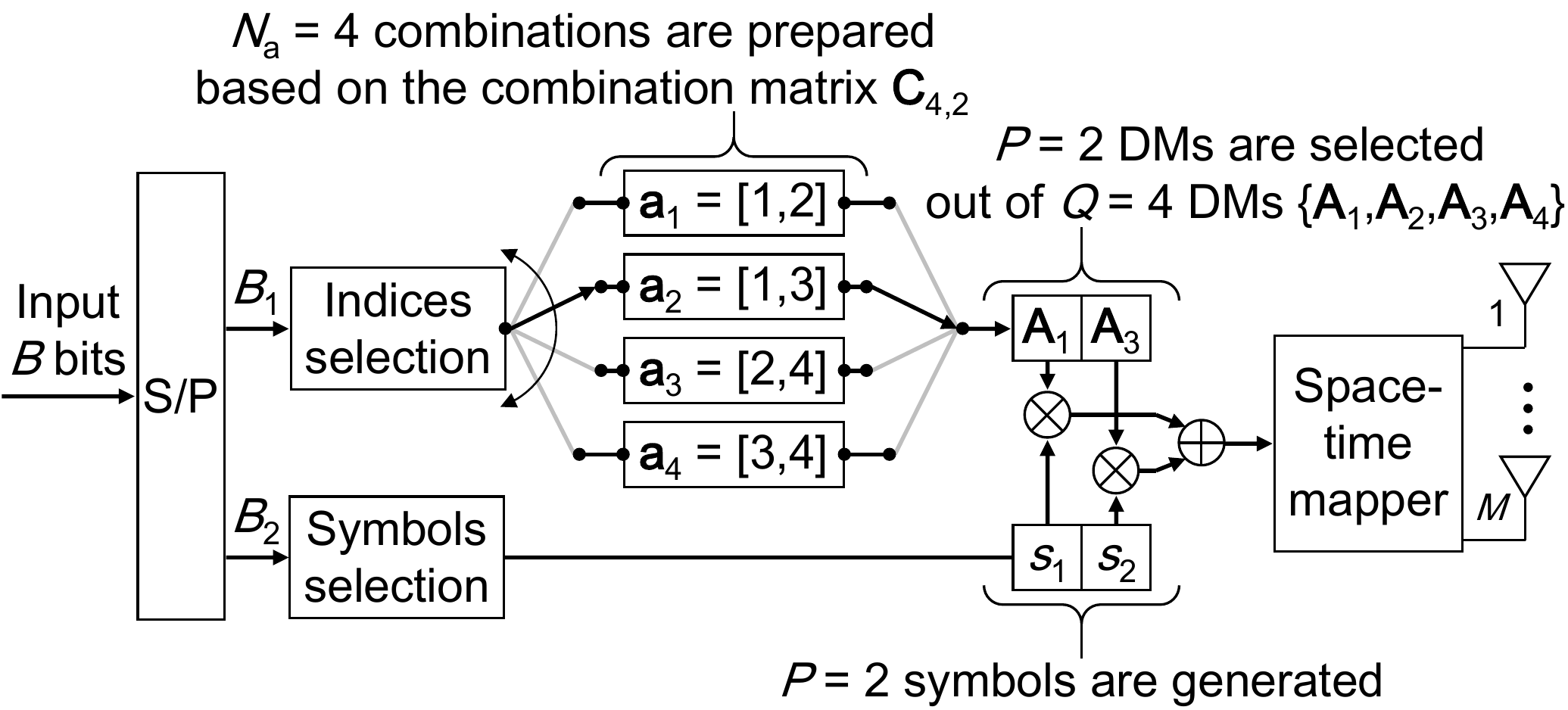}%
  \caption{Transmitter example of the GSTSK scheme for $Q=4$ and $P=2$, where LUT-based DM activation is considered. By changing the DMs, the GSTSK scheme becomes equivalent to BLAST, SM, GSM and ASTSK. \label{fig:GSTSK:structure}}
\end{figure}
Let us examine a detailed example.
Fig.~\ref{fig:GSTSK:structure} shows the transmitter example of the GSTSK scheme, where $P=2$ DMs are selected out of $Q=4$ DMs $\A_1, \cdots ,\A_4 \in \mathbb{C}^{M \times T}$.
Here, we have $\Na=2^{\left\lfloor \log_2 {4 \choose 2} \right\rfloor} = 2^2 = 4$ number of DM-activation patterns $\a_1,\cdots,\a_4 \in \mathbb{Z}^P$.
The combination matrix $\C_{Q,P} = \C_{4,2}$ is given by Eq.~\eqref{eq:combination-matrix-42} as follows:
\begin{align}
  \C_{4,2} =
  \left[
  \begin{array}{cccc}
	1&1&0&0\\
    1&0&1&0\\
    1&0&0&1\\
    0&1&1&0\\
    0&1&0&1\\
    0&0&1&1\\
  \end{array}
  \right]
  \begin{array}{l}
    \rightarrow \a_1=[1,2]\\
    \rightarrow \a_2=[1,3]\\
    \\
    \\
    \rightarrow \a_3=[2,4]\\
    \rightarrow \a_4=[3,4]\\
  \end{array}
  \label{eq:combination-matrix-42-inds}
\end{align}
Here, if we use the NBC method \cite{frenger1999pcofdm}, the DM-activation vectors are given by $\a_1=[1,2],~\a_2=[1,3],~\a_3=[1,4]$, and $\a_4=[2,3]$ based on the first, second, third, and fourth rows of $\C_{4,2}$.
By contrast, if we use the LUT method, the DM-activation vectors are given by $\a_1=[1,2],~\a_2=[1,3],~\a_3=[2,4]$, and $\a_4=[3,4]$ based on the first, second, fifth, and sixth rows of $\C_{4,2}$.
These activation patterns affect the achievable performance.
Specifically, the coding gain is maximized when each index is selected with equal probability \cite{wen2016equiprobable}.
Some beneficial LUT construction algorithms are detailed in \cite{savage1997combinatorial,wen2016equiprobable}.

In the following, we introduce the conventional schemes, including SM, SSK, GSM, \abbr{generalized space shift keying}{Generalized Space Shift Keying}{GSSK} and \abbr{asynchronous space-time shift keying}{Asynchronous Space-Time Shift Keying}{ASTSK}, by invoking the GSTSK$(M,N,T,Q,P)$ framework.

\paragraph*{SM/SSK \cite{chau2001sm,mesleh2008spatial,jeganathan2008spatial}}
The SM scheme is a member of the PM family, where the PM codeword is expanded into the spatial axis.
The SM scheme activates a single antenna out of multiple transmit antennas at any transmission time instant.
Similarly, the SSK scheme is a special form of SM, where no modulation constellation is used.
The SM encoding principle is represented by the GSTSK$(M,N,1,M,1)$ having the following DMs $\A_m \in \mathbb{C}^{M \times 1}~(1 \leq m \leq M)$:
\begin{align}
  \{\A_1,\A_2,\cdots,\A_M \}
  =\left\{
     \begin{bmatrix} 1 \\ 0 \\ \vdots \\ 0 \end{bmatrix},~
  \begin{bmatrix} 0 \\ 1 \\ \vdots \\ 0 \end{bmatrix},~\cdots,~
  \begin{bmatrix} 0 \\ 0 \\ \vdots \\ 1 \end{bmatrix}
     \right\},
  \label{eq:GSTSK:sm-dm}
\end{align}
where each DM is also calculated by the combination matrix $\C_{M,1} = \I_M$.
The $m$th index vector $\a_m$ is defined by the length-one vector of $\a_m = [m] \in \mathbb{Z}^1$.
Finally, the SM codeword is given by
\begin{align*}
  \S = s_1 \A_{\a_m(1)} = s_1 \A_m = [0 ~ \cdots ~ 0 ~ \underbrace{s}_{m\mathrm{th~row}} ~ 0 ~ \cdots ~ 0]^\Trm.
\end{align*}

\paragraph*{GSM/GSSK \cite{jeganathan2008generalized,younis2010gsm}}
The GSM and GSSK schemes are extensions of the SM and SSK schemes, where an arbitrary number of transmit antennas are activated simultaneously \cite{jeganathan2008generalized,younis2010gsm}.
The GSM encoding principle is represented by GSTSK$(M,N,1,M,P)$, where we have $Q=M$ number of DMs given by Eq.~\eqref{eq:GSTSK:sm-dm} divided by $\sqrt{P}$ and $\Na=2^{\left\lfloor \log_2 {M \choose P} \right\rfloor}$ number of DM-activation patterns $\a_1,\cdots,\a_{\Na} \in \mathbb{Z}^P$ defined by the combination matrix $\C_{M,P}$.
In the rest of this paper, we use the notation of GSM($M,P$), where $M$ is the number of transmit antennas and $P$ is the number of activated antennas.
Note that the GSM($M,P$) scheme having $\L=1$ is equivalent to the GSSK scheme.
Table \ref{table:gsm-mapping} shows the bit mapping example for the LUT method.
\begin{table}[tbp]
  \centering
  \small
  \caption{Bit mapping example of the BPSK-aided  GSM(4,2).}
  \label{table:gsm-mapping}
  \begin{tabular}{|r:l||c|c|c|}
    \hline
    \multicolumn{2}{|c||}{Source}  & Indices & Symbols & GSM \\
    \multicolumn{2}{|c||}{(4 bits)} & $\a_i$ & $s_1,s_2$ & codeword $\S$ \\
    \hline \hline
    $0~0$ & $0~0$ & $\a_1=[1,2]$ &  $+1,+1$ & $[+1,+1,~~0,~~0]^\Trm/\sqrt{2}$ \\
    $0~0$ & $0~1$ & $\a_1=[1,2]$ &  $+1,-1$ & $[+1,-1,~~0,~~0]^\Trm/\sqrt{2}$ \\
    $0~0$ & $1~0$ & $\a_1=[1,2]$ &  $-1,+1$ & $[-1,+1,~~0,~~0]^\Trm/\sqrt{2}$ \\
    $0~0$ & $1~1$ & $\a_1=[1,2]$ &  $-1,-1$ & $[-1,-1,~~0,~~0]^\Trm/\sqrt{2}$ \\
    $0~1$ & $0~0$ & $\a_2=[1,3]$ &  $+1,+1$ & $[+1,~~0,+1,~~0]^\Trm/\sqrt{2}$ \\
    $0~1$ & $0~1$ & $\a_2=[1,3]$ &  $+1,-1$ & $[+1,~~0,-1,~~0]^\Trm/\sqrt{2}$ \\
    $0~1$ & $1~0$ & $\a_2=[1,3]$ &  $-1,+1$ & $[-1,~~0,+1,~~0]^\Trm/\sqrt{2}$ \\
    $0~1$ & $1~1$ & $\a_2=[1,3]$ &  $-1,-1$ & $[-1,~~0,-1,~~0]^\Trm/\sqrt{2}$ \\
    $1~0$ & $0~0$ & $\a_3=[2,4]$ &  $+1,+1$ & $[~~0,+1,~~0,+1]^\Trm/\sqrt{2}$ \\
    $1~0$ & $0~1$ & $\a_3=[2,4]$ &  $+1,-1$ & $[~~0,+1,~~0,-1]^\Trm/\sqrt{2}$ \\
    $1~0$ & $1~0$ & $\a_3=[2,4]$ &  $-1,+1$ & $[~~0,-1,~~0,+1]^\Trm/\sqrt{2}$ \\
    $1~0$ & $1~1$ & $\a_3=[2,4]$ &  $-1,-1$ & $[~~0,-1,~~0,-1]^\Trm/\sqrt{2}$ \\
    $1~1$ & $0~0$ & $\a_4=[3,4]$ &  $+1,+1$ & $[~~0,~~0,+1,+1]^\Trm/\sqrt{2}$ \\
    $1~1$ & $0~1$ & $\a_4=[3,4]$ &  $+1,-1$ & $[~~0,~~0,+1,-1]^\Trm/\sqrt{2}$ \\
    $1~1$ & $1~0$ & $\a_4=[3,4]$ &  $-1,+1$ & $[~~0,~~0,-1,+1]^\Trm/\sqrt{2}$ \\
    $1~1$ & $1~1$ & $\a_4=[3,4]$ &  $-1,-1$ & $[~~0,~~0,-1,-1]^\Trm/\sqrt{2}$ \\
    \hline
  \end{tabular}
\end{table}
Note that GSM($M,M$) is equivalent to the conventional BLAST scheme, where $M$ number of independent symbols are embedded in a codeword.

In 2013, Khandani proposed a media-based modulation (MBM) concept \cite{khandani2013mbm,khandani2014mbm}, which conveys data by changing radio propagation.
In a theoretical system model, the MBM scheme is similar to the SSK signaling.
The key contribution of MBM is the higher capacity achieved by RF mirrors.
As we reviewed in this section, the transmission rate $R$ of the SSK scheme is limited by the number of transmit antennas.
In contrast, the transmission rate of the MBM scheme increases with the number of scattering patterns, where its capacity is equivalent to the AWGN channel while assuming the Rayleigh fading channel.
Motivated by this attractive nature, the MBM has gained attention in the wireless community \cite{khandani2013mbm,khandani2014mbm,naresh2017mbm,seifi2016mbm,yapeng2016mbm,bandi2017mbm,yildirim2017mbm,basar2017mbm}.

\paragraph*{ASTSK \cite{sugiura2012universal}}
The ASTSK scheme is an extension of the SM scheme. In the ASTSK scheme, the number of symbol intervals per block is increased to $T \geq 2$ \cite{sugiura2012universal}, which is represented by GSTSK$(M,N,T,Q,1)$.
Each DM $\A_q \in \mathbb{C}^{M \times T}~(1 \leq q \leq Q)$ has a single non-zero element in its column as well as row, and has the constraint of $\mathrm{rank}(\A_q)=\min(M,T)$.
For example, if we consider the $(M,T,Q)=(3,3,4)$ case, the DMs are given by
\begin{align*}
  &\{\mathbf{A}_1,\mathbf{A}_2,\mathbf{A}_3,\mathbf{A}_4 \}\\
  = &\left\{
  \begin{bmatrix}
    a_{11}&0&0\\
    0&a_{12}&0\\
    0&0&a_{13}
  \end{bmatrix},
  \begin{bmatrix}
    a_{21}&0&0\\
    0&0&a_{23}\\
    0&a_{22}&0
  \end{bmatrix},\right. \\
  &~~\left.
  \begin{bmatrix}
    0&a_{32}&0\\
    a_{31}&0&0\\
    0&0&a_{33}
  \end{bmatrix},
  \begin{bmatrix}
    0&0&a_{43}\\
    0&a_{42}&0\\
    a_{41}&0&0
  \end{bmatrix}
  \right\},
\end{align*}
where $a_{qm}$ represents a complex value.
Furthermore, for the $(M,T,Q)=(4,2,4)$ case, we have
\begin{align*}
  &\{\mathbf{A}_1,\mathbf{A}_2,\mathbf{A}_3,\mathbf{A}_4 \}\\
  = &\left\{
  \begin{bmatrix}
    a_{11}	&	0 \\
    0	&	a_{12} \\
    0	&	0\\
    0	&	0
  \end{bmatrix},
  \begin{bmatrix}
    0 &	a_{22}\\
    a_{21}	&	0\\
    0	&	0\\
    0	&	0
  \end{bmatrix},
  \begin{bmatrix}
    0&0\\
    0&0\\
    a_{31}&0\\
    0&a_{32}
  \end{bmatrix},
  \begin{bmatrix}
    0&0\\
    0&0\\
    0&a_{42}\\
    a_{41}&0
  \end{bmatrix}
  \right\}.
\end{align*}
Because the $P=1$ modulated APSK symbol is spread over $T$ time slots, the ASTSK scheme at most achieves the diversity order of $T$.

\subsection{PM-Based Differential MIMO\label{sec:app:diff}}
In Section~\ref{sec:app:coh}, we reviewed the PM schemes proposed for coherent MIMO systems, which require accurate estimates of the channel matrix $\H$ at the receiver.
In this section, we continue by reviewing the differentially encoded and non-coherently detected counterparts of the coherent PM schemes, which dispense with the channel estimation overhead.
\begin{table*}[t]
  \centering
  \small
  \caption{Contributions to the PM-based differential MIMO schemes.\label{table:survey-diffmimo}}
  \begin{tabularx}{\linewidth}{|c|l|X|}
    \hline
    Year & Authors & Contribution \\
    \hline
    \hline
    2000& Hughes \cite{hughes2000dstm} & Proposed a differential MIMO scheme relying on diagonal and anti-diagonal matrices to support an arbitrary number of transmit antennas. If we limit the matrix ``$D$'' in \cite{hughes2000dstm} to the identity matrix, the space-time codewords result is sparse, which was not clearly mentioned.\\
    \cline{2-3} &Hochwald and Sweldens \cite{hochwald2000dstm} &Proposed a differential MIMO scheme relying on diagonal matrices to support an arbitrary number of transmit antennas. The proposed scheme achieves full diversity and enables single-RF transmission. \\
    \hline
    2007&Oggier \cite{oggier2007cyclic} &Proposed a permutation-matrix-based differential MIMO scheme relying on cyclic division algebra. This scheme has similar advantages to that of \cite{hochwald2000dstm}.\\
    \hline
    2010&Sugiura \etal \cite{sugiura2010cad}&Proposed a differential counterpart of the coherent \abbr{space-time shift keying}{Space-Time Shift Keying}{STSK} scheme. This scheme conveys information bits by selecting a single out of multiple DMs. The encoding principle is based on the differential \abbr{linear dispersion code}{Linear Dispersion Code}{LDC} scheme of \cite{hassibi2002cdu}.\\
     \hline
    2011&Sugiura \etal \cite{sugiura2011rcc}&Proposed a differential counterpart of the SM scheme, based on the unitary matrices, which subsumes most of the unitary-matrix-based DSM schemes~\cite{bian2013dsm,ishikawa2014udsm,rajashekar2017dsm} shown below.\\
    \hline
    2013&Bian \etal \cite{bian2013dsm}&Designed the space-time codewords of the DSM scheme, which are generated from the diagonal and anti-diagonal matrices.\\ %
    \hline
    2014&Wen \etal \cite{wen2014berdsm}&Derived a tight BER bound for the DSM scheme having $M=2$.\\
    \cline{2-3}
    &Ishikawa and Sugiura \cite{ishikawa2014udsm}&Proposed a DM-based counterpart of \cite{bian2013dsm} with the aim of striking the tradeoff between diversity and rate.\\
    \cline{2-3}
    &Bian \etal \cite{bian2015dsm}&Proposed a generalized DSM scheme based on \cite{bian2013dsm} to support an arbitrary number of transmit antennas. The space-time codewords are generated from permutation matrices.\\ 
    \hline
    2015&Wen \etal \cite{wen2015dsm}&Proposed a low-complexity detector for the DSM scheme of \cite{bian2013dsm}. The proposed scheme with this detector achieved near-optimal performance at high SNRs.\\ 
    \hline
    2016&Rajashekar \etal \cite{rajashekar2017dsm} & Proposed a field-extension-based DSM scheme, which alleviated the DM optimization problem of \cite{ishikawa2014udsm}. The proposed scheme can adjust the diversity and rate tradeoff.\\
    \cline{2-3}
    &Li \etal \cite{li2016dsm}&Proposed a general method to determine the non-zero positions in the DMs of the DSM scheme. This method adopted Trotter-Johnson ranking and unranking algorithms.\\ 
    \cline{2-3}
    &Zhang \etal \cite{zhang2016dsm} &Applied the precoding-aided SM and DSM schemes to dual-hop virtual-MIMO relaying networks. They proposed two low-complexity detectors.\\
    \hline
    2017&Ishikawa and Sugiura \cite{ishikawa2017rdsm} & Proposed a rectangular-matrix-based DSM concept, which can support the massive MIMO, e.g., the scenario of $M=1024$ antennas. The transmission rate linearly increases as the number of transmit antennas increases.\\
    \cline{2-3}
    &Xiao \etal \cite{xiao2017stdsm}& Combined the DSM scheme with the space-time block coded SM \cite{basar2011stsm} to replace non-zero elements in SM symbols with OSTBCs. The proposed scheme was applied to large-scale MIMO scenarios by reducing the detection complexity at the receiver. The authors assumed $M=32$ transmit antennas as a maximum with the corresponding transmission rate $R=2.62$ [bits/symbol]. Note that this scheme is different from that of \cite{ishikawa2017rdsm} because it depends on square matrices instead of rectangular matrices.\\
    \cline{2-3}
    &Rajashekar \etal \cite{rajashekar2017afedsm}& Proposed an enhanced DSM scheme based on \cite{rajashekar2017dsm}, which is capable of avoiding the issue raised in \cite{xu2018isk}. Two novel buffer-based low-complexity detectors were conceived, where the successive codewords were used to improve the error rate. The proposed schemes were shown to achieve the near-coherent performance.  \\ 
    \cline{2-3}
    &Xu \etal \cite{xu2018isk}& Raised a novel issue concerning the cardinality of the resultant constellation after differential encoding, and proposed a solution for this issue. The conventional differential MIMO may result in an infinite cardinality of constellation, which requires high-resolution analog-to-digital converters. Additionally, the constrained AMI for arbitrary differential MIMO was firstly derived. The proposed scheme was shown to achieve a high diversity order at a reduced complexity.\\ 
    \hline
  \end{tabularx}
\end{table*}
The major contributions to the PM-based differential MIMO schemes are summarized in Table~\ref{table:survey-diffmimo}.
We introduce two types of unitary matrix construction methods: the permutation-matrix-based method and the Cayley-transform-based method.

\subsubsection{Differential Spatial Modulation\label{sec:app:diff:dsm}}
Motivated by the SM concept \cite{chau2001sm,mesleh2008spatial}, the differential counterpart of the SM scheme was proposed \cite{sugiura2010cad,sugiura2011rcc}, which includes the so-called \abbr{differential spatial modulation}{Differential Spatial Modulation}{DSM} family~\cite{bian2013dsm,bian2015dsm}.
The DSM scheme was generalized to strike a diversity vs multiplexing gain tradeoff \cite{sugiura2012smchannel,ishikawa2014udsm,rajashekar2017dsm}, and later it was extended to support the large-scale MIMO system concept \cite{ishikawa2017rdsm}.
The space-time codeword of the DSM scheme has a single non-zero element in its column and row.
Hence, the DSM scheme is capable of enabling single-RF operation, as well as dispensing with the channel estimation overhead.
Furthermore, because the number of non-zero elements in each column is limited in the DSM codewords, the transmitter complexity can be further improved by limiting the phase of non-zero elements \cite{xu2018isk,xu2018tast,rajashekar2017afedsm}.
Note that the concept of sparse space-time codewords was proposed in \cite{oggier2007cyclic,sugiura2012universal} before the invention of the DSM concept.

Let us review the DSM encoding principle of \cite{ishikawa2014udsm}.
The DSM transmitter maps an input bit sequence of length $B$ onto an output space-time matrix $\S(i)$, where $i$ represents the transmission index.
In advance of the transmissions, $Q$ number of DMs $\A_q \in \mathbb{C}^{M\times M} \ (q=1,\cdots,Q)$ have to be prepared.
Each DM $\A_q$ has a single non-zero-unit-absolute-value element  in its column and row.
Here, we represent the non-zero element as $a_{q,m}~(1\le q\le Q,~ 1\le m\le M)$, where $q$ denotes the DM index and $m$ denotes the activated antenna index.
The norm of the non-zero element is constrained to be $|a_{q,m}|=1$ to maintain the unitary constraint.
The following examples are $Q=2$ DMs for the $M=2$ transmit antenna scenario:
\begin{align}
  &\left\{  \A_{1}, \A_{2} \right\} \nonumber\\
  =& \left\{
  \begin{bmatrix}
    e^{-j0.82\pi} & 0.00 \\
    0.00 & e^{+j0.42\pi}
  \end{bmatrix},
  \begin{bmatrix}
    0.00 & e^{-j0.01\pi} \\
    e^{+j0.10\pi} & 0.00
  \end{bmatrix}
  \right\}.
  \label{diff:eq:dm-M2}
\end{align}
The $2 \times 2$ permutation matrices are multiplied by complex-valued phase shifters.
As seen in Eq.~\eqref{diff:eq:dm-M2}, the norm of each non-zero element is constrained to be 1.
Hence, each DM $\A_1, \cdots, \A_Q$ is kept as a unitary matrix.

\begin{figure}[tb]
  \centering
  \includegraphics[clip,scale=0.43]{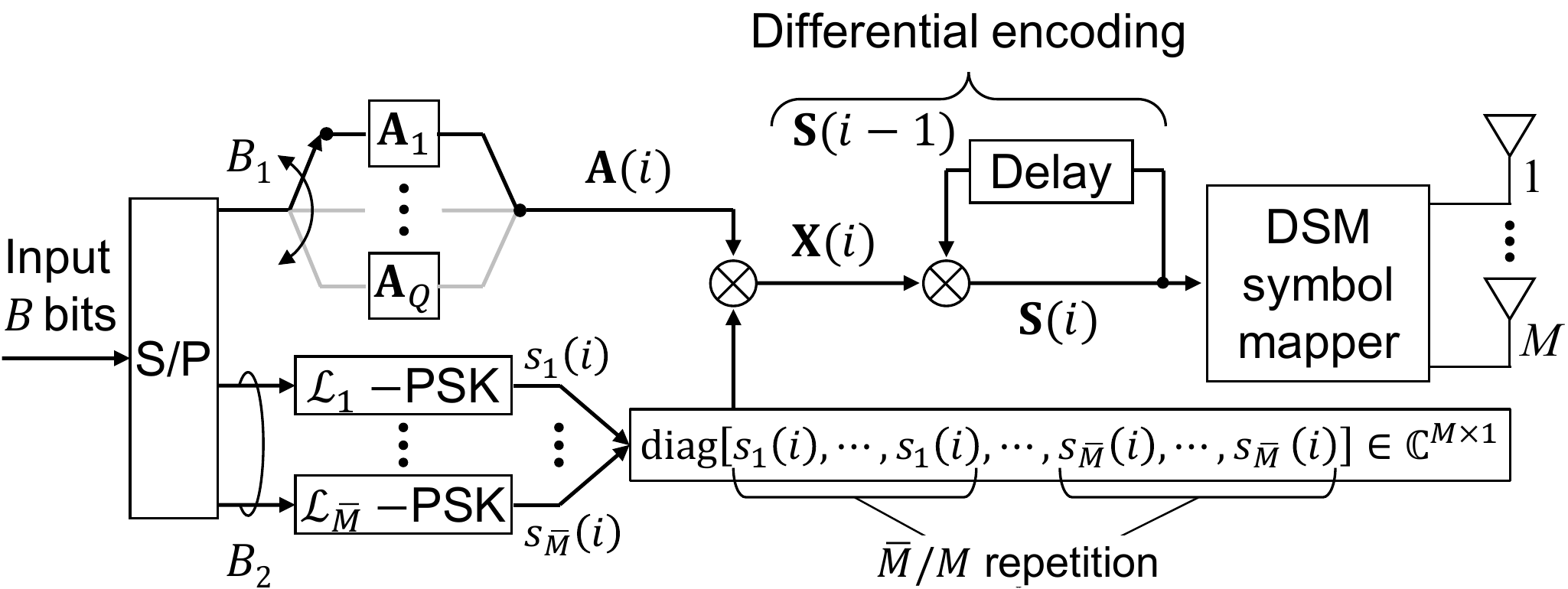}
  \caption{Schematic of the DSM transmitter. \copyright IEEE \cite{ishikawa2014udsm}\label{diff:fig:system}}%
\end{figure}
Fig.~\ref{diff:fig:system} shows the transmitter structure of the DSM scheme.
In Fig.~\ref{diff:fig:system}, the input bits are S/P converted to $B_1=\log_2(Q)$ bits and $B_2 = \log_2(\L_1 \cdot \L_2 \cdot \cdots \cdot \L_{\Mb})$ bits, where each of $\L_1, \cdots, \L_{\Mb}$ represents the constellation size.
The first $B_1$ bits are used for selecting a DM $\A_q(i)$ out of $Q$ number of DMs.
The second $B_2$ bits are mapped to $\Mb$ number of PSK symbols $s_1(i), s_2(i),\cdots,s_{\Mb}(i)$, which are packed into an $M \times 1$ vector as follows:
\begin{align}
	\mathbf{s}(i) = [
	\underbrace{s_1(i), \cdots, s_1(i)}_{M/\Mb~\mathrm{repetition}}
	, 
	\cdots, 
	\underbrace{s_{\Mb}(i), \cdots, s_{\Mb}(i)}_{M/\Mb~\mathrm{repetition}}
	] \in \mathbb{C}^{M \times 1}.
\end{align}
Here, $\Mb$ for $1 \leq \Mb \leq M$ represents the number of PSK symbols embedded into a space-time matrix, which determines the rate vs diversity factor.
For example, if we embed $\Mb=M$ symbols, then the maximum diversity order is $M/\Mb=1$, while the transmission rate is maximized.
Hence, the DSM scheme strikes a flexible tradeoff between the transmission rate and the achievable diversity order, which is often referred to as diversity vs multiplexing tradeoff.
A unitary matrix $\X(i) \in \mathbb{C}^{M\times M}$ is calculated as follows:
\begin{align}
  \X(i) = \mathrm{diag}\[\mathbf{s}(i)\]\A_q(i),
  \label{eq:diff:unitary}
\end{align}
which is associated with the input $B$ bits.
In Eq.~\eqref{eq:diff:unitary}, diag$\[\cdot\]$ denotes the diagonal operation that maps a vector to a diagonal matrix.
The data matrix $\X(i)$ is a sparse matrix, where each column and row has a single non-zero element.
The norm of each non-zero element in $\X(i)$ is also constrained to be 1, which is similar to the DM construction of $|a_{q,m}|=1$.
Finally, a space-time codeword $\mathbf{S}(i) \in \mathbb{C}^{M\times M}$ is differentially encoded, as given in Section~\ref{sec:sys:diff}.
The normalized transmission rate is given by
\begin{align}
  R = \frac{B}{M} = \frac{\log_2(Q\cdot \L_1\cdots \L_{\bar{M}})}{M}.
  \label{diff:eq:rate_imp}
\end{align}
For example, if we consider the $M=4$ and $\Mb=2$ scenarios, the embedded $\Mb=2$ BPSK symbols are represented as follows:
\begin{align}
  \mathrm{diag} \left[ \mathbf{s}(i) \right]
= \mathrm{diag} \left[ s_1(i),s_1(i),s_2(i),s_2(i) \right]. 
  \label{diff:eq:few-symbols}
\end{align}
The constellations are denoted by $\mathbf{L} = [(\mathcal{L}_1,\mathcal{L}_1), (\mathcal{L}_2,\mathcal{L}_2)] = [(2,2), (2,2)]$.
In Eq.~\eqref{diff:eq:few-symbols}, the pair of two BPSK symbols $s_1(i)$ and $s_2(i)$ are embedded into a space-time codeword.
The configuration in Eq.~\eqref{diff:eq:few-symbols} achieves a diversity order of $D=2$ because each symbol is spread over two successive symbols' transmissions.

\begin{figure}[tb]
  \centering
  \includegraphics[clip,scale=0.58]{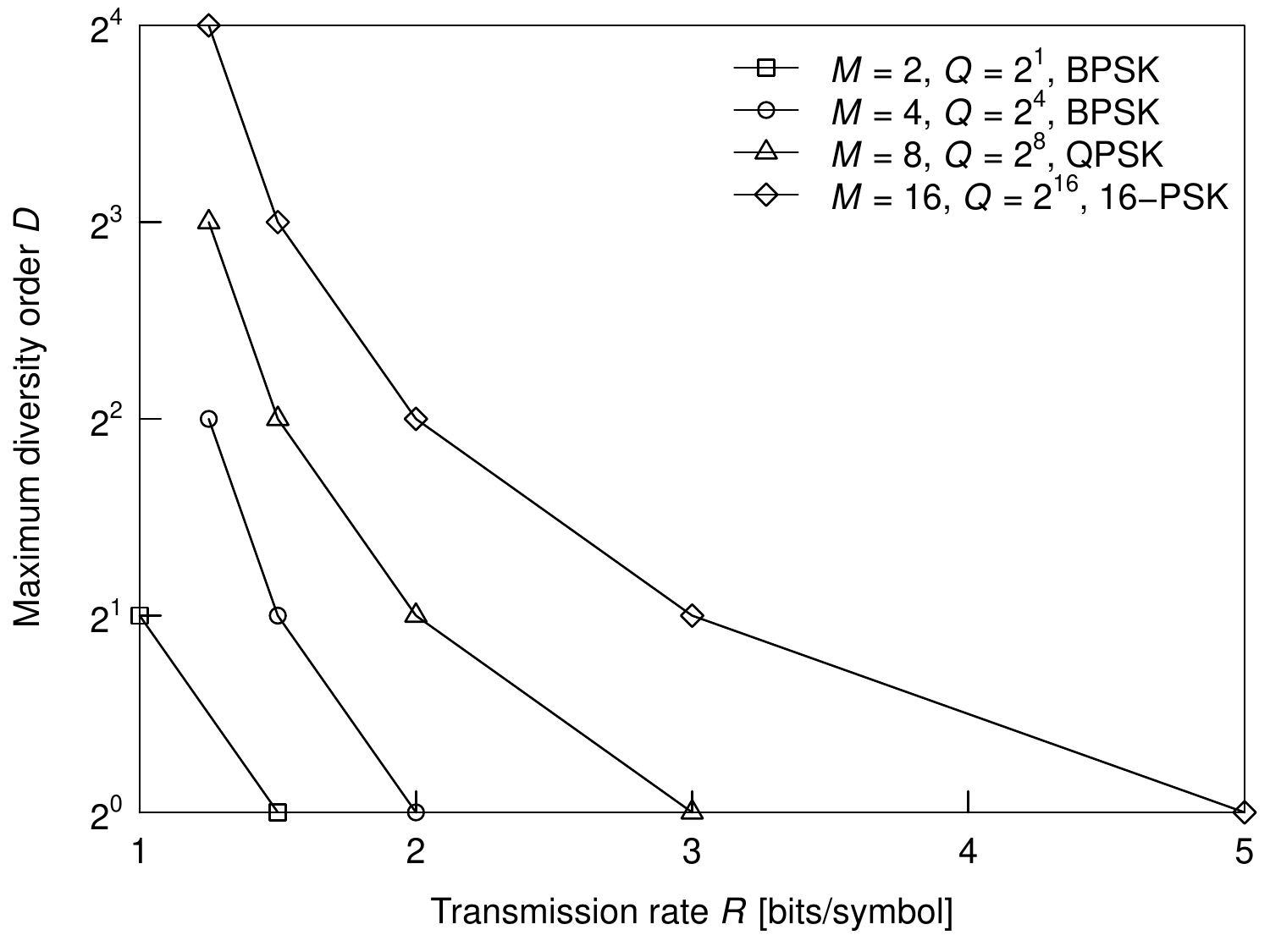}
  \caption{Tradeoff between transmission rate $R$ and maximum diversity order $D$ of the scheme of Fig.~\ref{diff:fig:system}, where the number of transmit antennas was $M=2,4,8,16$.\label{diff:fig:dmt}}%
\end{figure}
Fig.~\ref{diff:fig:dmt} shows the flexible rate vs diversity tradeoff of the DSM scheme.
Here, the setups of $\left(M, Q\right) = \left(2,2^1\right), \left(4,2^4\right), \left(8,2^8\right), \left(16,2^{16}\right)$ were considered.
The number of embedded symbols was in the range of $\bar{M}=2^0, 2^1, \cdots , 2^{\log_2(M)}$.
As shown in Fig.~\ref{diff:fig:dmt}, the maximum diversity order $D$ is reduced upon increasing the transmission rate $R$.

The DSM architecture of \cite{ishikawa2014udsm} subsumes the DSTSK  \cite{sugiura2010cad} and the \abbr{binary differential spatial modulation}{Binary Differential Spatial Modulation}{BDSM} \cite{bian2015dsm} schemes.
It was also be readily shown that a specific form of the DSM scheme is equivalent to DSTSK.
Specifically, the DSTSK modulation process embeds a single complex-valued symbol into a space-time codeword.
This configuration is equivalent to the DSM scheme having $\Mb=1$.
Here, the DSM scheme has no limitation in terms of $\Mb$ and therefore achieves a flexible rate vs diversity tradeoff.
Furthermore, the DSM scheme may be considered as a generalization of the conventional BDSM scheme proposed in \cite{bian2015dsm}.
The DMs of the BDSM have non-zero elements of one, formulated as $a_{q,m} = 1~(1 \leq q \leq Q,~ 1 \leq m \leq M)$.
Due to the $a_{q,m} = 1$ limitation, the number of DMs $Q$ is limited to $2^{\lfloor \log_2(M!) \rfloor}$, where we have $M! = M \cdot (M-1) \cdots 1$.
For example, if we consider the $M=3$ case, the number of DMs is defined by $Q = 2^{\lfloor \log_2(3!) \rfloor} = 2^{\lfloor 2.58... \rfloor} = 2^2 = 4$ and the DMs are given as follows:
\begin{align*}
  &\left\{ \A_1,\A_2,\A_3,\A_4  \right\} \\
  =&\left\{
     \begin{bmatrix}
       1	&	0	&	0	\\
       0	&	1	&	0	\\
       0	&	0	&	1
     \end{bmatrix},
                \begin{bmatrix}
                  1	&	0	&	0	\\
                  0	&	0	&	1	\\
                  0	&	1	&	0
                \end{bmatrix},
                          \begin{bmatrix}
                            0	&	1	&	0	\\
                            1	&	0	&	0	\\
                            0	&	0	&	1
                          \end{bmatrix},
                                    \begin{bmatrix}
                                      0	&	0	&	1	\\
                                      0	&	1	&	0	\\
                                      1	&	0	&	0
                                    \end{bmatrix}
                                              \right\}.
\end{align*}
Furthermore, in the BDSM codewords, the number of embedded symbols $\Mb$ is limited to $M$.
This limitation imposes the diversity order of $D=1$.
The DSM scheme supports the $Q > 2^{\lfloor \log_2(M!) \rfloor}$ case, because $a_{q,m}$ is a complex value.
As an example, we consider the DSM scheme having two DMs, $\A_1$ and $\A_2$, where each DM has the same positions of non-zero elements.
At the receiver, the scheme can differentiate the pair of DMs, if the phases of the non-zero elements are different, as follows:
\begin{align*}
  \left\{ \A_1, \A_2 \right\} =
\left\{
  \begin{bmatrix}
    e^{\frac{\pi}{2}j}& 0 &	0 \\
    0& 0 &	e^{\frac{\pi}{6}j}	\\
    0& e^{\frac{\pi}{3}j} &	0
  \end{bmatrix},
                            \begin{bmatrix}
                              e^{\frac{\pi}{4} j} & 0 &	0\\
                              0 & 0 & e^{\frac{\pi}{2} j}\\
                              0 & e^{\pi j} &	0
                          \end{bmatrix}
                                                    \right\}.
\end{align*}

\subsubsection{Non-Coherent Generalized Spatial Modulation\label{sec:app:diff:ncgsm}}
The \abbr{non-coherent generalized spatial modulation}{Non-Coherent Generalized Spatial Modulation}{NCGSM} \cite{ishikawa2014ncgsm} is the differential counterpart of the GSTSK scheme described in Section~\ref{sec:app:coh}.
The encoding principle of the NCGSM scheme is basically the same as that of the GSTSK scheme, with the following two exceptions:
\begin{itemize}
\item The embedded symbols $s_1, \cdots, s_P$ have to be real-valued, as is \abbr{pulse amplitude modulation}{Pulse Amplitude Modulation}{PAM}.
\item The DMs $\A_1, \cdots, \A_Q \in \mathbb{C}^{M \times M}$ are Hermitian matrices. Thus, each DM satisfies $\A_q = \A_q^\Hrm$.
\end{itemize}
The GSTSK space-time codewords $\S(i)$ are generated by the summation of the DMs, as given in Eq.~\eqref{GSTSK:eq:generateS}.
Similarly, the NCGSM space-time codewords $\Xt(i) \in \mathbb{C}^{M \times M}$ are generated as follows, when the $k$th DM-activation vector is selected:
\begin{align}
  \Xt(i) = \sum^P_{p=1}s_{p}\A_{\a_k(p)}.
  \label{NCGSM:eq:generateXt}
\end{align}
Because the DMs $\A_{\a_k(p)}$ are Hermitian and the PAM symbols $s_p$ are real-valued,  the summation of Eq.~\eqref{NCGSM:eq:generateXt} results in a Hermitian matrix.
Finally, a unitary matrix $\X(i) \in \mathbb{C}^{M \times M}$ is calculated by
\begin{align}
  \X(i)=\left(\I_M-j \tilde{\X}(i) \right) \left(\I_M+j \tilde{\X}(i) \right)^{-1} \equiv \mathrm{\zeta}\(\Xt(i)\),
  \label{NCGSM:eq:cayley}
\end{align}
which is referred to as the Cayley transform \cite{hassibi2002cdu}.
Explicitly, the Cayley transform is a mapping between a skewed-Hermitian matrix and a unitary matrix.
In Eq.~\eqref{NCGSM:eq:cayley}, the skewed counterpart of $\Xt(i)$, which is calculated by $j\Xt(i)$, is mapped to the unitary matrix $\X(i)$.
The Cayley transform of Eq.~\eqref{NCGSM:eq:cayley} is denoted by $\mathrm{\zeta}(\cdot)$.
The transmission rate of the NCGSM scheme is given by
\begin{align}
  R = \frac{\left\lfloor \log_2 {Q \choose P} \right\rfloor + P \log_2 (\L) }{M} \ \mathrm{[bits/symbol]}.
  \label{NCGSM:eq:rate}
\end{align}

As we mentioned in Section~\ref{sec:app:coh}, the GSTSK scheme subsumes the conventional GSM, ASTSK and LDC schemes.
Similar to the GSTSK scheme, the NCGSM architecture subsumes the conventional DSTBC schemes, which rely on the Cayley transform.
More specifically, the NCGSM scheme having $P = Q$ is equivalent to the differential LDC \cite{hassibi2002cdu}, which achieves a high transmission rate with the aid of the multiplexed PAM symbols.
Furethermore, the NCGSM scheme having $P=1$ is equivalent to \abbr{differential space-time shift keying}{Differential Space-Time Shift Keying}{DSTSK} \cite{sugiura2010cad}, which is the differential counterpart of the STSK scheme.

\subsection{PM-Based MIMO-MWC\label{sec:app:mwc}}
In Section~\ref{sec:sys:mwc}, we reviewed the hybrid BF technique conceived for MIMO-MWCs. The hybrid technique reduces the number of RF chains at the transmitter by using both analog and digital BF.
Although the hybrid BF technique significantly reduces the complexity of the transmitter, the complexity may still become excessive as the transmission rate increases, because the number of independent data streams also increases.
To maintain a high rate for MIMO-MWCs under practical resource constraints, a straightforward approach is to combine the PM concept with the MIMO-MWC concept.
Hence, PM-aided MIMO-MWC schemes have been proposed \cite{babakhani2008direct,valliappan2013asm,liu2015ssk,liu2016mmsm,ishikawa2017mmgsm,mesleh2017qsm,perovic2017mmgsm,perovic2017rsm,ding2017mmsm,hemadeh2016mmstsk} for reducing the hardware complexity of the transmitter, because the PM scheme transmits a reduced number of data streams, as mentioned in Section~\ref{sec:philo}.

Babakhani \etal \cite{babakhani2008direct} proposed an RF-switching-based modulation technique that
generates the conventional I/Q symbols by changing the electromagnetic boundary conditions.
Although the relationship between RF switching and the PM concept was not explicitly treated in \cite{babakhani2008direct}, the concept behind \cite{babakhani2008direct} is reminiscent of the PM philosophy.
Based on \cite{babakhani2008direct}, the RF-switching concept was extended to MIMO-MWCs, where the appropriate subarrays are switched on and off.
Similar to \cite{babakhani2008direct}, the MWC scheme of Valliappan \etal \cite{valliappan2013asm} achieves highly secured wireless communications.

\begin{figure}
  \centering
  \includegraphics[clip, scale=0.43]{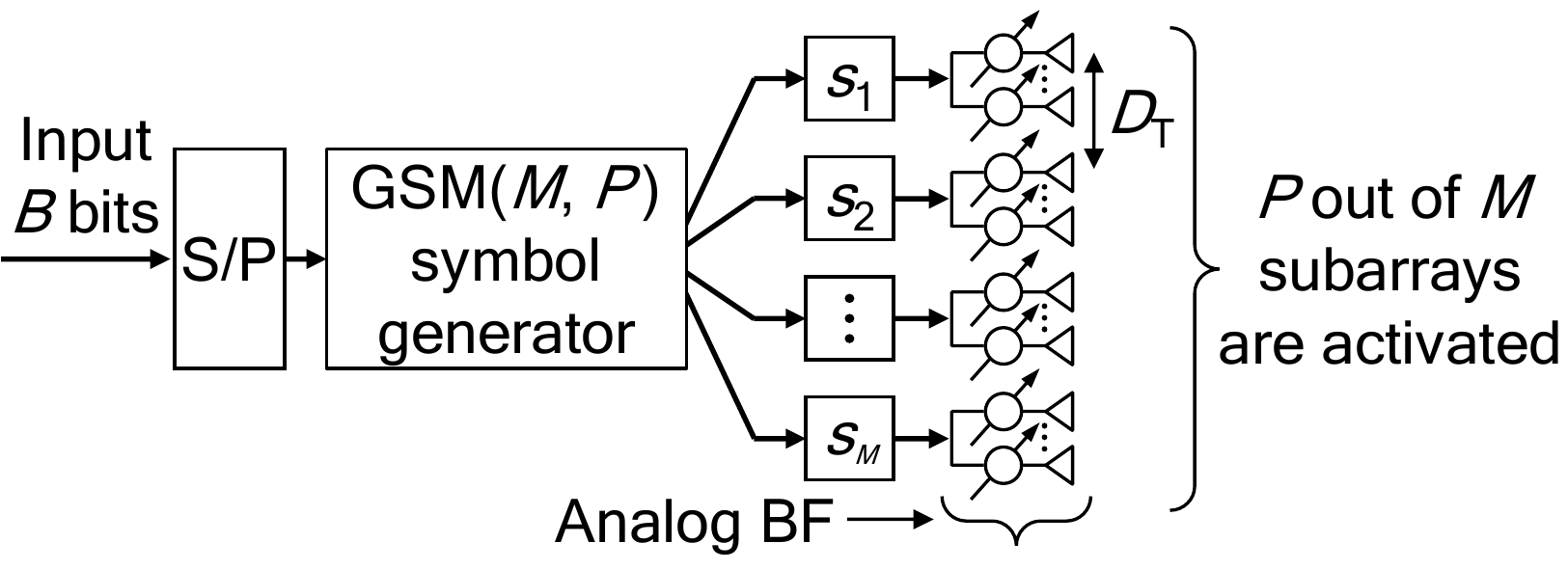}
  \caption{Schematic of the PM-based MIMO-MWC scheme.\label{fig:pm-based-mwc}}
\end{figure}
Apart from the RF-switching-based schemes \cite{babakhani2008direct,valliappan2013asm}, the SSK-based MWCs were first proposed by Liu and Springer in \cite{liu2015ssk}, which was the seminal research in this field.
In \cite{liu2015ssk}, the SSK modulation principle was directly applied in a MIMO-MWC system.
The SSK modulation principle was also extended to the GSM scheme by Liu \etal \cite{liu2016mmsm} and was combined with analog phase shifters \cite{ishikawa2017mmgsm,perovic2017mmgsm}.
Fig.~\ref{fig:pm-based-mwc} illustrates the schematic of the GSM-aided MWC transmitter of \cite{ishikawa2017mmgsm}, where the analog BF is considered.
The subarray separation $\DT$ is determined by the criterion of Eq.~\eqref{mm:eq:spacing}.
Quadrature SM, which becomes the family of GSM schemes, was also applied to MIMO-MWCs by Mesleh and Younis \cite{mesleh2017qsm}.
Hemadeh \etal have proposed the STSK-based MIMO-MWC system \cite{hemadeh2016mmstsk,hemadeh2016reduced,hemadeh2017multi,hemadeh2017reduced,hemadeh2017sfstsk}, where the STSK scheme is a generalization of GSM.
The achievable performance of PM-aided MIMO-MWC schemes has been evaluated in terms of the BER in uncoded scenarios \cite{liu2015ssk,liu2016mmsm,mesleh2017qsm,perovic2017mmgsm,perovic2017rsm,ding2017mmsm,hemadeh2016mmstsk} and the AMI in coded scenarios \cite{liu2016mmsm,mesleh2017qsm,ishikawa2017mmgsm}.
The major limitation of the PM-based MWC scheme is its reduced BF gain, where the number of activated subarrays is lower than that of the BLAST scheme \cite{ishikawa2017mmgsm}.

\begin{figure}
  \centering
  \subfigure[Full-RF-aided BLAST $\[s_1,s_2,s_3,s_4\]^\Trm$]{
    \includegraphics[clip, width=0.32\textwidth]{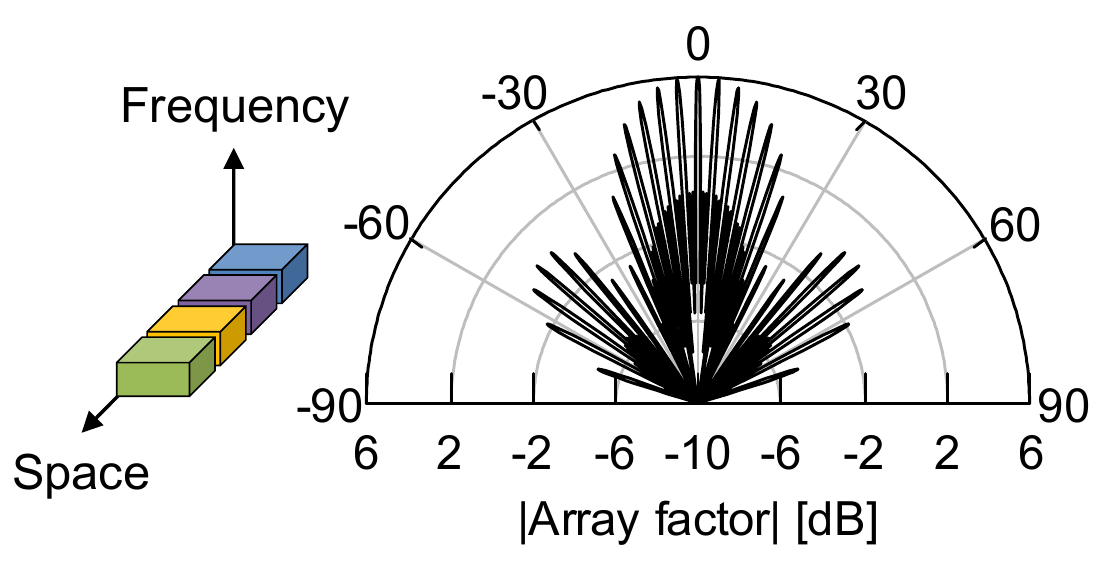} \label{fig:mmgsm0}
  }
  \subfigure[Reduced-RF-aided GSM $\[s_1,s_2,0,0\]^\Trm$]{
    \includegraphics[clip, width=0.32\textwidth]{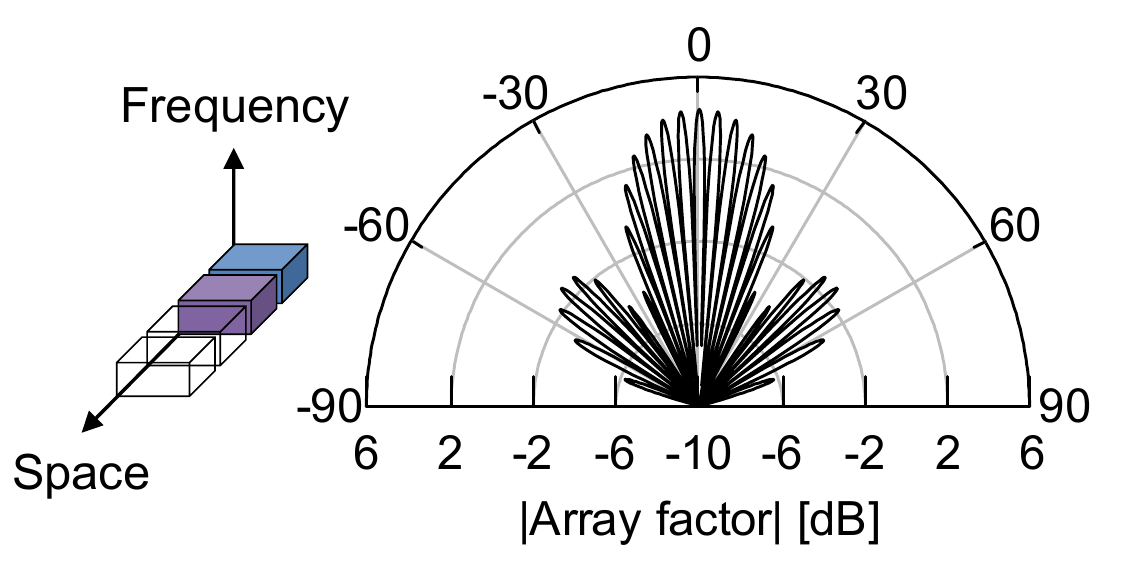} \label{fig:mmgsm1}
  }
  \subfigure[Reduced-RF-aided GSM $\[s_1,0,s_2,0\]^\Trm$]{
    \includegraphics[clip, width=0.32\textwidth]{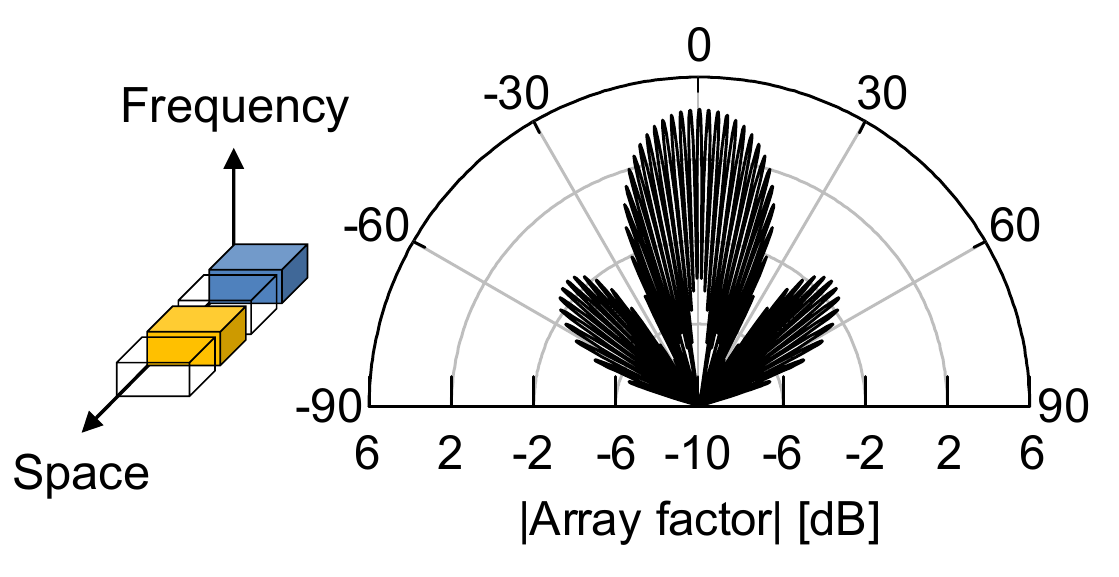} \label{fig:mmgsm2}
  }
  \caption{Directive gain of the conventional BLAST and the GSM-based MIMO-MWC schemes. The associated spatial domain symbols are also illustrated. The uniform linear array has 16 antenna elements that are separated into four subarrays. \label{mm:fig:directivity}}
\end{figure}
Fig.~\ref{mm:fig:directivity} illustrates the absolute values of the array factor with respect to the horizontal direction, where the optimal array alignment of Eq.~\eqref{mm:eq:spacing} was considered.
Fig.~\ref{mm:fig:directivity}(a) shows the achievable directional gain of the full-RF-aided BLAST scheme, while Figs.~\ref{mm:fig:directivity}(b) and (c) show that of the GSM-aided scheme.
Figs.~\ref{mm:fig:directivity}(a)~--~(c) demonstrate that the optimum alignment of Eq.~\eqref{mm:eq:spacing} changes the pattern of the directional BF gains.
We observe in Figs.~\ref{mm:fig:directivity}(a) and (b) that the measured BF gain of BLAST was 6.0 [dB], while that of GSM was 4.3 [dB].
This observation means that the GSM-based transmitter has a BF gain reduction.
We observe furthermore in Figs.~\ref{mm:fig:directivity}(b) and (c) that the beam pattern changes depending on the positions of non-zero elements in the codeword: $\x=[s_1,s_2,0,0]^\Trm$ and $[s_1,0,s_2,0]^\Trm$. 
Hence, the GSM-based scheme conveys the additional information bits by selecting a beam pattern.
\begin{table*}[t]
  \centering
  \small
  \caption{Contributions to the PM-based MWC\label{table:survey-mwc}}
  \begin{tabularx}{\linewidth}{|c|l|X|}
    \hline
    Year & Authors & Contribution \\
    \hline
    \hline
2008 & Babakhani \etal \cite{babakhani2008direct} & Proposed the near-field direct antenna modulation technique that generated an I/Q symbol by switching RF reflectors. The modulated I/Q symbol was scrambled for undesired directions. A proof-of-concept transmitter was implemented at 60 GHz to demonstrate the feasibility of the proposed technique. \\
    \hline
2013 & Valliappan \etal \cite{valliappan2013asm} & Proposed the secure wireless communications scheme based on \cite{babakhani2008direct}. In the proposed scheme, transmit antenna subarrays were switched on and off, and large sidelobes were suppressed by the simulated annealing algorithm.\\
    \hline
    2014 & Liu and Springer \cite{liu2015ssk} & Proposed the SSK-based MWC scheme, where antenna elements were properly placed to maintain channel orthogonality. This contribution \cite{liu2015ssk} is considered as a seminal research in the literature. \\
    \hline
    2016 & Hemadeh \etal \cite{hemadeh2016mmstsk} & Proposed the STSK-based MWC scheme, where the space-time codewords were used instead of the SSK-specific spatial codewords. In multiuser multicarrier downlink scenarios, the proposed scheme was shown to be capable of serving an increased number of users. \\
    \cline{2-3} & Ishikawa \etal \cite{ishikawa2017mmgsm} &Proposed the GSM-based MWC scheme with analog BF, where the constrained AMI was compared with both BLAST and GSM. The simulation results showed that the reduced-RF GSM scheme was equivalent to the full-RF-aided BLAST at the half-rate region.\\ 
    \cline{2-3} & Liu \etal \cite{liu2016mmsm} & Proposed the GSM-based MWC scheme, where the authors identified the channel conditions that minimize the error probability in uncoded scenarios. The unconstrained MI  was derived for the GSM scheme. Also, the authors considered a practical hardware implementation of the proposed scheme. The simulation results showed that the GSM scheme performed better than the BLAST scheme for two RF cases.  \\ 
    \cline{2-3} & Perovic \etal \cite{perovic2017rsm} & Applied the receive SM  concept \cite{yang2011rsm} to MIMO-MWCs. The authors identified the channel condition that minimizes the error probability of the proposed scheme. \\ 
    \hline
    2017 & Mesleh and Younis \cite{mesleh2017qsm} & Analyzed the constrained MI of the quadrature SM, where the channel coefficients were deterministic. The simulation results showed that the random alignment of antenna elements was effective in the overloaded scenario, i.e., $M > N$. \\ 
    \cline{2-3}
    & Perovic \etal \cite{perovic2017mmgsm} & Proposed phase-rotation-based precoding for the GSM-aided MWC. The proposed scheme was shown to be effective in the low modulation order  scenario. \\ 
    \cline{2-3}
    & Sacchi \etal \cite{sacchi2017mmstsk} & Proposed the STSK-based MWC scheme for the small-cell backhaul in a dense urban environment, where oxygen absorption, rain attenuation, and shadowing were considered. \\ 
    \cline{2-3}
    & Ding \etal \cite{ding2017mmsm} & Proposed spatial scattering modulation for the NLoS MWC-uplink, which was motivated by the SM concept. The proposed scheme conveyed additional bits by selecting a spatial direction for the scattering clusters. \\ 
    \cline{2-3}
    & Hemadeh \etal \cite{hemadeh2017reduced} & Proposed a reduced-RF-chain multi-set STSK-based MWC scheme, where OFDM and single-carrier frequency domain equalization were considered. The soft-decision single-carrier-based scheme with ABF was capable of achieving a near-capacity performance over dispersive MWC channels.\\ 
    \cline{2-3}
    & He \etal \cite{he2017mmgsm}& Proposed the phase-rotation-based precoding for GSM-aided MWCs. Different from \cite{perovic2017mmgsm}, the authors derived the lower bound of the spectral efficiency with a closed form, and then designed phase shifters to maximize the derived bound. \\
    \cline{2-3}
    & Botsinis \etal \cite{botsinis2017mm}& Proposed a joint-alphabet STSK for uplink non-orthogonal multiple access MWCs. The proposed scheme achieved a higher capacity than the conventional STSK. The authors also conceived a quantum-assisted low-complexity detector. \\ 
    \hline
  \end{tabularx}
\end{table*}
The contributions to the development of PM-based MWC schemes are summarized in Table~\ref{table:survey-mwc}.

\subsection{PM-Based MIMO-VLC\label{sec:app:vlc}}
As introduced in Section~\ref{sec:sys:mwc}, in PD-aided MIMO-VLCs, the channel relies on strong LoS elements.
In such an environment, the rank of the channel matrices becomes typically low.
Accordingly, both the MIMO diversity gain and the SMX gain are eroded, as described in Section~\ref{sec:sys:coh}.
To combat this limitation, the PM concept was first applied to the MIMO-VLCs by Mesleh \etal \cite{mesleh2011osm}. This scheme was referred to as \abbr{optical spatial modulation}{Optical Spatial Modulation}{OSM}.
The performance gain of the OSM scheme over the conventional single-stream scheme has been quantified in correlated channels \cite{mesleh2011osm,fath2010ssk,mesleh2011osm,fath2011osm,popoola2012ppm}, because the OSM scheme relies on a reduced number of data streams.
The high-frequency LED switching is feasible, and its bandwidth expansion is not a critical issue in the unlicensed VLC spectral band \cite{ishikawa2015osm}.
Thus, the OSM scheme is free from the SM antenna switching problem \cite{ishibashi2014sm} of the single-RF architecture.

In \cite{Popoola2013a,popoola2014merit}, the OSM scheme was evaluated by Popoola and Haas in a realistic LoS channel, where high correlations were observed.
It was shown by them in \cite{Popoola2013a,popoola2014merit} that it is difficult for the OSM scheme to attain a performance gain in these highly correlated channels.
Careful \abbr{power allocation}{Power Allocation}{PA} method was invoked for the OSM scheme. The new scheme was referred to as \abbr{power-imbalanced}{Power-Imbalanced}{PI} OSM \cite{fath2013comparison}.
The PA method of \cite{fath2013comparison} mitigated the channel  correlations with real-valued precoding associated with light sources.
In \cite{fath2013comparison}, the PA parameters were determined for the case of four light sources.

In most of the OSM studies \cite{mesleh2010osm,fath2010ssk,mesleh2011osm,fath2011osm,popoola2012ppm,popoola2014merit,popoola2013error,poves2012exp,fath2013comparison,fath2013color,ozbilgin2015osm}, the performance advantages have been demonstrated in terms of the BER in uncoded scenarios.
The information-theoretic analyses found in \cite{ishikawa2015osm,peppas2015osm,cai2016osm,wang2016osm} also demonstrated the benefits of the OSM scheme in coded scenarios, where it achieved a higher constrained mutual information (MI) in the low SNR region \cite{ishikawa2015osm,cai2016osm,ishikawa2015exitpiosm}.
\begin{table*}[t]
  \centering
  \small
  \caption{Contributions to OSM\label{table:survey-osm}}
  \begin{tabularx}{\linewidth}{|c|l|X|}
    \hline
    Year & Authors & Contribution \\
    \hline
    \hline
    2010 & Mesleh \etal \cite{mesleh2010osm} & Applied the SM concept \cite{mesleh2008spatial} to MIMO-VLCs, and the new scheme was termed OSM. The OSM scheme of \cite{mesleh2010osm} relies only on spatial domain symbols, where no constellations are considered.\\
    \cline{2-3}
    & Fath \etal \cite{fath2010ssk} & Studied the performance of the SSK-aided VLC scheme in measured channels, where the power imbalance technique was shown to be effective for improving error probability. \\ 
    \hline
    2011 & Mesleh \etal \cite{mesleh2011osm} & Analyzed the OSM scheme of \cite{mesleh2010osm} in terms of the channel alignment and the theoretical BER in uncoded and coded scenarios. The simulation results demonstrated the performance gain of the OSM scheme in the $M=4$ case. The benchmarks were conventional single-stream on-off keying and PAM schemes.\\
    \cline{2-3}
    & Fath \etal \cite{fath2011osm} & Combined the OSM scheme of \cite{mesleh2011osm} with PAM, which improved the spectral efficiency of OSM.\\ 
    \hline
    2012 & Popoola \etal \cite{popoola2012ppm} & Combined the OSM scheme of \cite{mesleh2011osm} with PPM to improve energy efficiency. Measured channel gains were considered.\\ 
    \cline{2-3} & Poves \etal \cite{poves2012exp} & Investigated the achievable performance of the OSM scheme \cite{mesleh2011osm} in a real environment, where the numbers of light sources and PDs were $(M,N)=(4,1)$. The transmission rate of 18 [Mbits/s] was achieved. \\ 
    \cline{2-3} & Fath and Haas \cite{fath2013comparison} & Compared the OSM scheme with the repetition coding and SMX schemes. In addition, they proposed a novel PI-OSM scheme, which allocated power imbalance factors to each transmit LED with the aim to combat highly correlated VLC channels. \\ 
    \hline
    2013 & Popoola \etal \cite{popoola2013error} & Applied the GSSK concept of \cite{younis2010gsm} to MIMO-VLCs, where multiple light sources were simultaneously activated.\\ 
    \cline{2-3}
    & Fath and Haas \cite{fath2013color} & Combined the OSM scheme with color-shift keying.\\ 
    \hline
    2014 & Popoola and Haas \cite{popoola2014merit} & Demonstrated the positive and negative effects of the GSSK-aided VLC scheme in a real environment. The transmission rate of 40 [Mbits/s] was achieved. \\
    \hline
    2015 & Ishikawa and Sugiura \cite{ishikawa2015osm} &Proposed a flexible PA method for the PI-OSM scheme of \cite{fath2013comparison}, where PA parameters were designed to maximize the constrained MI. The constrained MI comparisons showed that the PI-OSM scheme was beneficial over the conventional repetition coding and OSM schemes.\\ 
    \cline{2-3}
    & Ozbilgin and Koca \cite{ozbilgin2015osm} & Proposed a modulation scheme that combined OSM with both PPM and PAM for free space optical communications. The proposed scheme was shown to be capable of offering robustness against the scintillation effects of turbulence channels.\\ 
    \cline{2-3}
    & Peppas and Mathiopoulos \cite{peppas2015osm} & Analyzed the OSM scheme in terms of the average BER in uncoded and coded scenarios. The homodyned K distribution, which is a general free space optical channel model, was assumed in simulations. The analytical and numerical results verified the performance advantages of the OSM scheme over the conventional single-stream scheme. \\ 
    \hline
    2016 & Cai and Jiang \cite{cai2016osm} & Proposed a PPM-aided OSM scheme for multiuser MIMO-VLCs. The simulation results showed that the proposed scheme had advantages over conventional schemes at low illumination levels. \\    
    \cline{2-3}
    & He \etal \cite{he2016osm} & Compared the OSM scheme with the BLAST scheme in terms of the BER in uncoded scenarios. In the simulations, asymmetrically clipped optical OFDM signaling was assumed with an effective modulation bandwidth of 2 [MHz]. The BLAST scheme was shown to perform better than the OSM scheme for the transmission rate of 8 [bits/symbol] scenario. \\
    \cline{2-3}
    & Wang \etal \cite{wang2016osm} & Analyzed the OSM scheme in terms of the constrained MI. The simulation results revealed that the MED between the received symbols had a strong impact on the MI lower bound. The precoding parameters were designed to maximize the MED, which led to a higher MI.\\   
    \hline
  \end{tabularx}
\end{table*}
The important contributions of the family of PM-based MIMO-VLCs are summarized in Table~\ref{table:survey-osm}.

Before we review the PM schemes applied to MIMO-VLC, we revisit the simplest VLC scheme, which is referred to as PAM \abbr{repetition-code}{Repetition-Code}{RC}.
The conventional PAM-RC-based transmitter emits the same PAM symbol from all of the light sources.
It was shown by Safari and Uysal in \cite{safari2008needostbc} that the PAM-RC scheme is capable of outperforming the OSTBC scheme in free-space optical wireless communications.
The $\L$--PAM-RC symbols are defined by \cite{fath2011osm}
\begin{align}
  s = \frac{2(l-1)}{\L - 1} > 0,
\end{align}
where $l ~ (1 \leq l \leq \L)$ is the modulation index.
Then, the PAM-RC codeword is given by
\begin{align}
  \s^{\mathrm{PAM-RC}} = \left[s ~~ s ~~ \cdots ~~ s \right]^\Trm \in \mathbb{R}^{M},
  \label{eq:symbol-pam}
\end{align}
which consists of $M$ identical PAM symbols $s$.
We introduce two representative schemes each of which is the family of OSM schemes.

\subsubsection{Equal-Power OSM Scheme}
Again, the OSM concept in MIMO-VLC was proposed by Mesleh \etal in \cite{mesleh2010osm}, whose modulation principle is the same as that of the SM scheme, except for using multilevel modulation.
Here, $B=B_1+B_2$ input bits are partitioned into $B_1 = \log_2(M)$ and $B_2 = \log_2(\L)$ bits.
The first $B_1$ bits are used for selecting a single light source $q$ out of $M$ number of lights.
The second $B_2$ bits are mapped onto an $\L$--PAM symbol $s$ as follows: \cite{fath2011osm}
\begin{align}
  s = \frac{2l}{\L+1} \ (1\le l\le {\cal L}).
\end{align}
Then, the time-domain OSM symbols are generated by
\begin{align}
  \s^{\mathrm{OSM}} = [ \underbrace{0 \ \cdots \ 0}_{q-1~\mathrm{rows}} \ \underbrace{s}_{\substack{\uparrow \\ q\mathrm{th~row}}} \ \underbrace{0 \ \cdots \ 0}_{M-q~\mathrm{rows}}]^\Trm \in \mathbb{R}^{M}, \label{eq:symbol_osm}
\end{align}
with a single non-zero element.
The original OSM scheme proposed Mesleh \etal \cite{mesleh2011osm} conveys the input bits by selecting the transmit light source, and the PAM constellation size is constrained to be one, i.e., we have $s=1$.

\subsubsection{Power-Imbalanced OSM Scheme}
\begin{figure}
  \centering
  \includegraphics[clip, scale=0.43]{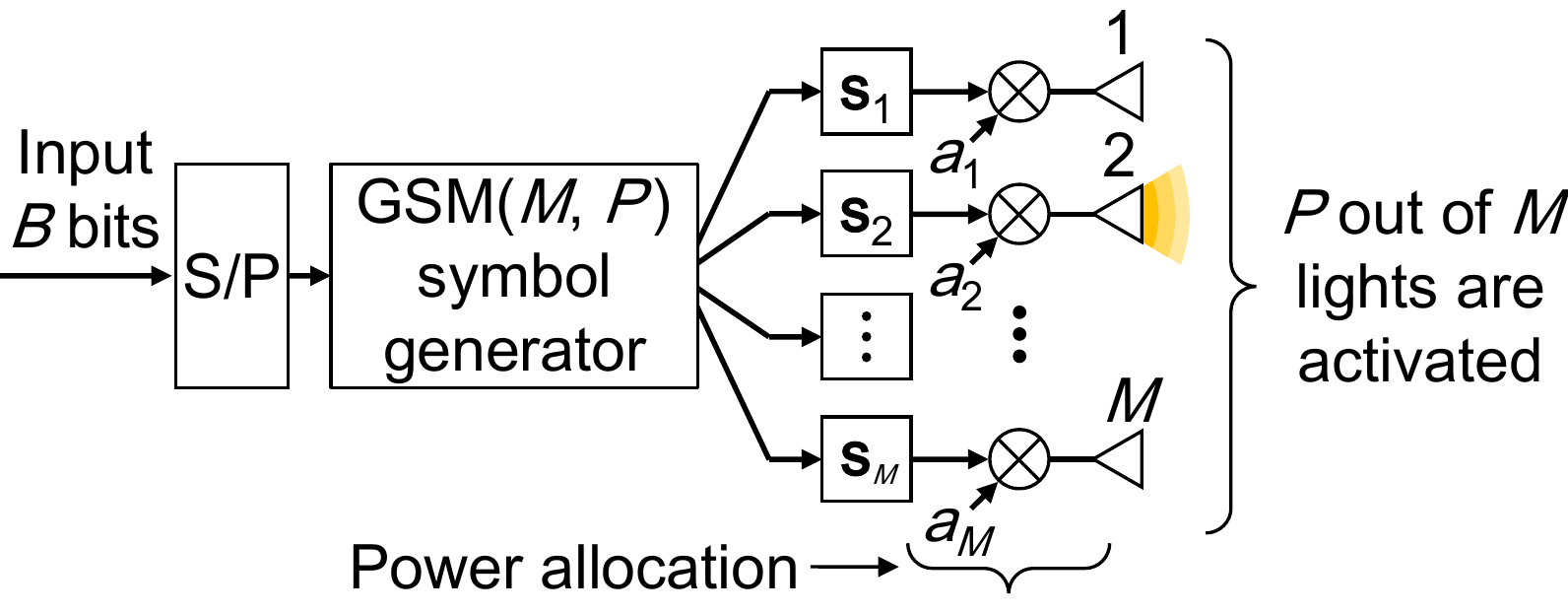}
  \caption{Schematic of the PM-based MIMO-VLC scheme.\label{fig:pm-based-vlc}}
\end{figure}
The \abbr{power-imbalanced OSM}{Power-Imbalanced OSM}{PI-OSM} concept was proposed by Fath and Haas \cite{fath2013comparison}, which is capable of communicating over VLC-specific LoS channels with the aid of PA.
It was reported by Poves \etal \cite{poves2012exp} that the performance of the OSM scheme depends on the power differences between channel paths.
Fig.~\ref{fig:pm-based-vlc} illustrates the schematic of the PI-OSM transmitter.
The encoding principle is basically the same as the GSM transmitter.
The transmitted symbols are generated by $\mathcal{L}$--PAM of~\eqref{eq:symbol-pam}, and then the symbols are multiplied by the PA factors $a_1, \cdots, a_M$.
Thus, the time-domain PI-OSM symbols are defined by \cite{fath2013comparison}
\begin{align}
  \s^{\mathrm{PI-OSM}} = \diag\(a_1, \cdots, a_M \) \s^{\mathrm{OSM}} \in \mathbb{R}^{M}, \label{eq:symbol_piosm}
\end{align}
where we have the constraint $\sum_{m=1}^{M}a_m=M$.
The PA factors $a_1, \cdots, a_M$ are designed to maximize a specific criterion, such as the constrained MI \cite{ishikawa2015osm} or the maximum achievable rate \cite{ishikawa2015exitpiosm}.
In \cite{fath2013comparison}, the PA factors are determined as follows:
\begin{align}
  a_m =
  \begin{dcases}
    \frac{M}{\sum^{M-1}_{i=0} \alpha^i} & (m=1) \\
    \alpha a_{m-1} & (2 \leq m \leq M) \\
  \end{dcases}.
  \label{vlc:eq:fathimbalance}
\end{align}
Here, the single parameter $\alpha$ determines the $M$ number of PA factors, where we have $\alpha = 10^{\frac{\beta}{10}}$.
In \cite{fath2013comparison}, the single parameter $\beta$ in dB was set to $\beta=1$, $3$, and $4$ [dB] for the $(M,N)=(4,4)$ setup.

\subsection{PM-Based Multicarrier Communications\label{sec:app:ofdm}}
The PM-aided multicarrier scheme, originally proposed by Schneider in 1967 \cite{schneider1968permutation}, simultaneously activated multiple frequencies.
In 1986, Padovani and Wolf proposed a modulation scheme that combined FSK and PSK \cite{padovani1986mod}.
Three years later, the PC-aided spread-spectrum concept was independently proposed by Sasaki \etal \cite{sasaki1991pcss}.
The PC-based scheme of \cite{sasaki1991pcss} conveys additional bits onto a set of spreading sequences, hence it achieves higher spectral efficiency while maintaining a low transmitter complexity.
This original concept was also evaluated in a channel-coded system \cite{sasaki1994performance} and a multiple-access system \cite{sasaki1995mapc}.
It was shown in \cite{sasaki1991pcss,sasaki1994performance,sasaki1995mapc} that the PM concept applied to the spread-spectrum system achieved a higher performance than the conventional direct sequence spread-spectrum system.

Motivated by the PC concept of \cite{sasaki1991pcss,sasaki1994performance,sasaki1995mapc},
the PM concept was also exported to OFDM by Frenger and Svensson \cite{frenger1999pcofdm}, where only a fraction of the subcarriers was activated.
Specifically, the PM-aided OFDM scheme of \cite{frenger1999pcofdm} conveyed the input bits with the same encoding principle as that of GSM \cite{younis2010gsm}.
Note that the PM scheme of \cite{frenger1999pcofdm} was proposed in 1999, while the GSM scheme was proposed in 2010.
The PM-aided OFDM scheme of \cite{frenger1999pcofdm} was later termed as \abbr{subcarrier index modulation}{Subcarrier Index Modulation}{SIM} \cite{abu2009sim}.
In \cite{frenger1999pcofdm}, it was shown that the SIM scheme is capable of striking a flexible tradeoff between the spectral efficiency, PAPR and reliability at the receiver.
The SIM scheme induces a loss in AMI \cite{wen2015mar,ishikawa2016sim} since the number of data streams is reduced.
The compressed-sensing-assisted SIM by Zhang \etal \cite{zhang2016cs} mitigates this IM-induced loss by compressing a high-dimensional sparse SIM into a low-dimensional dense codeword.
Apart from SIM, the subcarrier-hopping-based OFDM scheme was evaluated in multiple access scenarios \cite{hong2014fqam,hori2015hop,hori2016ma}.
\begin{table*}[tp]
  \centering
  \small
  \caption{Contributions to SIM\label{table:survey-sim}. A part of this table was imported from \cite{ishikawa2016sim}.}
  \begin{tabularx}{\linewidth}{|c|l|X|}
    \hline
    Year & Authors & Contribution \\
    \hline \hline
    1968 & Schneider \cite{schneider1968permutation} & Combined the PM concept with FSK, where $P$ out of $M$ frequencies were simultaneously activated.\\
    \hline
    1991 & Sasaki \etal \cite{sasaki1991pcss}
    & Proposed a PM-aided spread-spectrum scheme.\\
    \hline
    1994 & Sasaki \etal \cite{sasaki1994performance}
    & Evaluated the PM scheme of \cite{sasaki1991pcss} in terms of symbol error rate, while combining it with the selection diversity method and Reed-Solomon coding.\\
    \hline
    1995 & Sasaki \etal \cite{sasaki1995mapc}
    & Evaluated the PM scheme of \cite{sasaki1991pcss} in a multiple-access scenario.\\
    \hline
    1999 & Frenger and Svensson \cite{frenger1999pcofdm}
    & Proposed a subcarrier-based PM scheme in the OFDM context.\\
    \hline
    2005 & Kitamoto and Ohtsuki \cite{kitamoto2005pcoptical}
    & Analyzed the SIM of \cite{frenger1999pcofdm} in a VLC scenario.\\
    \hline
    2007 & Hou and Hamamura \cite{hou2007hc}
    & Developed a bandwidth-efficient SIM scheme by invoking a high-compaction multi-carrier concept.\\
    \hline
    2009 & Abu-Alhiga and Haas \cite{abu2009sim}
    & Evaluated the BER performance of an OFDM scheme in uncoded and coded scenarios, where its modulation concept was motivated by SM \cite{renzo2014spatial}. Note that the authors firstly coined the term `subcarrier-index modulation (SIM)'.\\
    \cline{2-3}
    & Hou and Hase \cite{hou2009papr}
    & Proposed a phase-rotation-based SIM scheme for reducing the PAPR.\\
    \hline
    2011 & Tsonev \etal \cite{tsonev2011sim}
    & Improved the SIM structure proposed in \cite{abu2009sim} and analyzed its PAPR.\\
    \hline
    2013 & Basar \etal \cite{basar2013sim}
    & Proposed an OFDM with IM scheme for both frequency-selective and time-varying fading channels, where the modulation principle was motivated by SM. In addition, the authors proposed a log-likelihood-ratio detector and provided a theoretical error performance analysis. The simulation results showed that the proposed scheme was capable of outperforming the classic OFDM scheme for ideal and realistic conditions. This paper \cite{basar2013sim} has been recognized as a worth-reading work in the literature, whose citation count is the largest among the SIM-related studies.\\
    \hline
    2014 & Xiao \etal \cite{xiao2014sim}
    & Proposed a subcarrier level interleaving method for the SIM scheme.\\
    \hline
    2015 & Basar \cite{basar2015interleave} & Proposed an interleaving method for the SIM scheme combined with space-time block codes.\\
    \cline{2-3}
    & Fan \etal \cite{fan2014generalizedsim} & Proposed the SIM scheme that supports an arbitrary number of selected subcarriers, and performs independent PM on the in-phase and quadrature components per subcarrier.\\
    \cline{2-3}
    & Zheng \etal \cite{zheng2015sim} & Proposed a low-complexity detector for the SIM scheme, and the detector performed independent PM on in-phase and quadrature components.\\
    \cline{2-3}
    & Basar \cite{basar2015mimosim} & Proposed a SIM scheme combined with SMX MIMO transmission and developed its low-complexity detector.\\
    \cline{2-3}
    & Datta \etal \cite{datta2015spacefreqindexmod} & Incorporated the PM concept with the space and frequency dimensions and developed a Gibbs-sampling-based detection algorithm.\\ 
    \cline{2-3}
    & Wen \etal \cite{wen2015mar} & Derived the maximum achievable rate of the SIM scheme and proposed an interleaved grouping method.\\ 
    \hline
    2016 & Basar \cite{basar2016simmimo} & Investigated the performance of the SIM scheme combined with MIMO and proposed low-complexity detectors. \\ 
    \cline{2-3} & Ma \etal \cite{ma2016sim} & Proposed a subcarrier-activation method for the SIM scheme, where the MED was maximized by an exhaustive search. \\ 
    \cline{2-3} & Ishikawa \etal \cite{ishikawa2016sim} &Analyzed the performance advantages of the SIM scheme over OFDM in terms of the MED and the constrained AMI. It was revealed that the SIM scheme is beneficial for a low rate region. \\ 
    \cline{2-3} & Wen \etal \cite{wen2016underwater} & Proposed a hybrid modulation scheme for underwater acoustic communications, where SIM and OFDM blocks were concatenated within a single frame.\\ 
    \cline{2-3} & Basar \cite{basar2016im} & Provided a tutorial on the SM and SIM schemes. \\ 
    \cline{2-3} & Wang \etal \cite{wang2016sim} & Proposed an OSTBC-based PM scheme in the space and the frequency domains that achieved the diversity order of two. \\ 
    \cline{2-3} & Mao \etal \cite{mao2016sim} & Proposed a dual-mode IM scheme that used two types of constellation sets. The type of constellation set corresponds to the on and off state of the conventional SIM scheme and leads to higher spectral efficiency. Coincidentally, this structure was similar to the original PM concept \cite{slepian1965pm}. The proposed scheme of \cite{mao2016sim} was generalized in \cite{mao2017gen}. \\ 
    \cline{2-3} & Zhang \etal \cite{zhang2016cs} & Proposed a compressed-sensing-assisted SIM system, which improved both spectral efficiency and energy efficiency at the same time. \\ 
    \hline
  \end{tabularx}
\end{table*}
The contributions to the development of SIM are summarized in Table~\ref{table:survey-sim}.

\begin{figure}
  \centering
  \includegraphics[clip, scale=0.43]{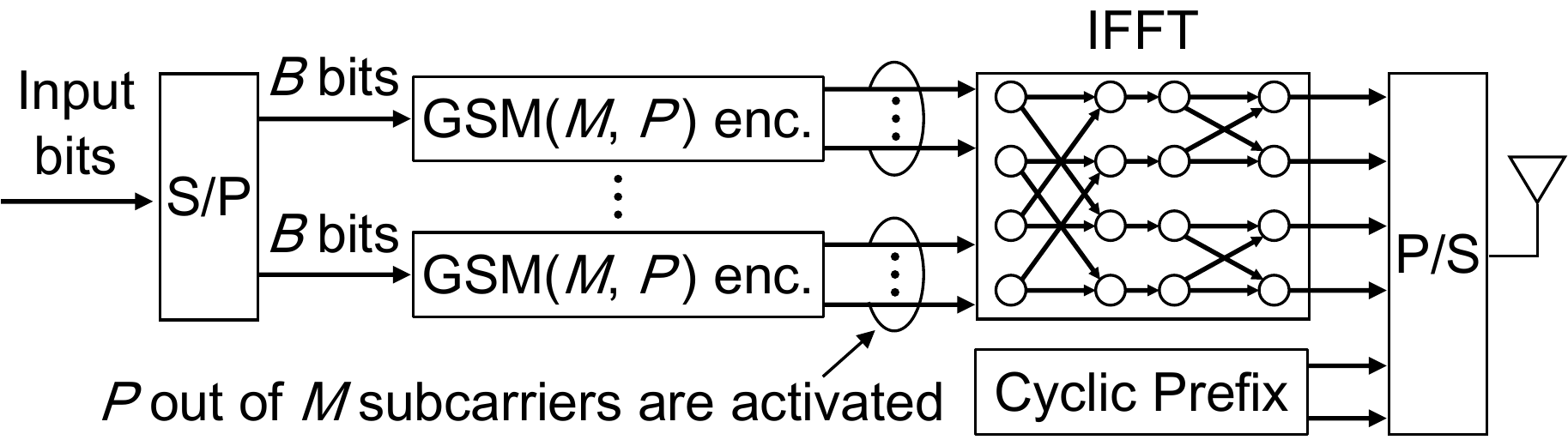}
  \caption{Schematic of the PM-based OFDM scheme.\label{fig:pm-based-ofdm}}
\end{figure}
The key feature of the SIM symbols is the sparsity of the frequency domain symbols, where $P$ number of subcarriers are activated out of $M$ subcarriers.
For simplicity, we represent the SIM system having the parameters $M$ and $P$ as ``SIM($M,P$)''.
Fig.~\ref{fig:pm-based-ofdm} illustrates the schematic of the SIM scheme.
The encoding principle of the SIM($M,P$) scheme is the same as that of the GSM($M,P$) scheme described in Section~\ref{sec:app:coh}.
Similar to the GSM($M,P$) principle, we have $\Na$ number of subcarrier-activation patterns, which are denoted by $\a_i$ for $1 \leq i \leq \Na$.
The encoded symbols are concatenated and converted to time-domain symbols by IFFT.
Finally, the cyclic prefix is inserted into the time-domain sequence to avoid the inter-channel interference.
\begin{figure}
  \centering
  \subfigure[Conventional OFDM]{
    \includegraphics[clip, width=0.36\textwidth]{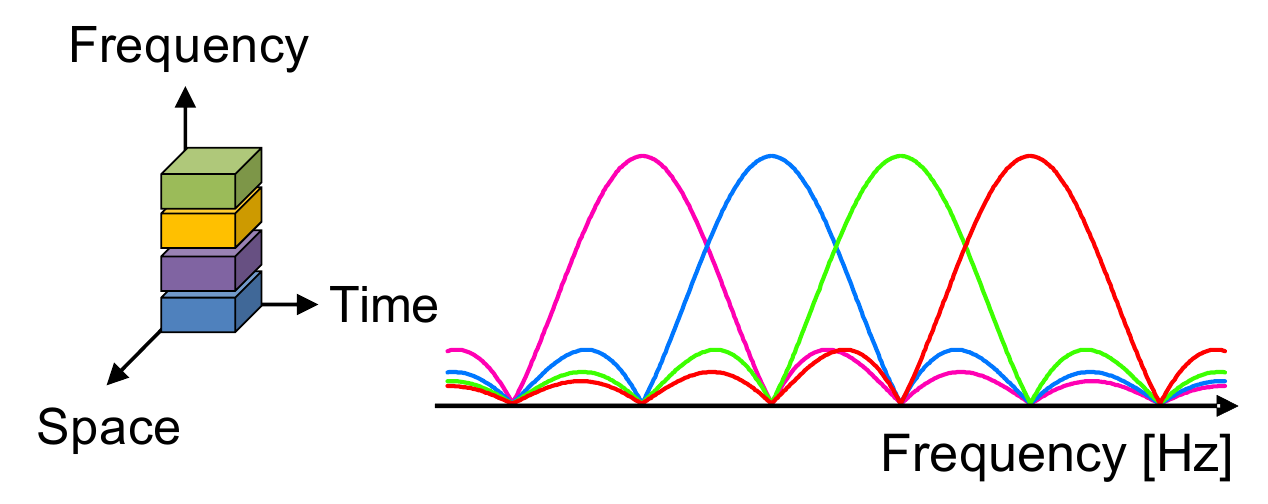} \label{fig:ofdm-conv}
  }
  \subfigure[SIM-based OFDM]{
    \includegraphics[clip, width=0.36\textwidth]{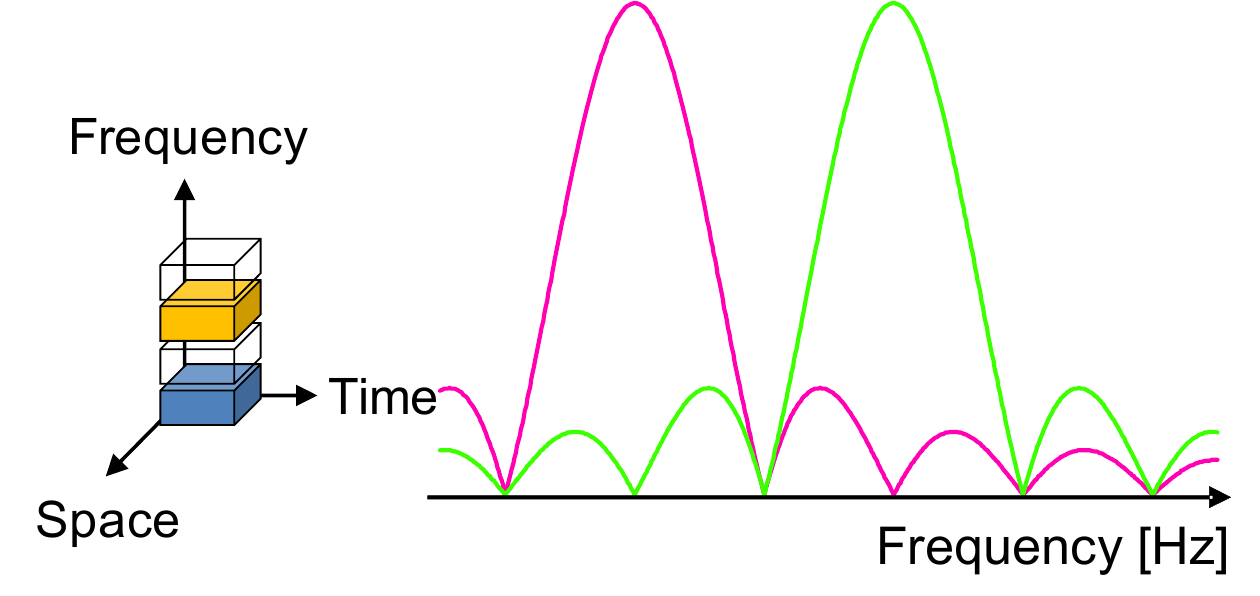} \label{fig:ofdm-pm}
  }
  \caption{Power spectral density of the conventional OFDM and SIM-aided OFDM schemes, where the associated frequency domain symbols are illustrated.\label{fig:pm-multicarrier-schemes}}
\end{figure}
Fig.~\ref{fig:pm-multicarrier-schemes} illustrates the frequency domain symbols of the conventional OFDM and the SIM-aided OFDM schemes.
The SIM scheme amplifies each symbol in order to maintain the same transmission power as OFDM.

\section{Performance Metrics \label{sec:metrics}}
In this section, we introduce the performance metrics for the general MIMO communication model.
We consider the following three metrics: AMI, MED and the decoding complexity at the receiver, which are described in Sections~\ref{sec:metrics:ami}, \ref{sec:metrics:rdc}, and \ref{sec:metrics:complexity}, respectively.

\subsection{Average Mutual Information (AMI)\label{sec:metrics:ami}}
The mutual information between the transmitted and received signals represents the maximum achievable rate, which is a maximum number of information bits that are successfully conveyed from the transmitter to the receiver per channel use.
In the literature, two popular AMI metrics are used: unconstrained AMI and constrained AMI.
Unconstrained AMI is derived by assuming continuous input symbols, where the input signal obeys a Gaussian distribution.
By contrast, constrained AMI is derived by assuming a finite number of discrete symbols.
The mutual information is averaged over each channel realization.
In the study of \cite{ishikawa2015osm}, the unconstrained and the constrained AMI of the PM system were referred to as the continuous-input continuous-output memoryless channel and the discrete-input continuous-output memoryless channel capacities, respectively.
It remains to show whether the Gaussian input maximizes the AMI of the PM system \cite{basnayaka2015massive}.
Hence, in this treatise, we use the terminology of {\it AMI} instead of {\it capacity}.

\subsubsection*{Unconstrained AMI\label{sec:metrics:ami:unconstrained}}
The unconstrained AMI $I_C$ of the general MIMO system model of Eq.~\eqref{COH:eq:blockmodel} is given by \cite{winters1987capacity,proakis2008}
\begin{align}
  I_C = \mathrm{E}_{\H}
  \left[ \sum^{\mathrm{rank}(\Q)}_{i=1} \log_2
    \left(1+ \mu_i \rho \right) \right] \ \textrm{[bits/symbol],}
  \label{eq:CCMC-final}
\end{align}
where we have a Hermitian matrix of
\begin{align}
  \Q = 
  \begin{dcases}
    \H^{\mathrm{H}} \H & (N \geq M) \\
    \H \H^{\mathrm{H}} & (N < M) \\
  \end{dcases},
\end{align}
with $\mathrm{E}_{\H}[\cdot]$ representing that the metric is averaged over the random channel matrices $\H \in \mathbb{C}^{N \times M}$ through a sufficiently high number of trials.
Here, $\mu_i$ and $\rho$ represent the $i$th eigenvalue of the Hermitian matrix $\Q$ and the received SNR.
Hence, the unconstrained AMI $I_C$ only depends on the channel matrix $\H$ and on the received SNR $\rho$.
Eq.~\eqref{eq:CCMC-final} is derived under the assumption that the input signals obey the complex-valued Gaussian distribution, and the signals are sampled at discrete intervals.
Moreover, the number of parallel streams is equal to $\mathrm{rank}(\Q)$.
By replacing the channel matrix $\H$ with an arbitrary channel matrix, Eq.~\eqref{eq:CCMC-final} becomes directly applicable to various channel models, such as the MWC channel.
The unconstrained AMI $I_C$ represents the upper bound of the constrained AMI, which is later denoted by $I_D$.
The constrained AMI $I_D$ is asymptotic to the unconstrained AMI $I_C$ at low SNRs, when increasing the transmission rate $R$.

\subsubsection*{Constrained AMI\label{sec:metrics:ami:constrained}}
The constrained AMI represents the effective upper bound of mutual information, where a finite number of input codewords is considered.
Here, we review its derivation in detail to highlight its background in an appropriate manner.
We assume that we have $N_c = 2^B$ number of space-time codewords  $\S^{(1)}, \cdots, \S^{\(N_c\)} \in \mathbb{C}^{M \times T}$, which are associated with the input bits of length $B$.
The constrained AMI of the general MIMO system model of Eq.~\eqref{COH:eq:blockmodel} is given by \cite{proakis2008,ng2006mimo}
\begin{align}
  I_D&=\frac{1}{T}
  \max_{\prm\(\S^{(1)}\), \cdots ,\prm\(\S^{\(N_c\)}\)}
  \sum^{N_c}_{f=1}
  \int_{-\infty}^{\infty} \cdots \int_{-\infty}^{\infty}
  \prm\(\Y|\S^{(f)}\) \nonumber \\
  &\cdot 
  \prm\(\S^{(f)}\)
  \log_2
  \left[
    \frac{
      \prm\(\Y|\S^{(f)}\)
    }{
      \sum^{N_c}_{g=1}
      \prm\(\Y|\S^{(g)}\)
      \prm\(\S^{(g)}\)
    }
  \right]
  \mathrm{d}\Y.
  \label{GSTSK:eq:DCMC1}
\end{align}
Eq.~\eqref{GSTSK:eq:DCMC1} is maximized when the codewords $\S^{(1)}, \cdots, \S^{\(N_c\)}$ are selected with the equal probability of $1 / N_c$.
This idealized assumption of $\prm\(\S^{(1)}\)= \cdots = \prm\(\S^{\(N_c\)}\) = 1/N_c$ simplifies Eq.~\eqref{GSTSK:eq:DCMC1} as follows:
\begin{align}
  I_D =
  \frac{1}{T}
  \left(
    B
    -
    \frac{1}{N_c}
    \sum^{N_c}_{f=1}
    \mathrm{E}
    \left[
      \log_2
      \sum^{N_c}_{g=1}
      \frac{
        \prm\(\Y|\S^{(f)}\)
      }{
        \prm\(\Y|\S^{(g)}\)
      }
    \right]
  \right).
  \label{GSTSK:eq:DCMC3}
\end{align}
We calculate $\frac{\prm\(\Y|\S^{(f)}\)}{\prm\(\Y|\S^{(g)}\)}$ in Eq.~\eqref{GSTSK:eq:DCMC3}
under the assumed relationship of $\Y = \H \S^{(f)} + \V$.
Based on Eq.~\eqref{GSTSK:eq:cond}, the likelihood ratio $\frac{\prm\(\Y|\S^{(f)}\)}{\prm\(\Y|\S^{(g)}\)}$ is given by
\begin{align*}
  \frac{\prm\(\Y|\S^{(f)}\)}{\prm\(\Y|\S^{(g)}\)}
  &=
  \exp
  \left(
    \frac{-\left\| \Y - \H \S^{(g)} \right\|_{\mathrm{F}}^2 + \left\| \Y - \H \S^{(f)} \right\|_{\mathrm{F}}^2}{\sigma_v^2}
  \right)\\
  &=
  \exp
  \left(
    \frac{-\left\| \H \( \S^{(f)} - \S^{(g)} \) + \V \right\|_{\mathrm{F}}^2 + \left\| \V \right\|_{\mathrm{F}}^2}{\sigma_v^2}
  \right).
\end{align*}
Finally, we arrive at the AMI of
\begin{align}
  I_D = \frac{1}{T} \left(B -\frac{1}{N_c}
    \sum^{N_c}_{f=1} \mathrm{E}_{\H, \V} \left[ \log_2 \sum^{N_c}_{g=1} e^{\eta[f,g]} \right] \right),
  \label{eq:DCMC-final}
\end{align}
where we have
\begin{align}
  \eta[f,g] =
  \frac{-\left\| \H \left( \S^{(f)} - \S^{(g)} \right) + \V \right\|_{\mathrm{F}}^2 
    + \left\| \V \right\|_{\mathrm{F}}^2}{\sigma_v^2}.
  \label{eq:DCMC:defeta}
\end{align}
The constrained AMI of Eq.~\eqref{eq:DCMC-final} is upper bounded by $I_D \leq B/T = R$ [bits/symbol].
Note that the general expression of Eq.~\eqref{eq:DCMC-final} is directly applicable to various channel models and codewords.
For example, Eq.~\eqref{eq:DCMC-final} supports the SISO symbols of $\S \in \mathbb{C}^{1 \times 1}$ and the SM symbols of $\S \in \mathbb{C}^{M \times 1}$.
The constrained AMI of $I_D$ estimates the turbo-cliff SNR of a channel-coded communication system, where the BER drops to an infinitesimal value with the aid of powerful channel coding schemes, such as turbo codes and low-density parity-check codes \cite{hanzo2009nearcapacity}.
This estimation procedure is described in Section~\ref{sec:comp}.
Recently, the constrained AMI was derived for the general differential MIMO \cite{nguyen2015capacity,xu2018isk}. The numerical results showed that the constrained AMI in a differential scenario can be approximated by the 3-dB shifted counterpart of the associated coherent scenario.

\subsection{Reliability\label{sec:metrics:rdc}}
A reliability evaluation between a transmitter and a receiver was the most common performance metric in previous studies.
Due to its simple formulation, the powerful analytical framework by Goldsmith \cite{goldsmith2005wireless} enables us to estimate the tight bound of error probabilities in uncoded scenarios, where no channel coding scheme is considered.
The \abbr{pairwise-error probability}{Pairwise-Error Probability}{PEP}, where a transmitted symbol $\S^{(f)}$ is decoded as a wrong symbol $\S^{(g)}$ at the receiver, is defined by \cite{sugiura2013jointopt}
\begin{align}
  &\mathrm{PEP}\(\S^{(f)} \rightarrow \S^{(g)} | \H\)\nonumber\\
  =& \prm\( \left\|\Y - \H \S^{(g)} \right\|_\mathrm{F}^2 < \left\|\Y - \H \S^{(f)} \right\|_\mathrm{F}^2 \)\nonumber\\
  =& \mathrm{Q}\( \sqrt{\frac{\|\H\D\|_\mathrm{F}^2}{2\sigma_v^2}} \),
  \label{eq:COH:pep-H}
\end{align}
where we have the channel matrix $\H$ and $\D = \(\S^{(f)} - \S^{(g)}\)\(\S^{(f)} - \S^{(g)}\)^\mathrm{H}$.
Note that $\mathrm{Q}(\cdot)$ denotes the Q-function\footnote{More specifically, the Q-function is defined by $\mathrm{Q}(\cdot) = \frac{1}{2}\(1-\mathrm{erf}\(\frac{x}{\sqrt{2}}\)\)$, where $\mathrm{erf}(x) = \frac{2}{\sqrt{\pi}} \int_{0}^{\pi} e^{-t^2} \mathrm{d}t$.}.
Then, averaging Eq.~\eqref{eq:COH:pep-H} over the legitimate channel matrices yields \cite{sugiura2013jointopt}
\begin{align}
  \mathrm{PEP}\(\S^{(f)} \rightarrow \S^{(g)}\) = \frac{1}{\pi} \int_0^{\pi/2} \prod_{m=1}^M \left( 1+\frac{\mu_m}{4\sigma_v^2 \sin^2 \theta} \right)^{-N} \mathrm{d} \theta.
  \label{eq:COH:pep}
\end{align}
Here, $\mu_m$ represents the $m$th eigenvalue of $\D$.
Based on Eq.~\eqref{eq:COH:pep}, the BER is upper bounded by \cite{sugiura2013jointopt}
\begin{align}
  \mathrm{BER} \leq \frac{1}{B2^B} \sum_{f} \sum_{f \neq g} d_\mathrm{H}\(\b^{(f)}, \b^{(g)}\) \mathrm{PEP}\(\S^{(f)} \rightarrow \S^{(g)}\),
  \label{eq:COH:ber}
\end{align}
where $d_\mathrm{H}\(\b^{(f)}, \b^{(g)}\)$ represents the Hamming distance between the bit sequences $\b^{(f)}$ and $\b^{(g)}$, which are associated with $\S^{(f)}$ and $\S^{(g)}$.
For example, $\mathrm{d_H}([0~0],[0~1])$ is calculated as $1$.

The rank and determinant criteria  maximize the coding gain, while maintaining the maximum diversity order.
The diversity order represents the slope of the error probability curve at high SNRs in uncoded scenarios.
The higher the coding gain is, the lower the achievable BER becomes for the communication system.
At high SNRs, Eq.~\eqref{eq:COH:pep} is upper bounded by \cite{tarokh1998space,heathjr2002ldc}
\begin{flalign}
  \mathrm{PEP}(\S^{(f)} \rightarrow \S^{(g)}) \leq
  \underbrace{
    \frac{1}{\prod_{m=1}^{m'} \mu_m^N}
  }_{\text{{\small coding gain}}}
  \left(
    \frac{1}{4\sigma_v^2}
  \right)^
  {
    -
    \overbrace{
      m'N
    }^{\mathrlap{\text{{\small diversity gain}}}}
  },
  \label{sec:eq:rdc-upper}
\end{flalign}
where $m'$ represents the minimum rank of $\D$, namely, $m'= \min_{\S^{(f)},\S^{(g)}} |\mathrm{rank}(\D)|$.
In addition, $\mu_m$ represents the $m$th eigenvalue of $\D$.
In Eq.~\eqref{sec:eq:rdc-upper}, the diversity order is given by $D = m'N$ and the coding gain is given by $\prod_{m=1}^{m'} \mu_m^N$.
The coding gain $\prod_{m=1}^{m'} \mu_m^N$ is determined by the MED between the codewords $\S^{(f)}$ and $\S^{(g)}$.
Typically, the MIMO codewords are designed by the rank and determinant criteria in uncoded scenarios, where both the diversity and the coding gains are maximized.

\subsection{Complexity\label{sec:metrics:complexity}}
In this treatise, we characterize the PM family in terms of the computational complexity at the receiver.
Due to the reduced number of data streams, the PM-based scheme has a lower complexity than that of the conventional multiplexing schemes.
In this treatise, the computational complexity is approximated by the number of real-valued multiplications of the detection process,\footnote{The complex addition cost is negligible against the multiplication cost \cite{brent2010arith}.} which is nearly equal to the number of multipliers in the receiver circuits.
	It was shown by Cavus and Daneshrad in \cite{cavus2006low} that both the power consumption as well as the hardware cost are increased as the number of multipliers is increased.
Specifically, if we have two complex-valued numbers $(a+bj) \in \mathbb{C}$ and $(c+dj) \in \mathbb{C}$, the total of the real-valued calculations of $(a+bj)(c+dj)$ is four, because we have $(a+bj)(c+dj)=ac-bd+(ad+bc)j$.

For example, if we consider the SM scheme designed for RF communications, as detailed in Section~\ref{sec:app:coh}, the computational complexity at the receiver is lower bounded by $\Omega(2^{R} N)$.\footnote{We use Donald Knuth's big Omega $\Omega(\cdot)$ notation \cite{knuth1976bigomega}.}
Here, the ML detection criterion of the SM scheme is given by
\begin{align}
  \hat{\S} = \arg \min_{\S} \left\| \Y - \H \cdot \S \right\|_{\mathrm{F}}^2.
  \label{sec:eq:sm-mld}
\end{align}
The receiver estimates the transmitted symbol $\S$ through $2^{R}$ number of trials in Eq.~\eqref{sec:eq:sm-mld}.
In each trial, the matrix multiplication of $\H \cdot \S$ includes $N$ number of complex-valued multiplications, which is equivalent to $4N$ real-valued multiplications.
In addition, the Frobenius norm calculation of $\left\| \Y - \H \cdot \S \right\|_{\mathrm{F}}^2$ includes $2N$ real-valued multiplications.
Hence, the computational complexity of the SM scheme is given by $2^{R} \cdot 6N$, which is lower bounded by $\Omega(2^{R} N)$.
\begin{table*}[tbp]
  \centering
  \footnotesize
  \caption{Diversity, transmission rate, and complexity comparisons for the coherent and non-coherent schemes.\label{table:comp}}
  \begin{tabular}{|l|c|c|c|ll|}
    \hline
    Scheme & Section & Diversity $D$ & Rate $R$ & \multicolumn{2}{c|}{ML complexity} \\
    \hline \hline
    Coherent APSK & \ref{sec:sys:coh} & $N$ & $\log_2{\mathcal{L}}$ & $2^{R+1}(2N+1)$ & $\geq \Omega(2^R N)$ \\
    SM \cite{mesleh2008spatial} & \ref{sec:app:coh} & $N$ & $\log_2{\mathcal{L}} + \log_2{M}$ & $2^{R+1}3N$ & $\geq \Omega(2^R N)$ \\
    GSM \cite{jeganathan2008generalized} & \ref{sec:app:coh} & $N$ & $P \cdot \log_2{\mathcal{L}} + \left\lfloor \log_2{{M \choose P}} \right\rfloor$ & $2^{R+1}(2P+1)N$ & $\geq \Omega(2^R P N)$ \\
    Rectangular ASTSK \cite{sugiura2012universal} & \ref{sec:app:coh} & $T \cdot N$ & $\left( \log_2{\mathcal{L}} + \log_2{Q} \right) / T$ & $2^{RT+1}3N$ & $\geq \Omega(2^{RT} N)$ \\
    BLAST \cite{foschini1996layered} & \ref{sec:sys:coh} & $N$ & $M \cdot \log_2{\mathcal{L}}$ & $2^{R+1}(2M+1)N$ & $\geq \Omega(2^R M N)$ \\
    Square GSTSK \cite{sugiura2012universal} & \ref{sec:sys:coh} & $M \cdot N$ & $\left( P \cdot \log_2{\mathcal{L}} + \left\lfloor \log_2{{Q \choose P}} \right\rfloor \right) / M$ & $2^{RM+1}(2M+1)N$ & $\geq \Omega(2^{RM} M N)$ \\
    \hline
    \hline
    Differential APSK & \ref{sec:sys:diff} & $N$ & $\log_2{\mathcal{L}}$ & $2^{R+2}(N+1)+8$ & $\geq \Omega(2^{R}N)$ \\
    Rectangular DSM \cite{ishikawa2017rdsm} & \ref{sec:comp:diff} & $T \cdot N$ & $\left( \log_2{\mathcal{L}} + \log_2{Q} \right) / T$ & $2^{RT+1}3N+4N(M/T+1)$ & $\geq \Omega(2^{RT}N)$ \\
    Square DSM \cite{ishikawa2014udsm} & \ref{sec:comp:diff} & $N \leq D \leq M \cdot N$ & $\left( \frac{N \cdot M}{D} \cdot \log_2{\mathcal{L}} + \log_2{Q} \right) / M$ & $2^{RM+1}3N$ & $\geq \Omega(2^{RM}N)$ \\
    Square NCGSM \cite{ishikawa2014ncgsm} & \ref{sec:sys:diff} & $M \cdot N$ & $\left( P \cdot \log_2{\mathcal{L}} + \left\lfloor \log_2{{Q \choose P}} \right\rfloor \right) / M$ & $2^{RM+1}(2M+1)N$ & $\geq \Omega(2^{RM}MN)$ \\
    \hline
  \end{tabular}
\end{table*}
In the same manner, we derived the ML complexities for the coherent and non-coherent schemes introduced in this treatise, such as single-stream APSK, SM \cite{mesleh2008spatial}, GSM \cite{jeganathan2008generalized}, rectangular ASTSK \cite{sugiura2012universal}, BLAST \cite{foschini1996layered}, square GSTSK \cite{sugiura2012universal}, differential APSK, rectangular DSM \cite{ishikawa2017rdsm}\footnote{The basic modulation concept is the same with the ASTSK scheme.}, square DSM \cite{ishikawa2014udsm}, and square NCGSM \cite{ishikawa2014ncgsm}.
Here, the ML complexity was divided by the codeword's time slots $T$.
The derived complexities are summarized in Table~\ref{table:comp}.
In addition, the corresponding diversity and transmission rate are also summarized in Table~\ref{table:comp}.

\section{Performance Comparisons\label{sec:comp}}
In this section, we provide performance comparisons between the PM-based schemes and the conventional multiplexing schemes, where coherent MIMO, differential MIMO, MIMO-MWC, MIMO-VLC and multicarrier systems are considered.
We used the performance metrics described in Section~\ref{sec:metrics}.
In our comparisons, the total transmit power was fixed to unity for all schemes.

\subsection{PM-Based Coherent MIMO\label{sec:comp:coh}}
First, we investigated the achievable performance of the PM-based coherent MIMO scheme in terms of its AMI, reliability and complexity.
We illustrate the relationship between the AMI and reliability.

\begin{figure}[tbp]
  \centering
  \subfigure[Small-scale MIMO scenario ($M=2$)]{
    \includegraphics[clip, scale=0.58]{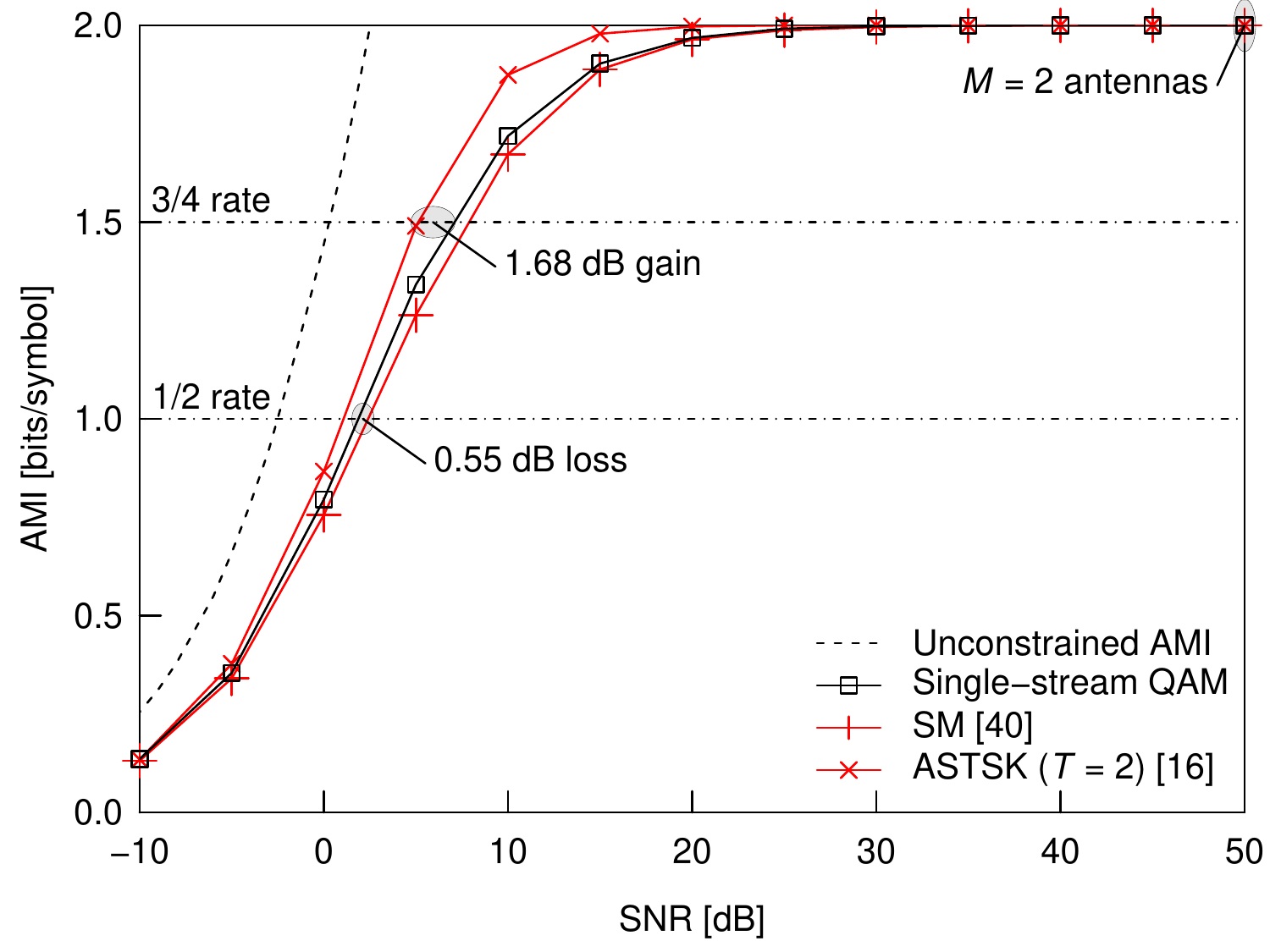}
  }
  \subfigure[Large-scale MIMO scenario ($M=4,64,1024$)]{
    \includegraphics[clip, scale=0.58]{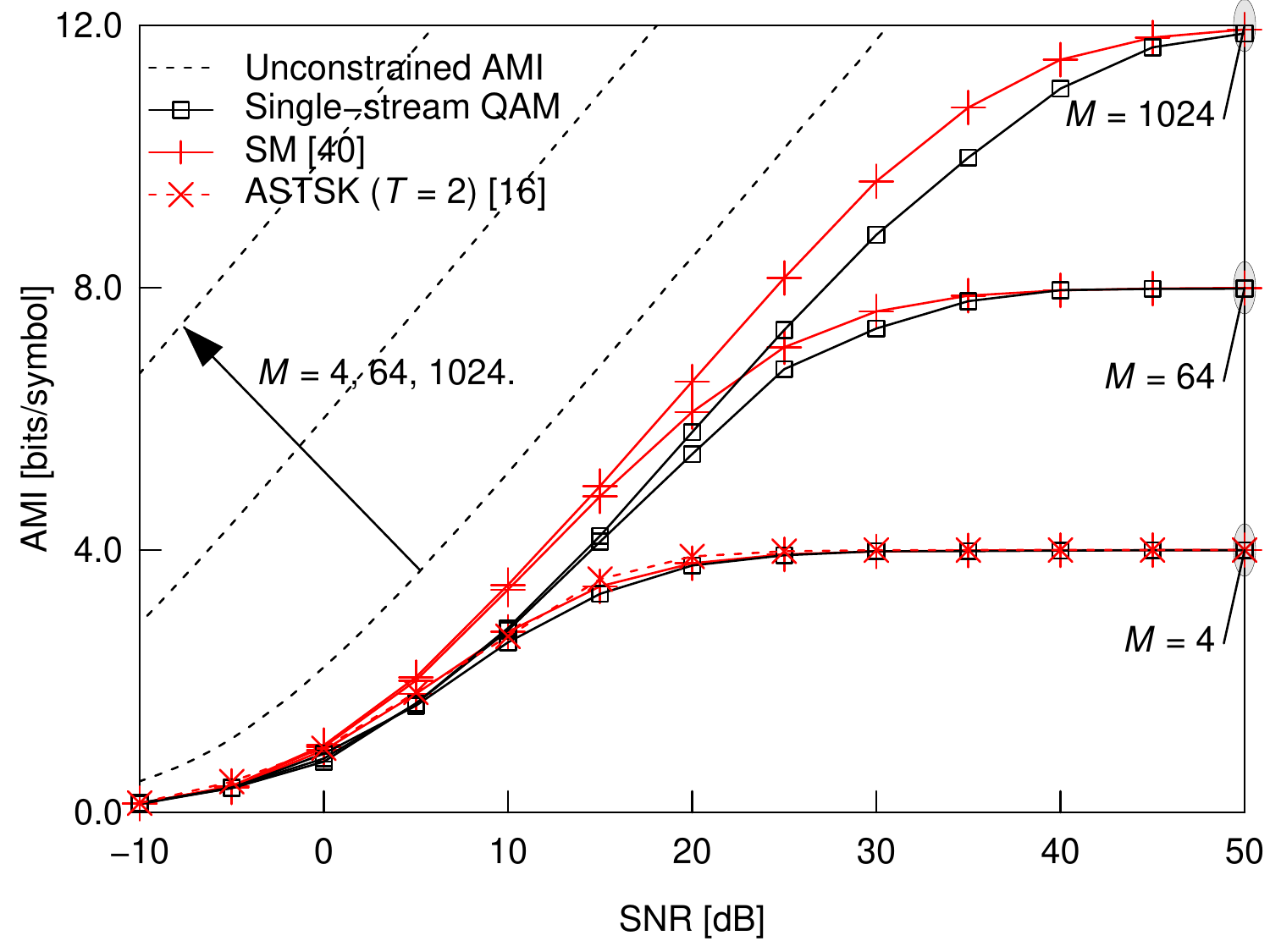}
  }
  \caption{Constrained AMI comparisons between the SM and ASTSK arrangements of Fig.~\ref{fig:GSTSK:structure}, where the number of transmit antennas was increased from $M=2$ to $1024$. The number of receive antennas was set to $N=1$.\label{fig:dcmc}}
\end{figure}
\begin{figure}[tbp]
  \centering
  \includegraphics[clip,scale=0.58]{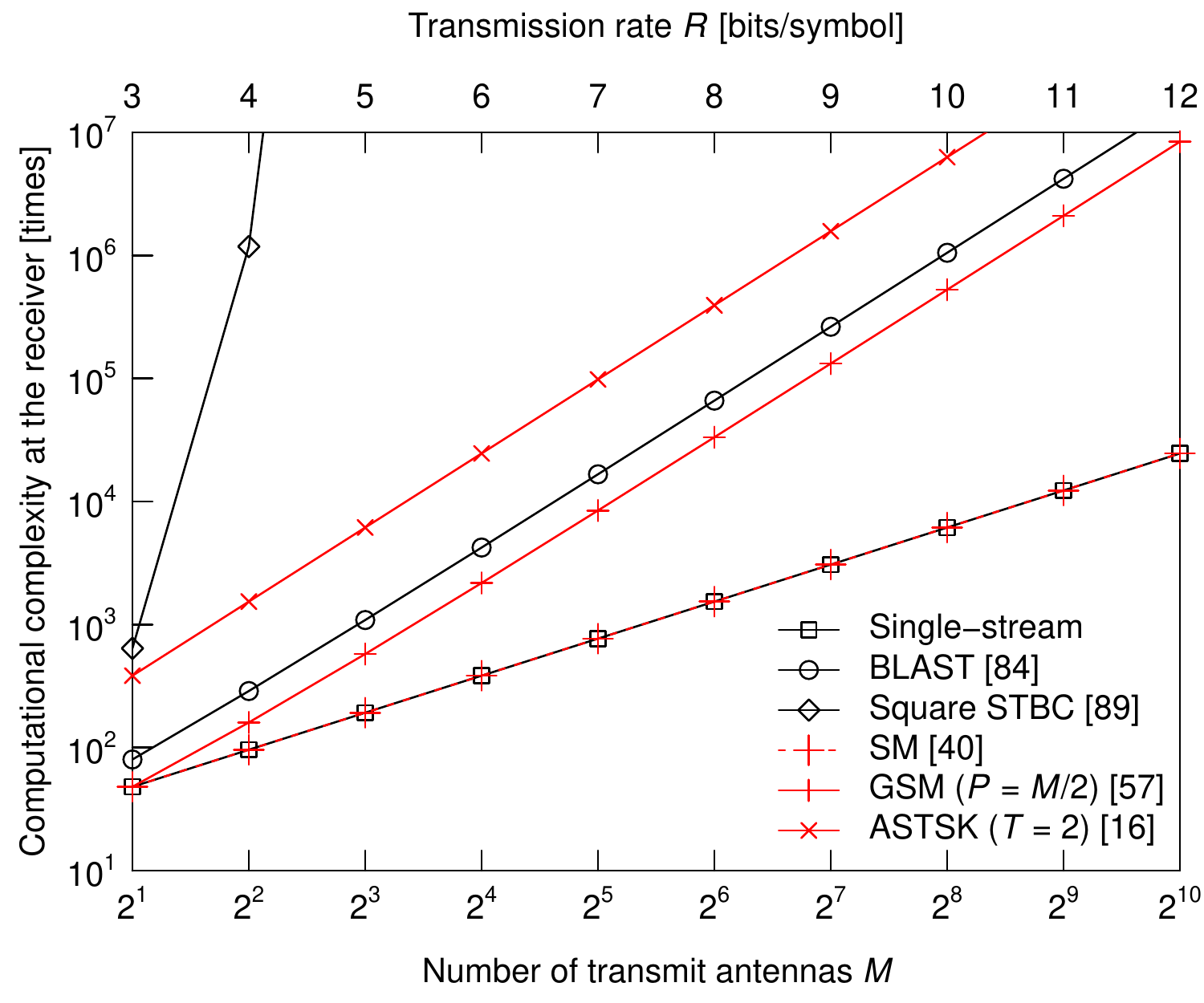}
  \caption{ML decoding complexities of the coherent MIMO schemes introduced in Section~\ref{sec:sys:coh}. The number of transmit antennas was changed from $M=2$ to $1024$, and the associated transmission rate corresponded to $R=3, \cdots, 12$ [bits/symbol]. The number of receive antennas was $N=1$. \label{fig:txvsmulti}}	
\end{figure}

Fig.~\ref{fig:dcmc} compares the constrained AMI of the SM and of the ASTSK schemes.
The single-stream-based QAM scheme was also included for reference.
Fig.~\ref{fig:dcmc}(a) characterizes a small number of transmit antennas, namely $M=2$.
By contrast,  Fig.~\ref{fig:dcmc}(b) characterizes a large number of transmit antennas, namely $M=1024$.
The associated curves of the unconstrained AMI were also plotted.
Note that the unconstrained AMI only depends on the transmit power and on the numbers of transmit and receive antennas, as defined in Eq.~\eqref{eq:CCMC-final}.
The DMs of the ASTSK scheme were designed based on the rank and determinant criteria described in Section~\ref{sec:app:coh}.
We observe in Fig.~\ref{fig:dcmc}(a) that the SM scheme exhibited a slight performance loss as compared to the single-stream 4--QAM scheme.
Here, the ASTSK scheme of Fig.~\ref{fig:GSTSK:structure} alleviated the performance gap and achieved 1.68 dB gain over the single-stream scheme at the 3/4 rate region.
Next, in Fig.~\ref{fig:dcmc}(b), we compared the schemes used in Fig.~\ref{fig:dcmc}(a) for large-scale MIMO scenarios, where the number of transmit antennas was set to $M=4$, $64$, and $1024$.
Note that the ASTSK scheme cannot be simulated for the $M=64$ and $1024$ scenarios due to its excessive complexity, which will be discussed later in the context of Fig.~\ref{fig:txvsmulti}.
It was shown in Fig.~\ref{fig:dcmc}(b) that the SM, the ASTSK and the single-stream schemes exhibited similar AMI for $M=4$ antennas.
The performance gain of SM over the single-stream scheme increased, as the number of transmit antennas increased.
This observation implies that the SM scheme is beneficial for open-loop massive MIMO scenarios.
However, when we consider closed-loop massive MIMO scenarios, the conjugate BF scheme of Marzetta \cite{marzetta2010noncooperative} achieves near-capacity performance and a large performance gap exists between the BF and the SM schemes \cite{ishikawa2017rdsm}.

\begin{figure}[tb]
  \centering
  \includegraphics[clip,scale=0.58]{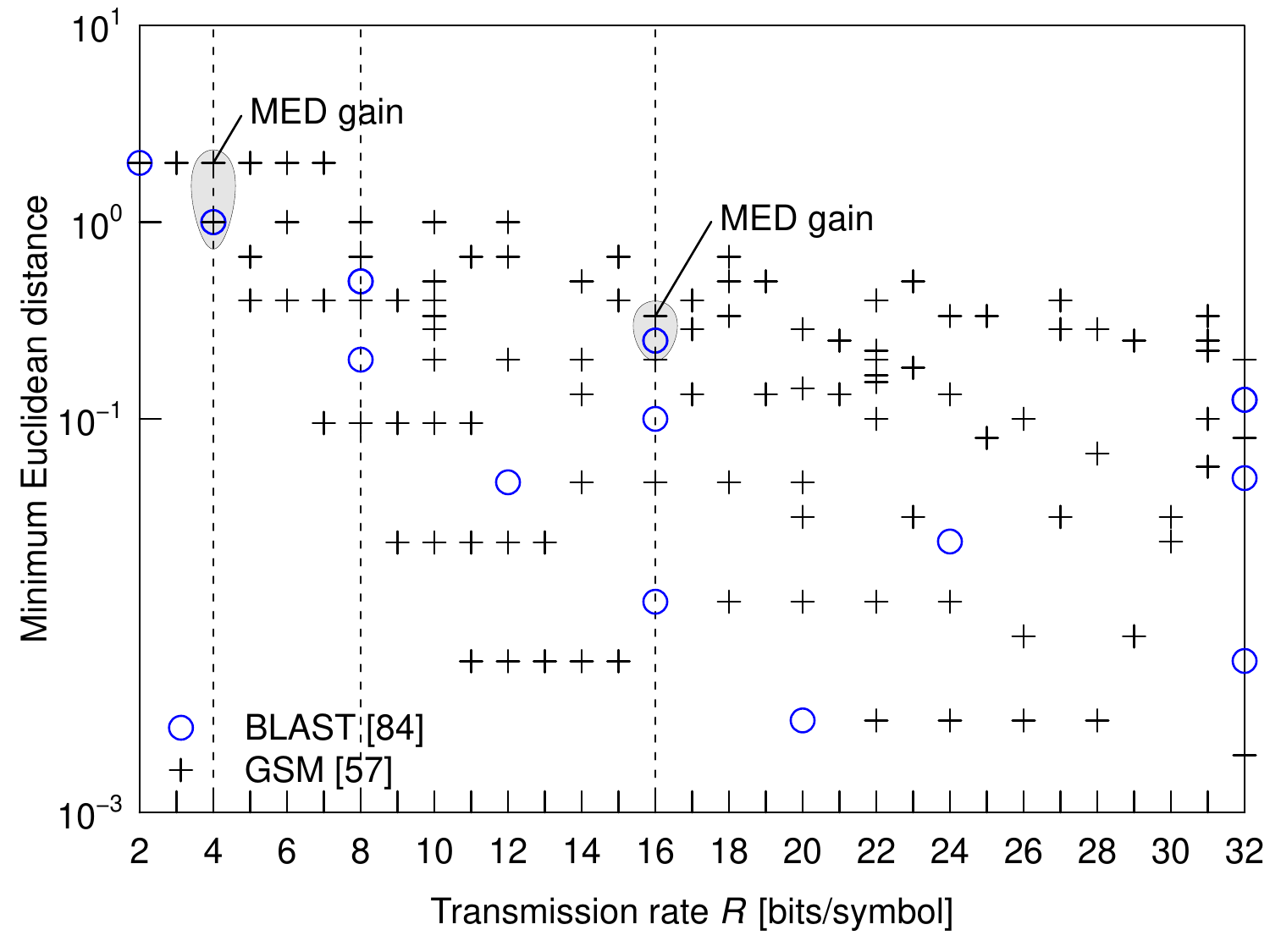}
  \caption{MEDs of the BLAST and GSM schemes of Fig.~\ref{fig:GSTSK:structure}, where the number of transmit antennas was set to $M=2,~4,~8,~16,~32$, and the number of activated antennas was $P=1,\cdots,M$. We used the constellation of $\mathcal{L}=2$--BPSK and $\mathcal{L}=4,~16,~64,~256,~1024$--QAM.\label{fig:mindet-comparison}}%
\end{figure}
Fig.~\ref{fig:mindet-comparison} shows an MED comparison for the BLAST and GSM schemes of Fig.~\ref{fig:GSTSK:structure}, where the number of transmit antennas was varied from $M=2$ to $32$, and the constellation size was increased from $\mathcal{L}=2$ to $1024$.
The transmission power was set to a constant value for all the scenarios.
As shown in Fig.~\ref{fig:mindet-comparison}, the GSM scheme of Fig.~\ref{fig:GSTSK:structure} achieved MED gains right across the whole transmission rate region, i.e., $2 \leq R \leq 32$ [bits/symbol].
For example, for the $R=16$ [bits/symbol] case, the QPSK-aided BLAST scheme having $M=8$ achieved the MED of $0.25$, while the QPSK-aided GSM scheme having $(M,P)=(8,6)$ achieved the MED of $0.33$.\footnote{Note that the MED comparison is valid only when the transmission rate is the same.}
Fig.~\ref{fig:mindet-comparison} demonstrates the scalability of different-throughput PM-aided coherent MIMO schemes under diverse practical constraints.
Note that this scalability cannot be achieved by the frequency- or temporal-domain PM scheme because its MED gain typically diminishes as the effective transmission rate increases \cite{ishikawa2016sim}.

\begin{figure}[tbp]
  \centering
  \includegraphics*[clip,scale=0.58]{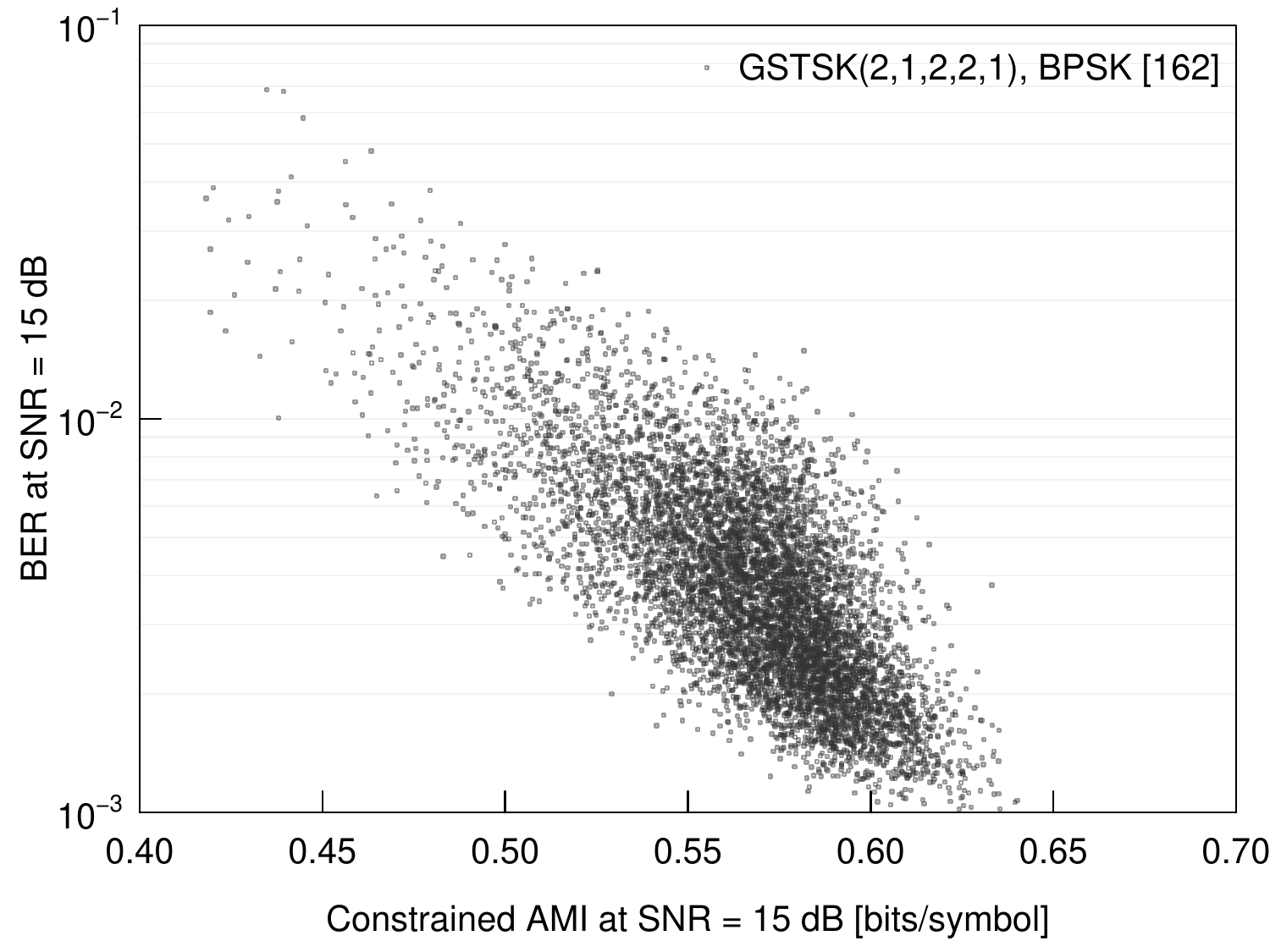}
  \caption{Relationship between the constrained AMI and the BER of the BPSK-aided GSTSK($2,1,2,2,1$) scheme of Fig.~\ref{fig:GSTSK:structure}, where SNR was set to 15 dB, and 7187 DMs were randomly generated.}
  \label{fig:coh-gstsk-dcmc-vs-ber}
\end{figure}
Fig.~\ref{fig:coh-gstsk-dcmc-vs-ber} shows the relationship between the constrained AMI and the BER at $\mathrm{SNR}=15$ [dB].
For our general discussions, we have employed the GSTSK scheme, which is capable of subsuming the family of conventional MIMO schemes.
A total of 7187 random DMs were generated, and then the associated constrained AMI and BER were simulated for each DM set.
We observe in Fig.~\ref{fig:coh-gstsk-dcmc-vs-ber} that the BER and the constrained AMI exhibited a certain relationship.
Specifically, in the constrained AMI calculation of Eq.~\eqref{eq:DCMC:defeta},
approximating $\V$ by $\0$ yields
\begin{align}
  \eta[f,g] = -\frac{1}{\sigma_v^2} \left\| \H \left( \S^{(f)} - \S^{(g)} \right)\right\|_{\mathrm{F}}^2,
  \label{eq:DCMC:defeta-simplified}
\end{align}
which contains the MED of $\left\| \H \left( \S^{(f)} - \S^{(g)} \right)\right\|_{\mathrm{F}}^2$ at the receiver's output.
Thus, again, the constrained AMI and the MED exhibit a correlation at high SNRs, due to Eq.~\eqref{eq:DCMC:defeta-simplified}.

Fig.~\ref{fig:txvsmulti} shows the decoding complexities for the coherent MIMO schemes, including BLAST, square STBC, SM, GSM and ASTSK schemes of Fig.~\ref{fig:GSTSK:structure}. The single-stream APSK scheme was considered as a benchmark.
In Fig.~\ref{fig:txvsmulti}, the number of transmit antennas was varied from $M=2$ to $1024$, where the transmission rate corresponding to the number of transmit antennas $M$ ranged from $R=3$ to $12$ [bits/symbol].
Here, QPSK signaling was considered for the SM scheme, i.e., $\mathcal{L}=4$, while the other schemes used the $N_c= (M \cdot \mathcal{L})^T=(4M)^T$-element arbitrary constellation to maintain the same transmission rate\footnote{This arbitrary constellation setup does not affect the decoding complexity.}.
Note that state-of-the-art low-complexity detectors were not considered in this comparison.
As shown in Fig.~\ref{fig:txvsmulti}, the SM scheme exhibited the same complexity as the single-stream APSK scheme\footnote{Single-stream APSK exhibits the lowest complexity in general, and the gap between APSK and SM increases with increasing $N$.}.
The ASTSK scheme was capable of achieving a lower complexity than the square STBC scheme, while its complexity trend was similar to that of the conventional BLAST scheme.
The ML complexity of the square STBC scheme was prohibitively high, especially for $M>5$ scenarios, because the number of transmission symbol intervals per block $T$ was the same as the number of transmit antennas $M$.
For example, the ML decoding complexity of square STBC was $3.74 \times 10^{13}$ for $M=8$ antennas and $5.23 \times 10^{30}$ for $M=16$ antennas.

\subsection{PM-Based Differential MIMO\label{sec:comp:diff}}
Next, we investigated the achievable performance of the PM-based differential MIMO scheme in terms of its BER, where we considered Rayleigh fading channels as well as the effects of channel estimation errors.
For simplicity, we represent the binary-valued-DM-aided and the complex-valued-DM-aided DSM schemes as BDSM \cite{bian2015dsm} and UDSM \cite{ishikawa2014udsm}, respectively.

\begin{figure}[tbp]
  \centering
  \includegraphics*[clip,scale=0.58]{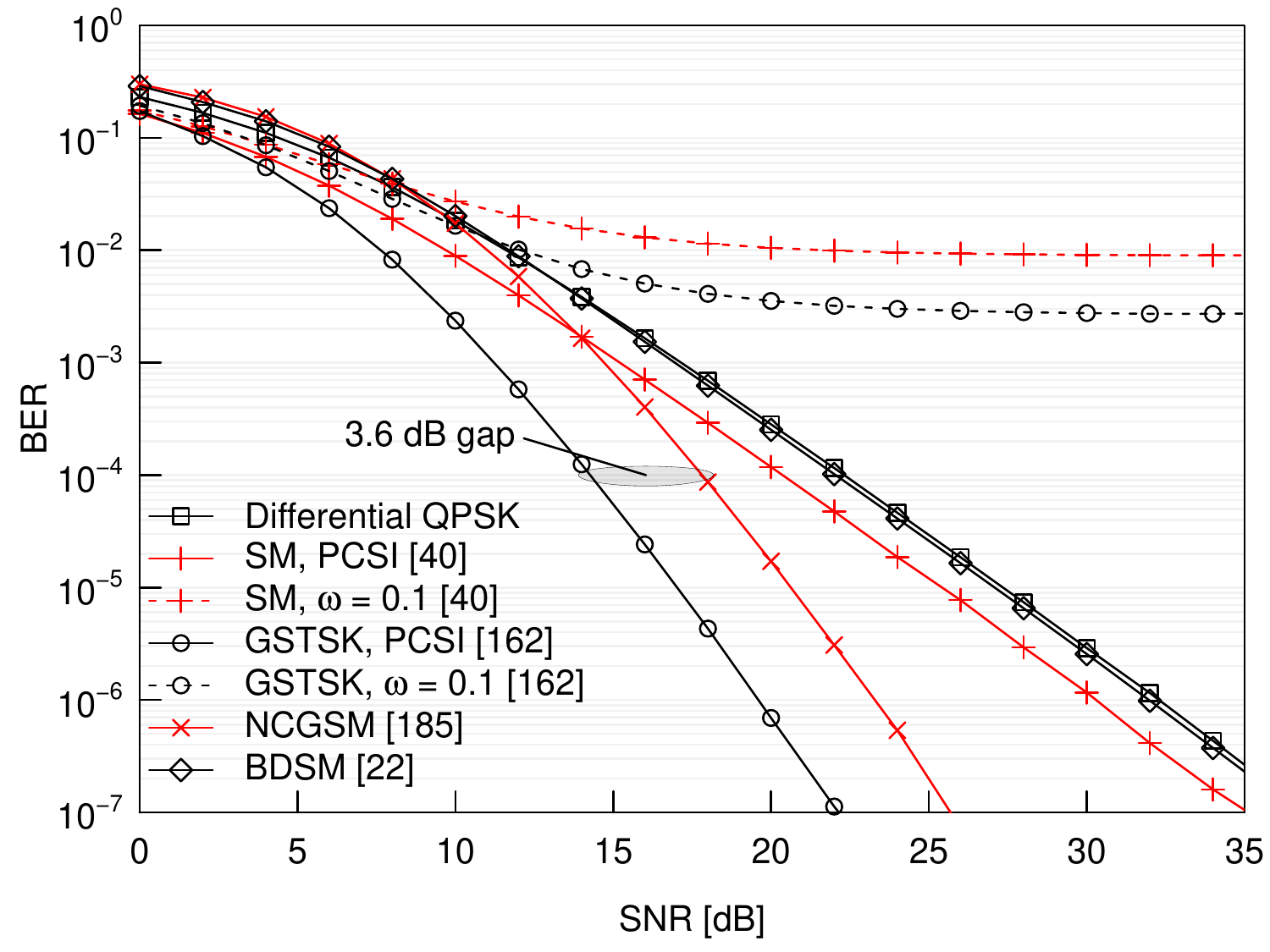}%
  \caption{BER comparisons of differential QPSK, BPSK-aided SM, BPSK-aided GSTSK(2,2,2,4,2), 2--PAM-aided NCGSM(2,2,2,4,2), and (BPSK,QPSK)-aided BDSM where the channel estimation errors were considered. The numbers of transmit and receive antennas were $(M,N)=(2,2)$. The transmission rate was $R=2.0$ [bits/symbol].\label{NCGSM:fig:coherent_vs_noncoherent}}
\end{figure}
Fig.~\ref{NCGSM:fig:coherent_vs_noncoherent} shows the BER of the NCGSM(2,2,2,4,2) and BDSM, both of which are introduced in Section~\ref{sec:app:diff:ncgsm}.
The BER curves of the coherent schemes and the differential QPSK were also plotted for reference.
The transmission rate was $R=2.0$ [bits/symbol].
Again, in Fig.~\ref{NCGSM:fig:coherent_vs_noncoherent}, we considered the effects of the channel estimation errors.
Specifically, for the \abbr{perfect CSI}{Perfect CSI}{PCSI} scenario, we assumed that the receiver had a perfect estimate of the channel matrix $\H(i)$.
By contrast, for the imperfect CSI scenario we assumed that the receiver had a realistic estimate of $\H(i)$, where the channel matrix was contaminated by the complex-valued AWGN of $\mathcal{CN}(0, \omega)$.
We observe in Fig.~\ref{NCGSM:fig:coherent_vs_noncoherent} that the coherent GSTSK achieved the best performance for PCSI scenario, but it exhibited an error floor for the imperfect CSI scenario characterized by a channel error variance of $\omega = 10$.
We observed the same trend for the SM scheme.
Hence, it is difficult for coherent MIMO schemes to attain a low BER, when the estimated channel matrix is inaccurate.
The differential MIMO schemes are free from the channel estimation errors.
It is shown in Fig.~\ref{NCGSM:fig:coherent_vs_noncoherent} that both the NCGSM and BDSM schemes were capable of operating without an error floor.
The performance gap between the GSTSK and NCGSM was 3.5 [dB], which was higher than 3.0 [dB].

\begin{figure}[tbp]
  \centering
  \includegraphics[clip,scale=0.58]{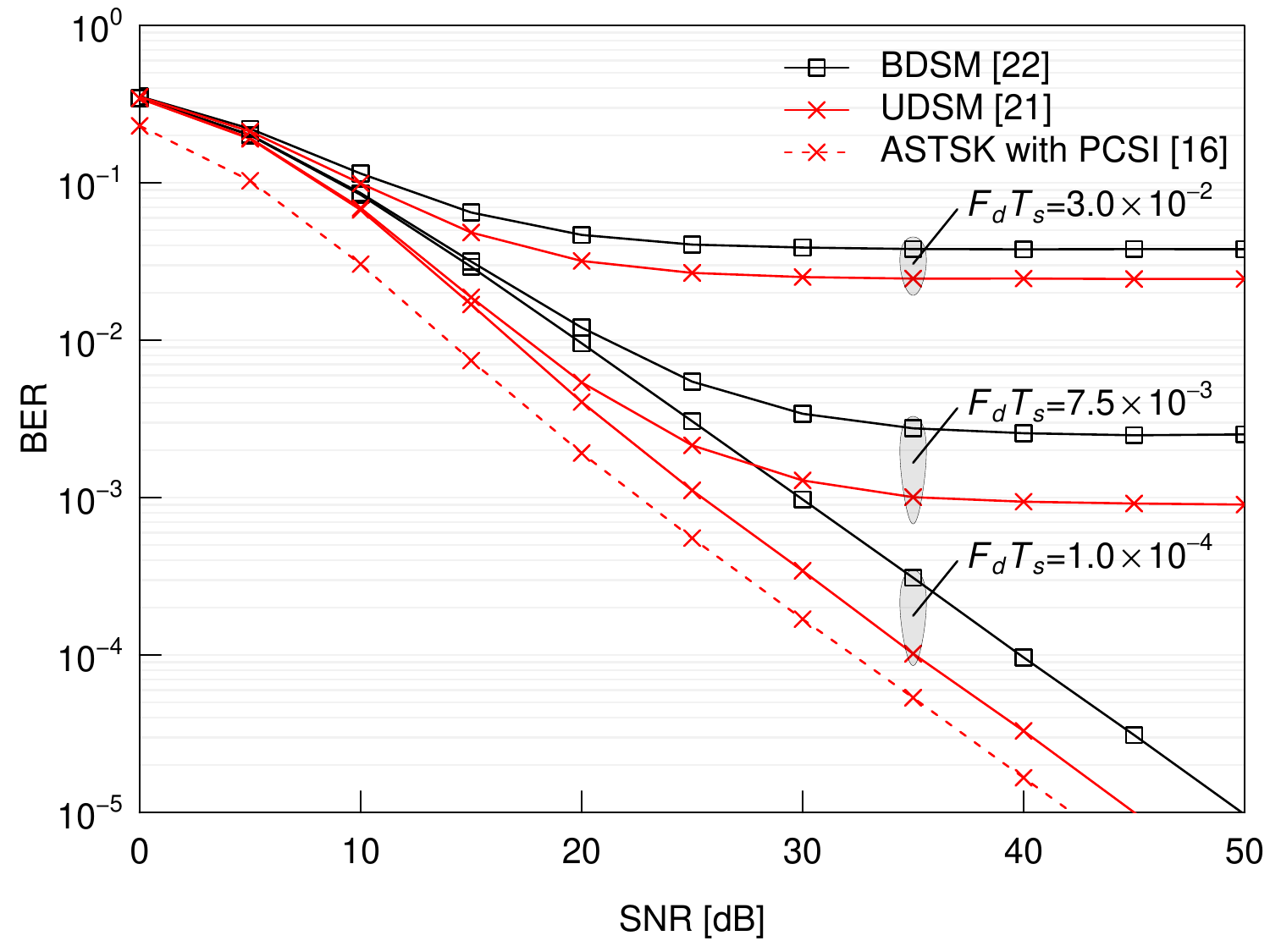}
  \caption{Effects of Jakes channels on BPSK-aided BDSM and UDSM having $(M,Q)=(2,2)$, where the number of scatterers was $8$. The normalized Doppler frequency was set to $F_dT_s = 1.0 \times 10^{-4}$, $7.5 \times 10^{-3}$, and $3.0 \times 10^{-2}$. The transmission rate was $R=1.5$ [bits/symbol].}
  \label{diff:fig:ber-M2-rate15-jakes}
\end{figure}
Fig.~\ref{diff:fig:ber-M2-rate15-jakes} shows the effects of Doppler frequency on the DSM schemes, where we considered the Jakes channel model of the Rayleigh fading to have $8$ scatterers.
In Fig.~\ref{diff:fig:ber-M2-rate15-jakes}, the normalized Doppler frequency was varied from $F_dT_s = 1.0 \times 10^{-4}$ to $3.0 \times 10^{-2}$ \cite{schober1999dpsk,hwang2005differential}.
As shown in Fig.~\ref{diff:fig:ber-M2-rate15-jakes}, both the BDSM and UDSM schemes of Fig.~\ref{diff:fig:system} exhibited error floors for the $F_dT_s \geq 7.5 \times 10^{-3}$ scenarios.
As the normalized Doppler frequency increased, the performance advantages of the UDSM scheme were maintained, as a benefit of its appropriately designed DMs.

\begin{figure}[tbp]
  \centering
  \includegraphics[clip,scale=0.58]{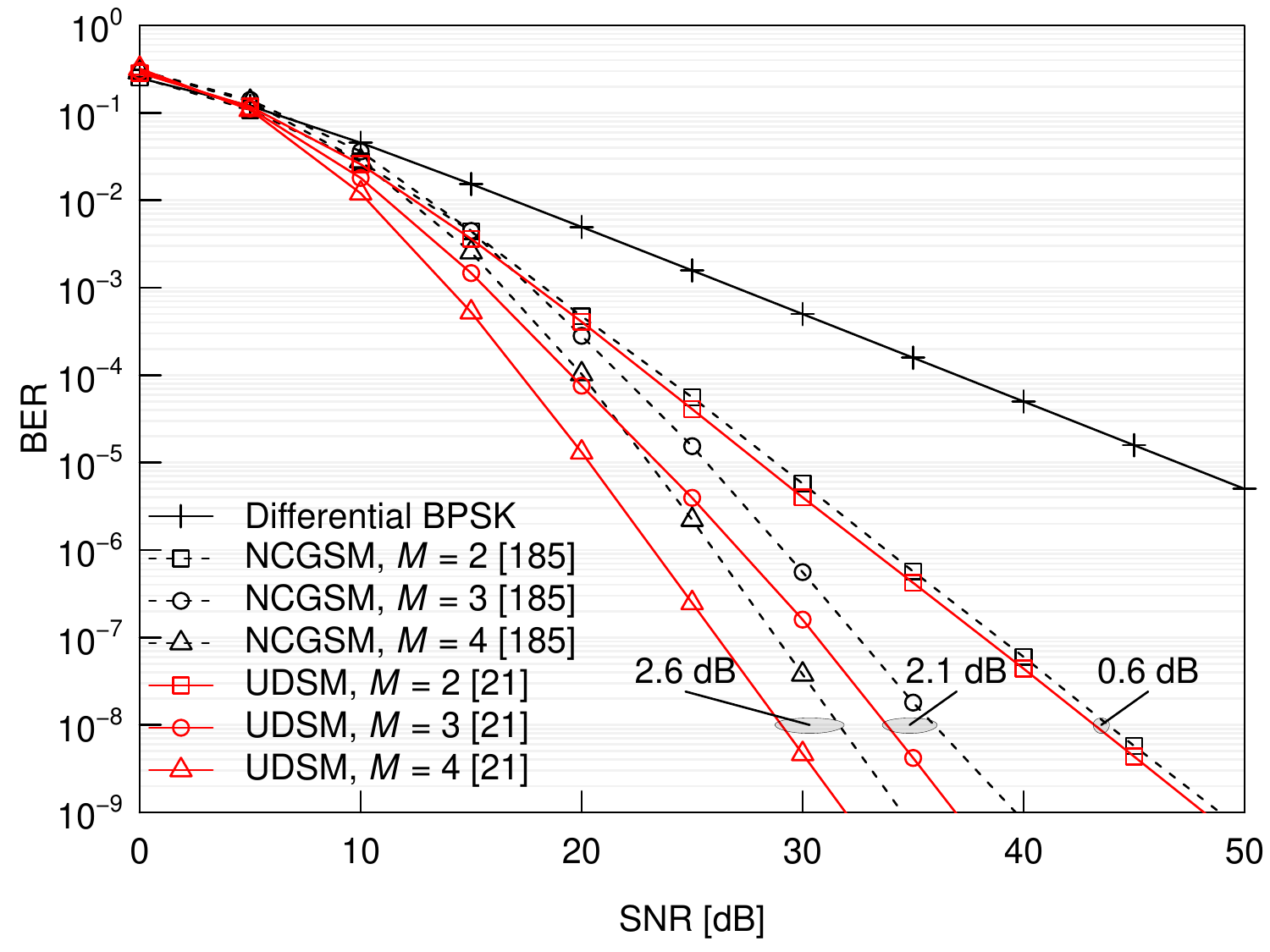}
  \caption{BER comparisons between the proposed UDSM and the conventional single-RF NCGSM, where $(M,Q)=(2,2),(3,4),(4,8)$. The BER curve of differential BPSK was plotted for reference. The transmission rate was set to $R=1.0$ [bits/symbol]. 
}
  \label{diff:fig:ber-compared-with-singlerfncgsm}
\end{figure}
Fig.~\ref{diff:fig:ber-compared-with-singlerfncgsm} shows the BER comparisons between UDSM and single-RF NCGSM detailed in Section~\ref{sec:app:diff}.
Both the UDSM and the single-RF NCGSM schemes were designed for a reduced-RF-chain transmitter.
The difference between both these two schemes is highlighted by the structure of the codewords in the diagonal and the permutation matrices, which were described in Section~\ref{sec:app:diff}.
In Fig.~\ref{diff:fig:ber-compared-with-singlerfncgsm}, the number of transmit antennas was set to $M=2,3,4$, and the number of receive antennas was constrained to $N=1$.
The BER curves of the BDSM of Fig.~\ref{diff:fig:system} were not listed because its transmission rate exceeds $R=1.0$ [bits/symbol].
It is shown in Fig.~\ref{diff:fig:ber-compared-with-singlerfncgsm} that the diversity orders of both UDSM and single-RF NCGSM improved upon increasing the number of transmit antennas $M$.
Furthermore, the performance gaps between UDSM and single-RF NCGSM increased upon increasing $M$, where the SNR gaps associated with $M=2$, 3, and 4 were $0.6$, $2.1$, and $2.6$ [dB], respectively, at $\mathrm{BER}=10^{-8}$.

\subsection{PM-Based MIMO-MWC\label{sec:comp:mwc}}
Third, we investigated the achievable performance of the PM-based MIMO-MWC scheme of Fig.~\ref{fig:pm-based-mwc} in terms of its AMI, where we considered both perfect and imperfect BF scenarios.

\begin{figure}[tbp]
  \centering
  \includegraphics[clip,scale=0.58]{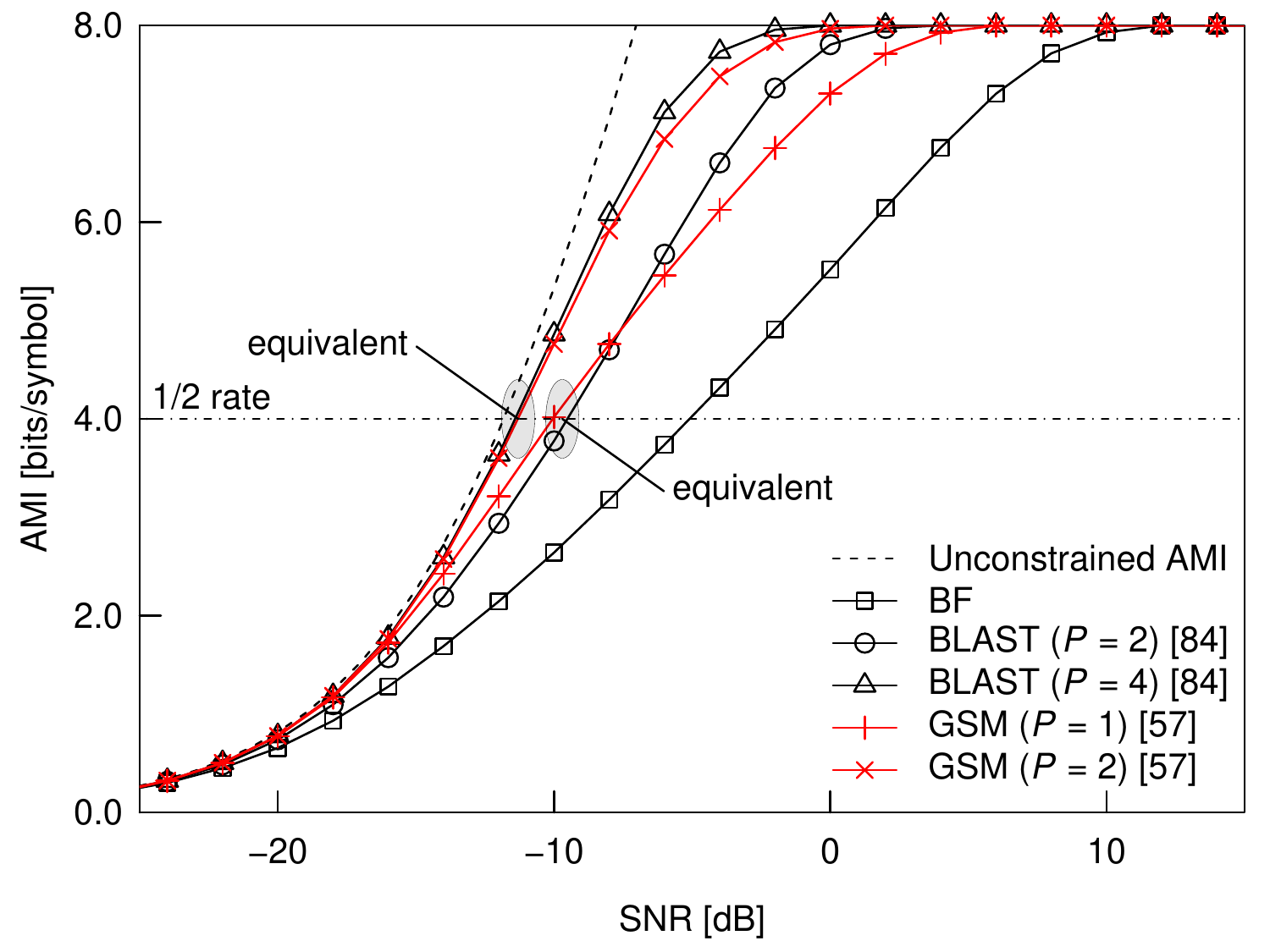}
  \caption{Constrained AMI comparisons between the GSM and the benchmark schemes. The transmitter and the receiver had $M_e=N_e=16$ antenna elements and $M=N=4$ separated ABFs. The transmission rate was $R=8.0$ [bits/symbol]. \copyright IEEE \cite{ishikawa2017mmgsm}\label{mm:fig:dcmc4x4-rate8}}
\end{figure}
Fig.~\ref{mm:fig:dcmc4x4-rate8} shows the constrained AMI comparisons of GSM, single-stream BF, and BLAST, where the number of antenna elements was $M_e=N_e=16$, and the number of subarrays was $M=N=4$.
The separation between the ABFs was set to $\DT=\DR=7.91$ [cm] based on Eq.~\eqref{mm:eq:spacing}.
In this scenario, the mean rank of the channel matrix was $\mathrm{rank}(\H) = \min(4, 4) = 4$.
In Fig.~\ref{mm:fig:dcmc4x4-rate8}, we considered 256--QAM-aided BF and BLAST associated with $P=2$ or 4, and GSM with $P=1$ or 2.
It is shown in Fig.~\ref{mm:fig:dcmc4x4-rate8} that the constrained AMI of the single-RF GSM scheme was nearly the same as that of the BLAST scheme for $P=2$ at the half-rate point of 4.0 [bits/symbol].
Furthermore, the constrained AMI of the GSM scheme associated with $P=2$ was similar to that of the BLAST scheme associated with $P=4$.
Hence, the performances of both GSM and BLAST are equivalent, if we assume the use of half-rate channel coding.

\begin{figure}[tbp]
	\centering
	\includegraphics[clip,scale=0.58]{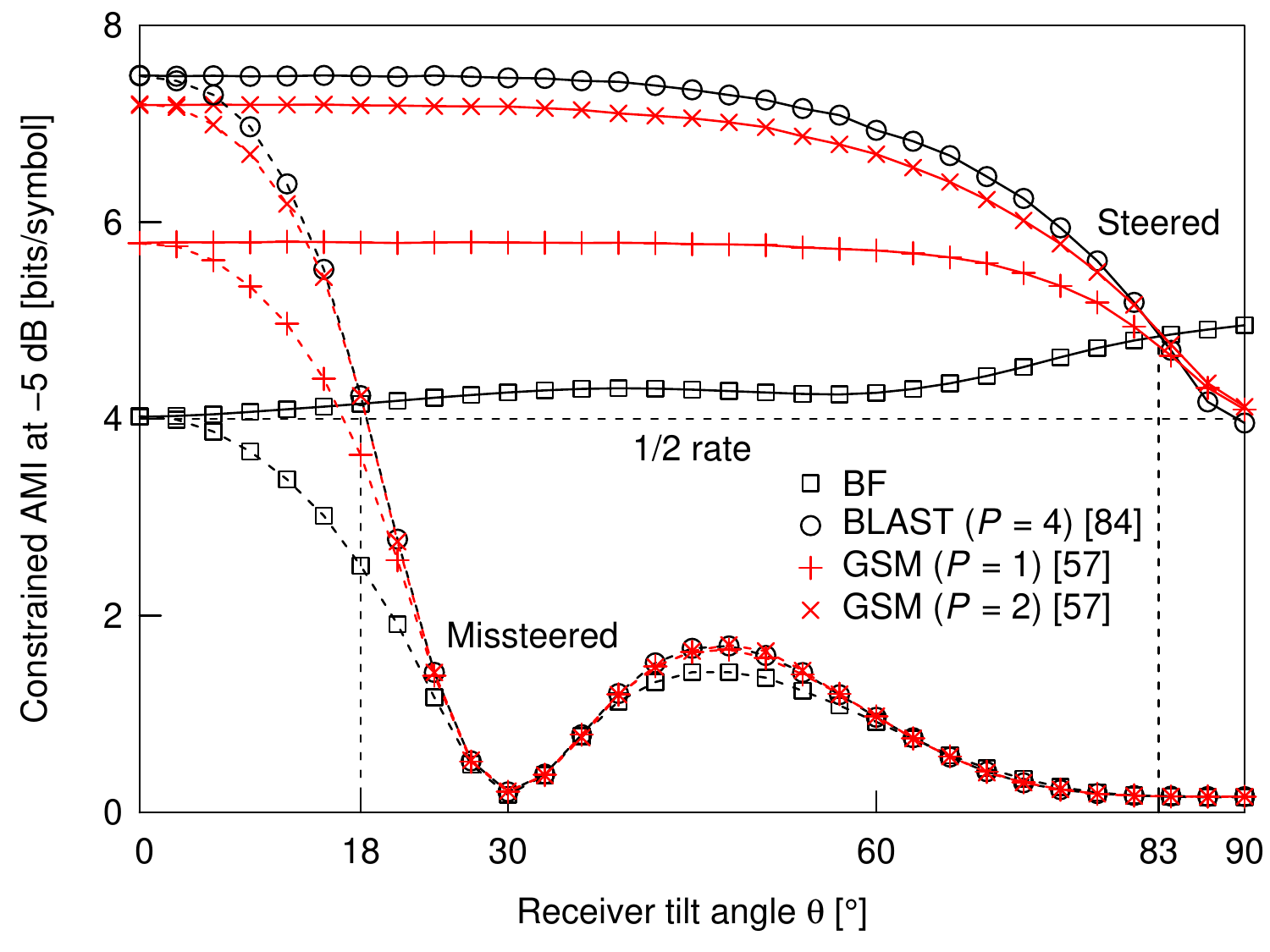}
	\caption{Effects of receiver tilt on the constrained AMI at the SNR of $-5$ [dB]. The system parameters are the same as in Fig.~\ref{mm:fig:dcmc4x4-rate8}. The constrained AMI at $\theta = 0^\circ$ corresponded to that shown in Fig.~\ref{mm:fig:dcmc4x4-rate8}, where $\mathrm{SNR}=-5$ [dB]. \copyright IEEE \cite{ishikawa2017mmgsm}}
	\label{mm:fig:tilt}
\end{figure}
Fig.~\ref{mm:fig:tilt} shows the effects of receiver tilt, defined by $\theta$ in Section~\ref{sec:sys:mwc}.
In Fig.~\ref{mm:fig:tilt}, the system parameters and the considered schemes were the same as those used in Fig.~\ref{mm:fig:dcmc4x4-rate8}.
The constrained AMI was calculated at $\mathrm{SNR}=-5$ [dB].
As shown in Fig.~\ref{mm:fig:tilt}, the constrained AMI of all schemes decreased from $\theta=0^\circ$ to $30^\circ$, which corresponds to the main lobe of the directional BF gain shown in Fig.~\ref{mm:fig:directivity}.
Fig.~\ref{mm:fig:tilt} also shows that the constrained AMI of the BLAST and GSM schemes were higher than the half-rate $4.0$ [bits/symbol] within the range of $0^\circ \leq \theta \leq 18^\circ$.
For the steered case, $\theta_\mathrm{AoD}$ and $\theta_\mathrm{AoA}$ were accurately adjusted based on perfect estimates of $\theta$.
As shown in Fig.~\ref{mm:fig:tilt}, the constrained AMI of the BLAST and of the GSM schemes remained high within the range of $0\le \theta < 83^\circ$.

\subsection{PM-Based MIMO-VLC\label{sec:comp:vlc}}
Fourth, we investigated the achievable performance of the PM-based MIMO-VLC scheme of Fig.~\ref{fig:pm-based-vlc} in terms of its MI and BER.
We also illustrate constellation examples of the OSM family.
Because the channel coefficients are static, we use the term {\it MI} instead of {\it AMI}.  

\begin{figure}
  \centering
  \includegraphics[clip,scale=0.58]{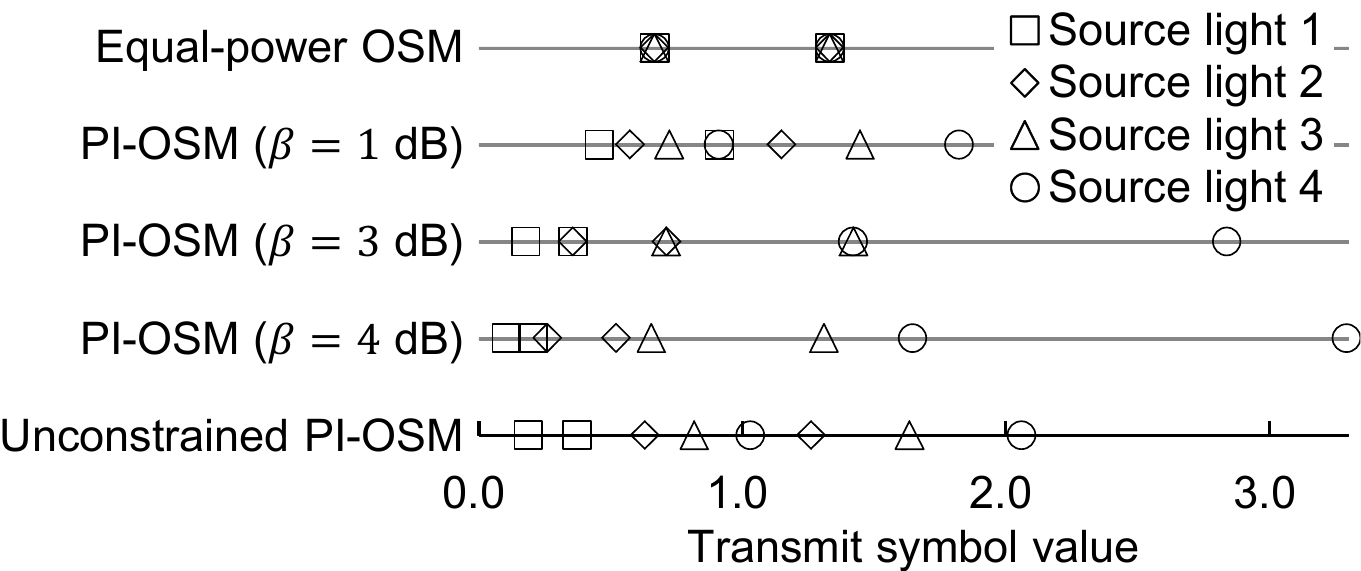}%
  \caption{Constellation examples of the equal-power OSM, the PI-OSM having a single parameter $\beta$ and the unconstrained PI-OSM. The number of transmit light sources was $M=4$. The size of the PAM constellation was $\mathcal{L}=2$. Each mark corresponds to the emitted index of the light source. \label{vlc:fig:constellation}}
\end{figure}
Fig.~\ref{vlc:fig:constellation} compares the constellations of the equal-power OSM scheme~\cite{mesleh2011osm}, of the PI-OSM \cite{fath2013comparison} having $\beta=1,~3,~4$ [dB] and of the unconstrained PI-OSM \cite{ishikawa2015osm}, where the number of transmit light sources was $M=4$, and the size of PAM symbols was $\mathcal{L}=2$.
We observe in Fig.~\ref{vlc:fig:constellation} that the constellations of the single-parameter PI-OSM scheme having $\beta=3$ and $4$ [dB] were biased.
This is because Eq.~\eqref{vlc:eq:fathimbalance} exponentially increased power upon increasing the single parameter $\beta$.
By contrast, the unconstrained PI-OSM scheme had a higher degree of freedom for designing the PA parameters.
As shown in Fig.~\ref{vlc:fig:constellation}, the constellations of the unconstrained PI-OSM scheme were uniformly distributed from 0.0 to 2.0, and the constellations were designed to maximize the AMI at the received SNR of $25$ [dB].

\begin{figure}[tbp]
  \centering
  \subfigure[MI comparison \copyright IEEE \cite{ishikawa2015osm}]{
    \includegraphics[clip,scale=0.58]{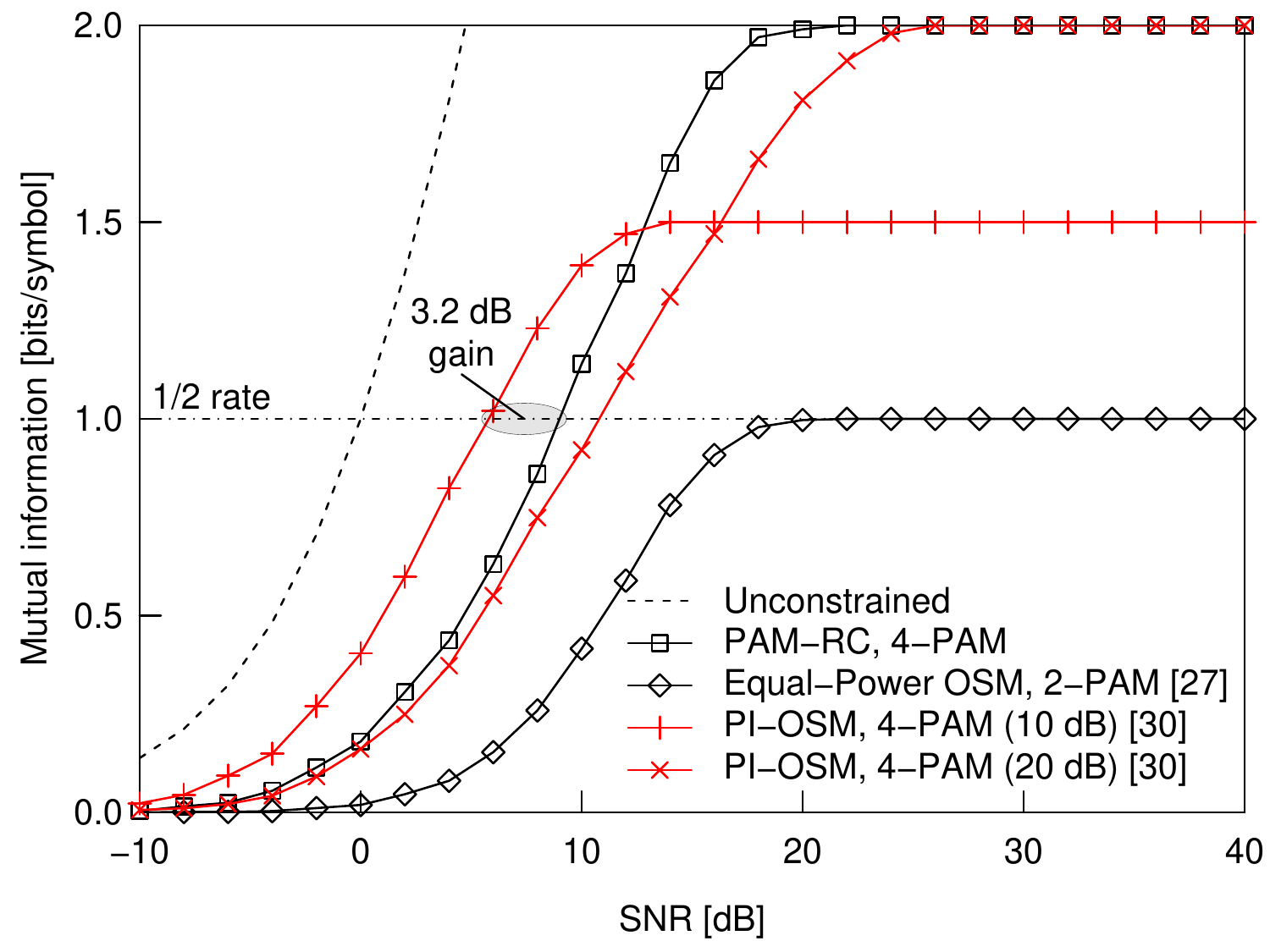}%
    \label{vlc:fig:dcmc-smallm}
  }
  \subfigure[BER comparison in uncoded scenarios]{
    \includegraphics[clip,scale=0.58]{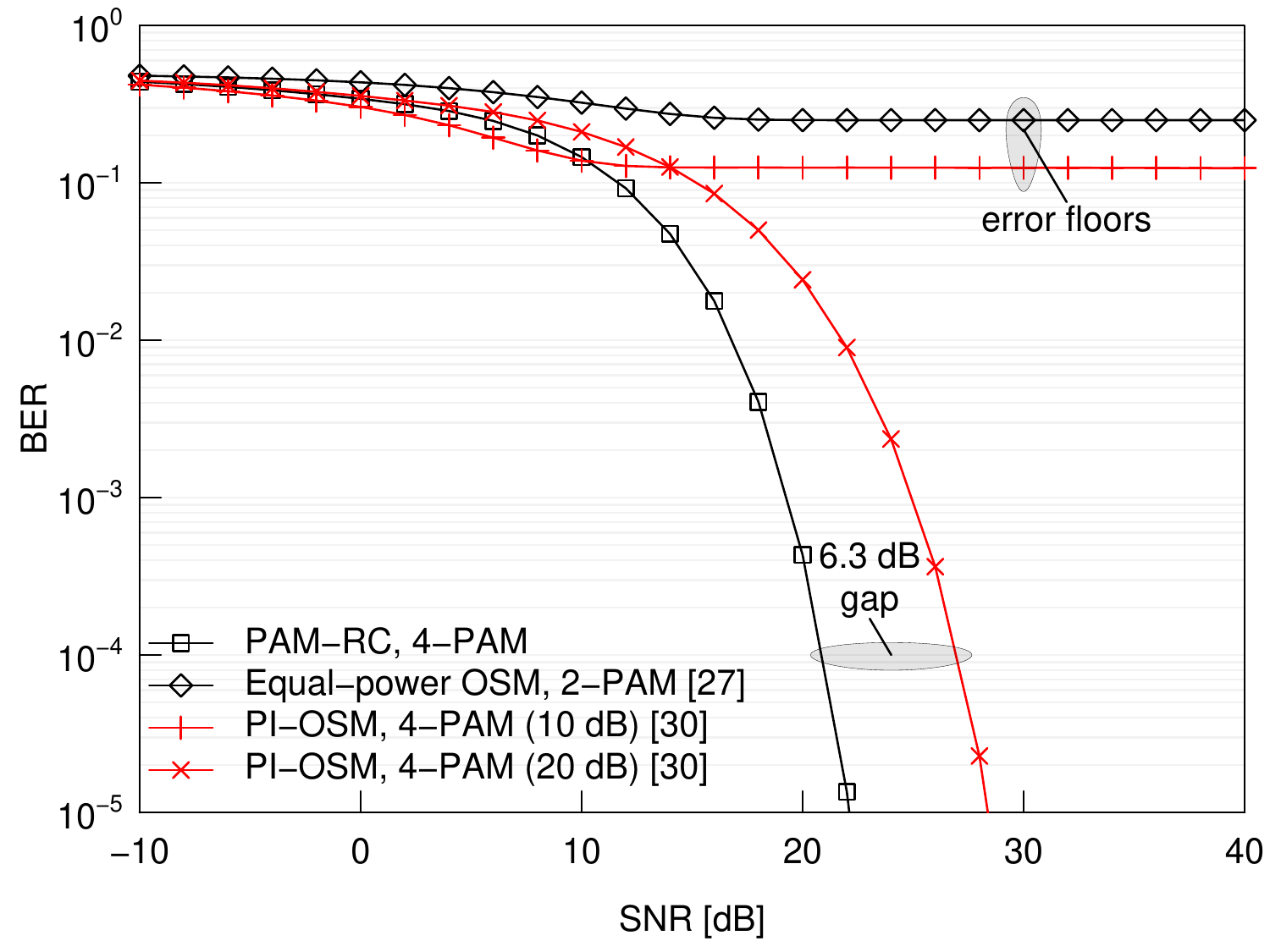}%
    \label{fig:survvlc:ber-m2}
  }
  \caption{BER and MI comparisons of the PAM-RC and the OSM schemes of Fig.~\ref{fig:pm-based-vlc}, where the transmission rate was $R=2.0$ [bits/symbol]. The numbers of light sources and receive PDs were $(M,N)=(2,1)$. The distance between the light sources and the PDs was $d_{\mathrm{Tx}}=d_{\mathrm{Rx}}=0.1$ [m].\label{fig:survvlc:ber-mi}}
\end{figure}
Fig.~\ref{vlc:fig:dcmc-smallm} shows the constrained MI of the $\mathcal{L}=4$--PAM-aided RC, of the $\mathcal{L}=2$--PAM-aided equal-power OSM and of the $\mathcal{L}=2$--PAM-aided unconstrained PI-OSM schemes, where $(M,N)=(2,1)$ and $R=2.0$ [bits/symbol].
We designed the PA matrix of the PI-OSM scheme both at the low SNR of 10 [dB] and at the high SNR of 20 [dB] in order to investigate the effects of the target SNR.
It is shown in Fig.~\ref{vlc:fig:dcmc-smallm} that the PI-OSM scheme designed at $\mathrm{SNR}=10$ [dB] achieved the best constrained MI at the effective throughput of 1.0 [bits/symbol], when using half-rate coding.
Note that the PAM-RC scheme achieved the best performance in the SNR region between 13 and 27 [dB], where the PI-OSM scheme designed to operate at $\mathrm{SNR}=20$ [dB] performed worse than the PAM-RC scheme.
Due to the high correlations between the channel coefficients, the conventional equal-power OSM scheme only achieved a throughput of 1.0 [bits/symbol] at high SNRs.
In Fig.~\ref{fig:survvlc:ber-m2}, we compared the BER of the schemes considered in Fig.~\ref{vlc:fig:dcmc-smallm}. The simulation parameters were the same as those used in Fig.~\ref{vlc:fig:dcmc-smallm}.
We observe in Fig.~\ref{fig:survvlc:ber-m2} that the PAM-RC scheme achieved the BER of $10^{-4}$ at 20.97 [dB], whereas the PI-OSM scheme designed for operation at $\mathrm{SNR}=20$ [dB] achieved it at 27.05 [dB], where a 6.3 [dB] SNR gap existed.
Furthermore, the PI-OSM scheme designed for operation at $\mathrm{SNR}=10$ [dB] exhibited an error floor.
Hence, the PAM-RC scheme was superior to PI-OSM in this uncoded scenario.
However, as shown in the MI comparison of Fig.~\ref{vlc:fig:dcmc-smallm}, the PI-OSM scheme exhibited a 3.2 dB gain over PAM-RC, which implies that the PI-OSM scheme is expected to be superior to PAM-RC in the channel-coded scenarios.

\subsection{PM-Based Multicarrier Communications\label{sec:comp:ofdm}}
Finally, we investigated the achievable performance of the PM-based OFDM scheme of Fig.~\ref{fig:pm-based-ofdm} in terms of its AMI and BER, where turbo coding was considered.

\begin{figure}[tbp]
  \centering
  \subfigure[The reciprocal of MED]{
    \includegraphics[clip,scale=0.58]{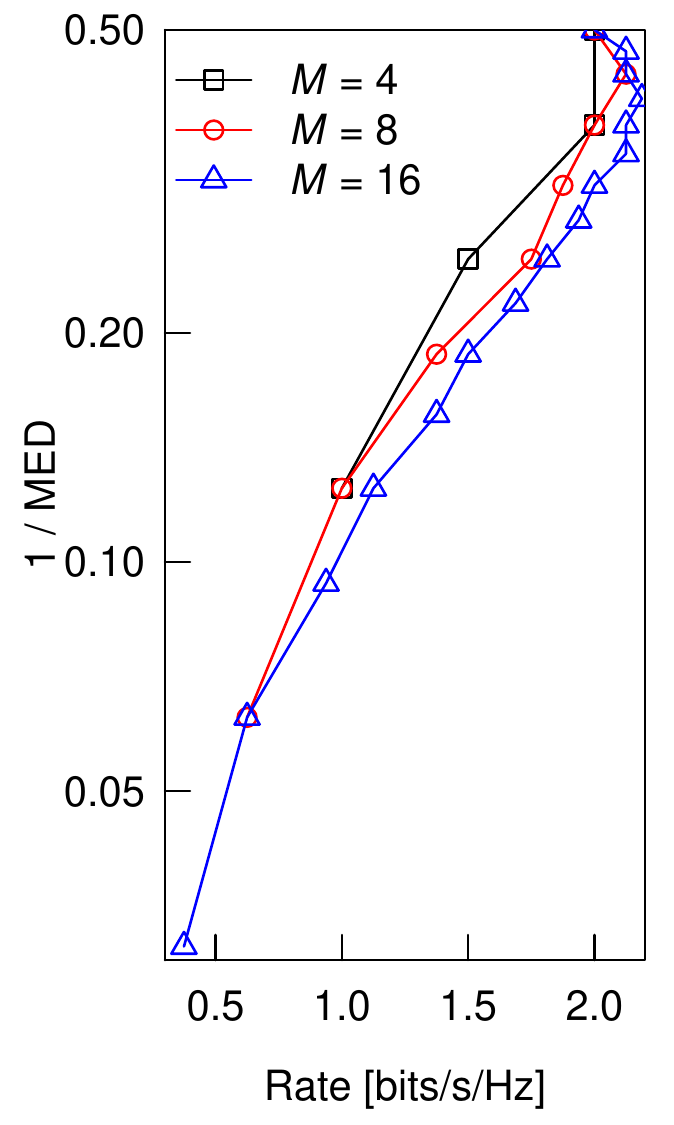}%
  }
  \subfigure[BER at 30 dB]{
    \includegraphics[clip,scale=0.58]{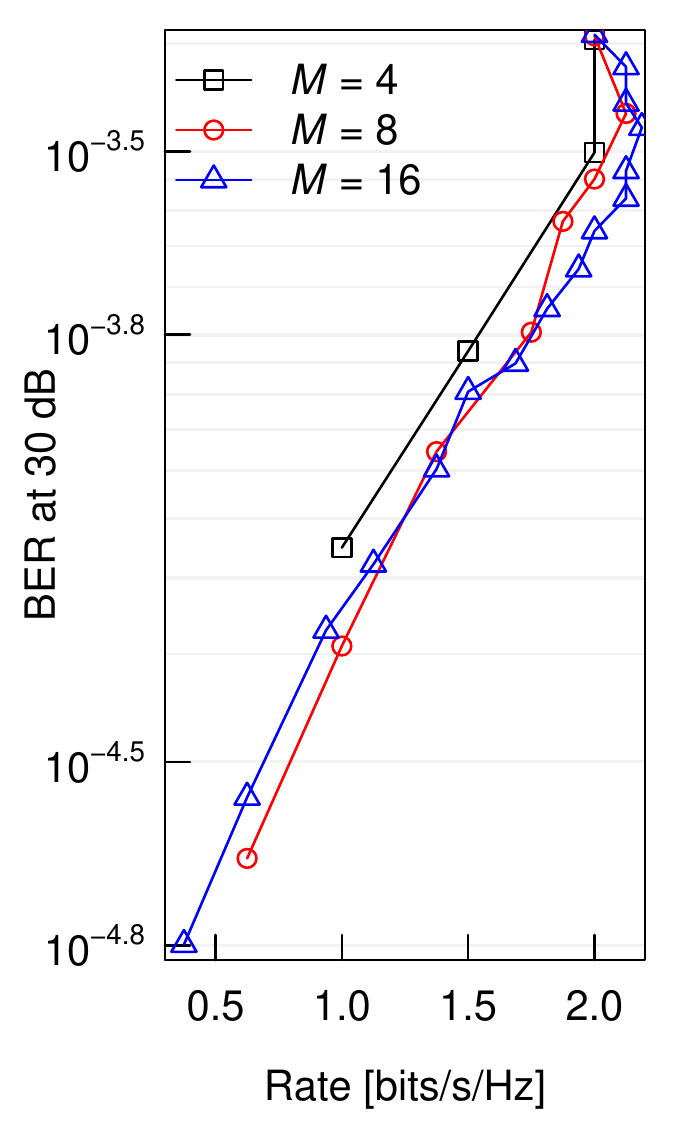}%
  }
  \caption{Correlation between MED and BER, where the constellation size was $\mathcal{L}=4$. The number of subcarriers was set to $M=4,~8,~16$, and the number of selected subcarriers was changed from $P=1$ to $M$.\label{sim:fig:corr-med-ber}}
\end{figure}
Fig.~\ref{sim:fig:corr-med-ber}(a) shows the MED comparison of the QPSK-aided SIM scheme having $M=4,~8,$ and $16$, where the number of selected subcarriers was increased from $P=1$ to $M$.
Fig.~\ref{sim:fig:corr-med-ber}(b) shows the simulated BER of the SIM setups considered in Fig.~\ref{sim:fig:corr-med-ber}(a), where the received SNR was 30 [dB].
Figs.~\ref{sim:fig:corr-med-ber}(a) and (b) show that a correlation existed between the reciprocal of the MED and of the BER. 
Thus, the MED is a useful metric for predicting the BER in uncoded scenarios.

\begin{figure}[tbp]
  \centering
  \subfigure[AMI comparison \copyright IEEE \cite{ishikawa2016sim}]{
    \includegraphics[clip,scale=0.58]{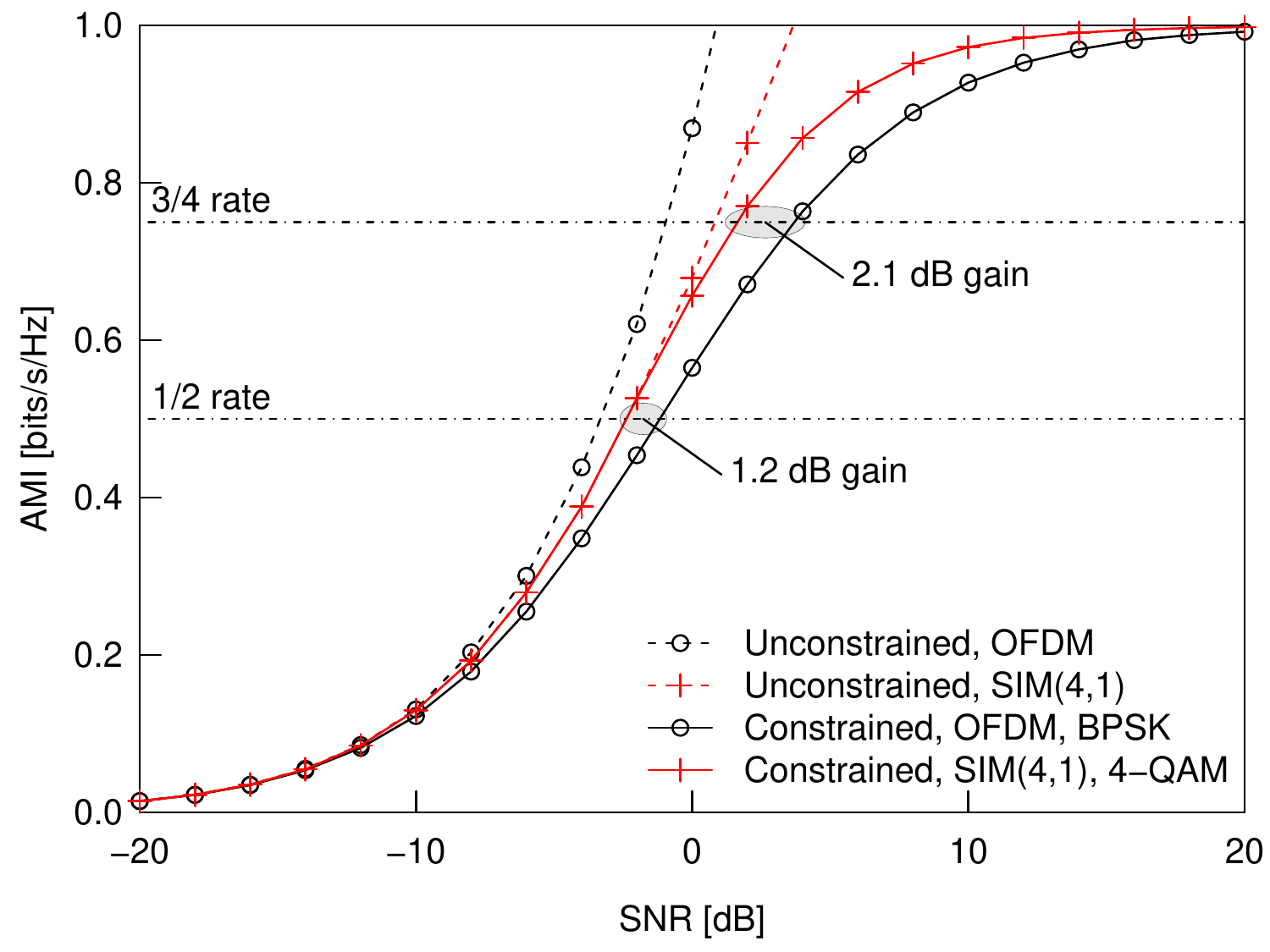}%
    \label{fig:ofdm:ami}
  }
  \subfigure[BER comparison in coded scenarios]{
    \includegraphics[clip,scale=0.58]{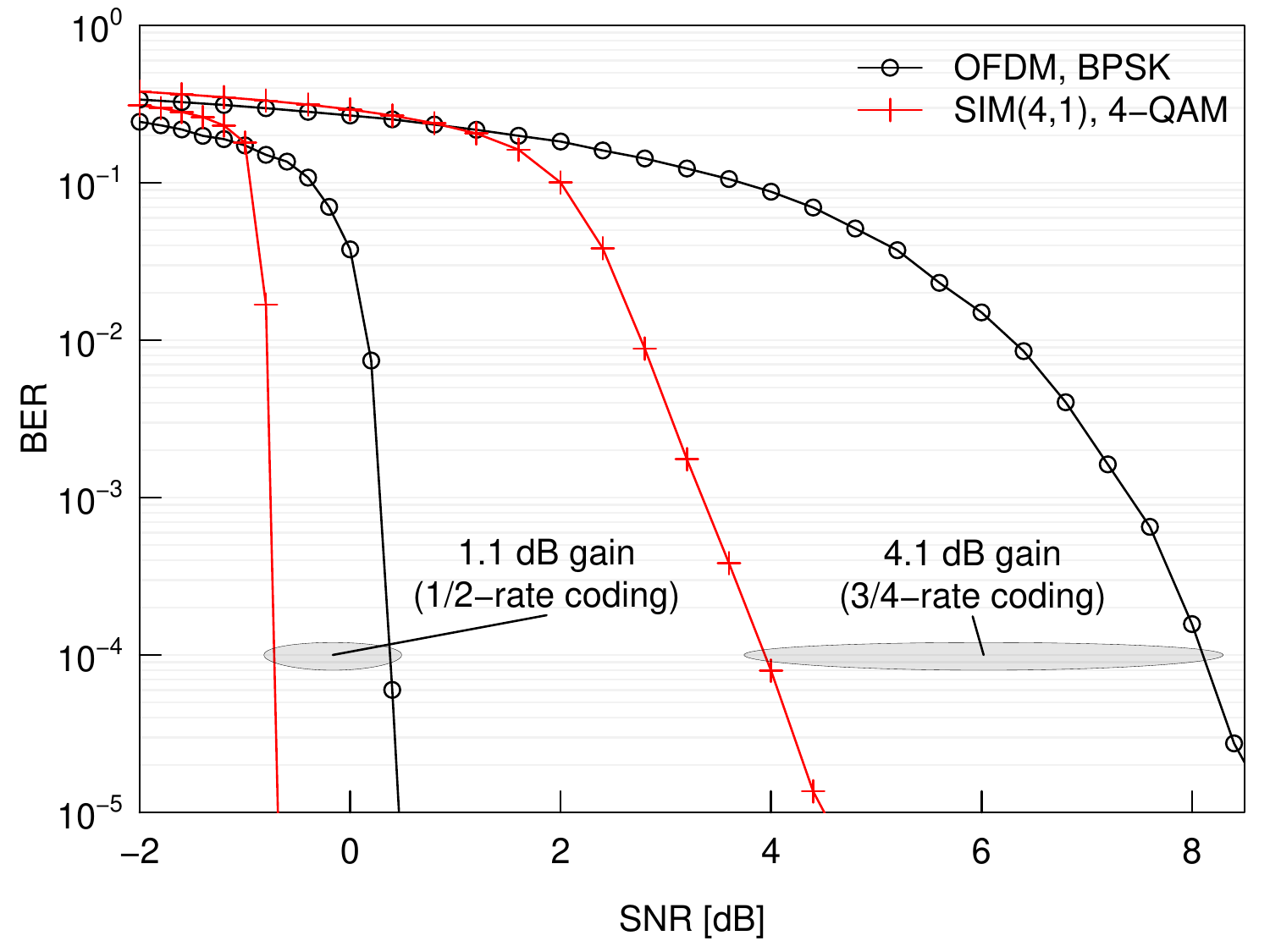}%
    \label{fig:ofdm:ber}
  }
  \caption{BER and AMI comparisons of the SIM and the OFDM schemes of Fig.~\ref{fig:pm-based-ofdm}.\label{fig:ofdm:ber-ami}}
\end{figure}
Fig.~\ref{fig:ofdm:ami} shows the unconstrained and constrained AMI comparisons of the BPSK-aided OFDM and 4--QAM-aided SIM(4,1) schemes in Rayleigh fading channels.
Fig.~\ref{fig:ofdm:ami} shows that the unconstrained AMI of OFDM exceeded that of SIM right across the entire SNR region.
The reason for this is that the SIM scheme transmitted a reduced number of Gaussian streams compared to OFDM.
However, as shown in Fig.~\ref{fig:ofdm:ami}, the constrained AMI of SIM was higher than that of OFDM.
Specifically, the performance gains of 1.2 and 2.1 [dB] were achieved at the coding rates of $1/2$ and $3/4$, respectively.
This observation implies that the performance gain of channel-coded scenarios increases with the increase of the coding rate.
Similarly, in Fig.~\ref{fig:ofdm:ber}, we investigated the BER performance of the three-stage turbo-coded SIM and OFDM schemes.
The schemes considered in Fig.~\ref{fig:ofdm:ber} were the same as those used in Fig.~\ref{fig:ofdm:ami}.
Fig.~\ref{fig:ofdm:ber} shows that the SIM scheme achieved lower turbo-cliff SNRs than the OFDM scheme, where gaps of 1.1 and 4.1 [dB] existed for the 1/2- and 3/4-rate coding cases, respectively. These performance gains were nearly the same as those anticipated in Fig.~\ref{fig:ofdm:ami}.

\section{Conclusions\label{sec:conc}}
\subsection{Summary and Design Guidelines}
In this treatise, we provided an interdisciplinary survey of the PM family that has been proposed for coherent MIMO, differential MIMO, MIMO-MWC, MIMO-VLC and multicarrier RF communications.
The PM concept was originally proposed by Slepian in 1965 and has since flourished in data storage research.
All of the conventional PM, PC, SM, and SIM schemes rely on a common structure that maps the input bits by selecting a permutation of an arbitrary set, which consists of ``on'' and ``off'' states, for example.
In data storage studies, this unique structure has achieved a storage capacity gain, while maintaining low-latency reading and writing with the aid of the reduced number of ``on'' states.
In digital communication studies, this structure has achieved low-complexity encoding and decoding with the aid of the reduced number of data streams, where the transmission rate is kept the same as that in the conventional schemes.
Our hope is that this interdisciplinary survey may inspire you new colleagues to join us in this research field.

\begin{table}[tbp]
  \centering
  \footnotesize
  \caption{Summary of our discussions.\label{table:discussion}}
  \begin{tabular}{|l|c|c|c|}
    \hline
    & Single-stream & PM family & Full-complexity \\
    \hline \hline
    MED & Fair & \bf{Best} & Good \\
    Unconstrained AMI & Fair & Good & \bf{Best}\\
    Constrained AMI & Fair & \bf{Best} & Good\\
    Transmitter complexity & \bf{Best} & Good & Fair\\ 
    Detection complexity & \bf{Best} & Good & Fair\\
    BF gain\footnotemark & \bf{Best} & Fair & Good\\
    \hline
  \end{tabular}
\end{table}
\footnotetext{This metric is only valid for coherent microwave communications.}
Our simulation results demonstrated the fundamental tradeoffs of the PM-aided systems, which determine the design guidelines, in terms of both the theoretical analyses and the numerical simulations.
Our discussions are summarised in Table~\ref{table:discussion}.
As described in Section~\ref{sec:philo}, the PM encoding process selects $P$ number of elements out of a $Q$-sized arbitrary set.
Here, the design of $P$ for $1 \leq P \leq Q$ provides additional flexibilities that affect the system transmission rate, reliability, AMI, hardware complexity at the transmitter and computational complexity at the receiver.
The transmission rate of the PM-aided coherent MIMO logarithmically increases upon increasing the number of transmit antennas, while maintaining a reduced complexity.
Furthermore, the PM-aided coherent MIMO achieves a higher coding gain than the conventional SMX scheme in the entire transmission rate region. 
In any scenario, the reduced number of data streams enables low-complexity detection at the receiver.
However, since the number of Gaussian streams is reduced, the unconstrained AMI is reduced.
In our simulations, the constrained AMI of the PM-aided family, in place of the unconstrained AMI, was shown to have advantages over the conventional schemes.

\subsection{Suggestions for Future Research}
\subsubsection{Cross-Pollination of SM and Data Storage Research}
The analogy between both the SM and PM schemes could certainly attract researchers to these fields.
Many attractive low-complexity detectors have been proposed for the SM scheme.
These schemes may be exported to PM-aided data storage systems to achieve low-latency reading.
Furthermore, the low-latency reading and writing algorithms of PM-aided data storage may also be exported to SM-aided MIMO systems.
This interaction would contribute to the development of both data storage and digital communications.

\subsubsection{Large-Scale High-Rate PM-Based Differential MIMO}
The PM-based differential MIMO communications have to rely on sparse square matrices.
This square constraint $M=T$ limits the design of space-time codewords.
It has been known that the square-matrix-based differential schemes achieve bad performances for large-scale high-rate scenarios.
At the time of writing, the only exception is found in \cite{ishikawa2017rdsm}, which proposed a nonsquare-matrix-based differential MIMO.
This nonsquare scheme  achieved a competitive performance for the $M=1024$ antennas and $R=12$ [bits/symbol] scenario.
However, the nonsquare scheme of \cite{ishikawa2017rdsm} suffers from an error-propagation issue when we consider a small number of receive antennas or a high Doppler shift.

\subsubsection{Striking the Tradeoff between Sparsity and Capacity}
The PM-based OFDM scheme transmits $P$ independent data streams over $M$ orthogonal subcarriers.
Thus, the unconstrained AMI of PM becomes inevitably smaller than the classic OFDM scheme, which transmits $M$ Gaussian streams over $M$ subcarriers.
Our simulation results showed that the constrained AMI of PM becomes better than OFDM for a specific scenario, but this achievable AMI gain may vanish for high-rate scenarios.
The sparsity of PM codewords improves the computational complexity at the receiver, while it also decreases the unconstrained AMI.
Hence, the multiple-mode IM \cite{mao2016sim,wen2017mode} and the compressed-sensing-assisted IM \cite{zhang2016cs} are promising for high-rate scenarios because these construct a dense symbol in the frequency domain.
This dense construction improves the corresponding unconstrained AMI, while it also induces a complexity issue at the same time.

\section*{Acknowledgment}
The authors are indebted to Prof. H. Ochiai, Yokohama National University, Japan for his comment on the parallel combinatory concept. Thanks are also due to the anonymous reviewers for their valuable suggestions and comments, which improved this treatise.

\section*{Glossary\label{sec:glo}} 

{\small
\begin{description}[\IEEEsetlabelwidth{so so long}\IEEEusemathlabelsep]
  
\item[ABF] Analog BeamForming
\item[AMI] Average Mutual Information
\item[APSK] Amplitude and Phase-Shift Keying
\item[ASTSK] Asynchronous Space-Time Shift Keying
\item[AWGN] Additive White Gaussian Noise
\item[BDSM] Binary Differential Spatial Modulation
\item[BER] Bit Error Ratio
\item[BF] BeamForming
\item[BLAST] Bell-Laboratories Layered Space-Time
\item[BPSK] Binary Phase-Shift Keying
\item[CDMA] Code-Division Multiple Access
\item[CMOS] Complementary Metal Oxide Semiconductor
\item[CSI] Channel State Information
\item[DBF] Digital BeamForming
\item[DM] Dispersion Matrix
\item[DSM] Differential Spatial Modulation
\item[DSTBC] Differential Space-Time Block Code
\item[DSTSK] Differential Space-Time Shift Keying
\item[FSK] Frequency Shift Keying
\item[GSM] Generalized Spatial Modulation
\item[GSSK] Generalized Space Shift Keying
\item[GSTSK] Generalized Space-Time Shift Keying
\item[IFFT] Inverse Fast Fourier Transform
\item[i.i.d.] independent and identically distributed
\item[IM] Index Modulation
\item[LDC] Linear Dispersion Code
\item[LED] Light Emitting Diode
\item[LoS] Line-of-Sight
\item[LUT] Look-Up Table
\item[MAP] Maximum {\it a Posteriori}
\item[MBM] Media-Based Modulation
\item[MED] Minimum Euclidean Distance
\item[MI] Mutual  Information
\item[MIMO] Multiple-Input Multiple-Output
\item[ML] Maximum-Likelihood
\item[MWC]  Millimeter Wave Communication
\item[NBC] Natural Binary Code
\item[NCGSM] Non-Coherent Generalized Spatial Modulation
\item[NLoS] Non-Line-of-Sight
\item[OFDM] Orthogonal Frequency-Division Multiplexing
\item[OSM] Optical Spatial Modulation
\item[OSTBC] Orthogonal Space-Time Block Code
\item[PAM-RC] PAM Repetition-Code
\item[PAM] Pulse Amplitude Modulation
\item[PA] Power Allocation
\item[PAPR] Peak-to-Average Power Ratio
\item[PC] Parallel Combinatory
\item[PCSI] Perfect Channel State Information
\item[PD] PhotoDetector
\item[PEP] Pairwise-Error Probability
\item[PI] Power-Imbalanced
\item[PM] Permutation Modulation
\item[PPM] Pulse Position Modulation
\item[QAM] Quadrature Amplitude Modulation
\item[QPSK] Quaternary Phase-Shift Keying
\item[RC] Repetition Code
\item[RF] Radio Frequency
\item[SIM] Subcarrier Index Modulation
\item[SISO] Single-Input Single-Output
\item[SM] Spatial Modulation
\item[SMX] Spatial MultipleXing
\item[SNR] Signal-to-Noise Ratio
\item[SSK] Space Shift Keying
\item[STBC]   Space-Time Block Code
\item[STSK] Space-Time Shift Keying
\item[VLC]   Visible Light Communication
\end{description}}

\bibliographystyle{IEEEtran}
\bibliography{library}

\begin{thebibliography}{100}
\providecommand{\url}[1]{#1}
\csname url@samestyle\endcsname
\providecommand{\newblock}{\relax}
\providecommand{\bibinfo}[2]{#2}
\providecommand{\BIBentrySTDinterwordspacing}{\spaceskip=0pt\relax}
\providecommand{\BIBentryALTinterwordstretchfactor}{4}
\providecommand{\BIBentryALTinterwordspacing}{\spaceskip=\fontdimen2\font plus
\BIBentryALTinterwordstretchfactor\fontdimen3\font minus
  \fontdimen4\font\relax}
\providecommand{\BIBforeignlanguage}[2]{{%
\expandafter\ifx\csname l@#1\endcsname\relax
\typeout{** WARNING: IEEEtran.bst: No hyphenation pattern has been}%
\typeout{** loaded for the language `#1'. Using the pattern for}%
\typeout{** the default language instead.}%
\else
\language=\csname l@#1\endcsname
\fi
#2}}
\providecommand{\BIBdecl}{\relax}
\BIBdecl

\bibitem{ishikawa2017thesis}
N.~Ishikawa, ``{Space-, time-, and frequency-domain permutation modulation
  designed for microwave and optical wireless communications},'' Ph.D.
  dissertation, Tokyo University of Agriculture and Technology, 2017.

\bibitem{renzo2011commag}
M.~D. Renzo, H.~Haas, and P.~Grant, ``{Spatial modulation for multiple-antenna
  wireless systems: A survey},'' \emph{IEEE Communications Magazine}, vol.~49,
  no.~12, pp. 182--191, 2011.

\bibitem{renzo2014spatial}
M.~{Di Renzo}, H.~Haas, A.~Ghrayeb, S.~Sugiura, and L.~Hanzo, ``{Spatial
  modulation for generalized MIMO: Challenges, opportunities, and
  implementation},'' \emph{Proceedings of the IEEE}, vol. 102, no.~1, pp.
  56--103, 2014.

\bibitem{yang2015smdesign}
P.~Yang, M.~{Di Renzo}, Y.~Xiao, S.~Li, and L.~Hanzo, ``{Design guidelines for
  spatial modulation},'' \emph{IEEE Communications Surveys {\&} Tutorials},
  vol.~17, no.~1, pp. 6--26, 2015.

\bibitem{yang2016sm}
P.~Yang, Y.~Xiao, Y.~L. Guan, K.~V. Hari, A.~Chockalingam, S.~Sugiura, H.~Haas,
  M.~{Di Renzo}, C.~Masouros, Z.~Liu, L.~Xiao, S.~Li, and L.~Hanzo,
  ``{Single-carrier SM-MIMO: A promising design for broadband large-scale
  antenna systems},'' \emph{IEEE Communications Surveys {\&} Tutorials},
  vol.~18, no.~3, pp. 1687--1716, 2016.

\bibitem{kadir2015stsk}
M.~I. Kadir, S.~Sugiura, S.~Chen, and L.~Hanzo, ``{Unified MIMO-multicarrier
  designs: A space–time shift keying approach},'' \emph{IEEE Communications
  Surveys {\&} Tutorials}, vol.~17, no.~2, pp. 550--579, 2015.

\bibitem{renzo2016smbook}
M.~Di~Renzo, H.~Haas, A.~Ghrayeb, L.~Hanzo, and S.~Sugiura, \emph{Spatial
  Modulation for Multiple-Antenna Communication}.\hskip 1em plus 0.5em minus
  0.4em\relax Wiley Encyclopedia of Electrical and Electronics Engineering,
  2016.

\bibitem{basar2017im}
E.~Basar, M.~Wen, R.~Mesleh, M.~{Di Renzo}, Y.~Xiao, and H.~Haas, ``{Index
  modulation techniques for next-generation wireless networks},'' \emph{IEEE
  Access}, vol.~5, pp. 16\,693--16\,746, 2017.

\bibitem{rajashekar2014reduced}
R.~Rajashekar, K.~V.~S. Hari, and L.~Hanzo, ``{Reduced-complexity ML detection
  and capacity-optimized training for spatial modulation systems},'' \emph{IEEE
  Transactions on Communications}, vol.~62, no.~1, pp. 112--125, 2014.

\bibitem{sugiura2011rcc}
S.~Sugiura, C.~Xu, S.~Ng, and L.~Hanzo, ``{Reduced-complexity coherent versus
  non-coherent QAM-aided space-time shift keying},'' \emph{IEEE Transactions on
  Communications}, vol.~59, no.~11, pp. 3090--3101, 2011.

\bibitem{wen2015dsm}
M.~Wen, X.~Cheng, Y.~Bian, and H.~V. Poor, ``{A low-complexity near-ML
  differential spatial modulation detector},'' \emph{IEEE Signal Processing
  Letters}, vol.~22, no.~11, pp. 1834--1838, 2015.

\bibitem{zhang2016dsm}
M.~Zhang, M.~Wen, X.~Cheng, and L.~Yang, ``{A dual-hop virtual MIMO
  architecture based on hybrid differential spatial modulation},'' \emph{IEEE
  Transactions on Wireless Communications}, vol.~15, no.~9, pp. 6356--6370,
  2016.

\bibitem{lee2015smsphere}
K.~Lee, ``{Doubly ordered sphere decoding for spatial modulation},'' \emph{IEEE
  Communications Letters}, vol.~19, no.~5, pp. 795--798, 2015.

\bibitem{xu2017survdec}
C.~Xu, S.~Sugiura, S.~X. Ng, P.~Zhang, L.~Wang, and L.~Hanzo, ``{Two decades of
  MIMO design tradeoffs and reduced-complexity MIMO detection in near-capacity
  systems},'' \emph{IEEE Access}, vol.~5, pp. 18\,564--18\,632, 2017.

\bibitem{basar2011stsm}
E.~Basar, U.~Aygolu, E.~Panayirci, and H.~Poor, ``{Space-time block coded
  spatial modulation},'' \emph{IEEE Trans. Communications}, vol.~59, no.~3, pp.
  823--832, 2011.

\bibitem{sugiura2012universal}
S.~Sugiura, S.~Chen, and L.~Hanzo, ``{A universal space-time architecture for
  multiple-antenna aided systems},'' \emph{IEEE Communications Surveys {\&}
  Tutorials}, vol.~14, no.~2, pp. 401--420, 2012.

\bibitem{helmy2016stm}
A.~Helmy, M.~Renzo, and N.~Al-Dhahir, ``{Enhanced-reliability cyclic
  generalized spatial-and-temporal modulation},'' \emph{IEEE Communications
  Letters}, vol.~20, no.~12, pp. 2374--2377, 2016.

\bibitem{le2014ostbc}
M.~T. Le, V.~D. Ngo, H.~A. Mai, X.~N. Tran, and M.~{Di Renzo}, ``{Spatially
  modulated orthogonal space-time block codes with non-vanishing
  determinants},'' \emph{IEEE Transactions on Communications}, vol.~62, no.~1,
  pp. 85--99, 2014.

\bibitem{bian2013dsm}
Y.~Bian, M.~Wen, X.~Cheng, H.~Poor, and B.~Jiao, ``{A differential scheme for
  spatial modulation},'' in \emph{IEEE Global Communications Conference},
  Atlanta, GA, USA, Dec. 9-13, 2013.

\bibitem{wen2014berdsm}
M.~Wen, Z.~Ding, X.~Cheng, Y.~Bian, H.~Poor, and B.~Jiao, ``{Performance
  analysis of differential spatial modulation with two transmit antennas},''
  \emph{IEEE Communications Letters}, vol.~18, no.~3, pp. 475--478, 2014.

\bibitem{ishikawa2014udsm}
N.~Ishikawa and S.~Sugiura, ``{Unified differential spatial modulation},''
  \emph{IEEE Wireless Communications Letters}, vol.~3, no.~4, pp. 337--340,
  2014.

\bibitem{bian2015dsm}
Y.~Bian, X.~Cheng, M.~Wen, L.~Yang, H.~V. Poor, and B.~Jiao, ``{Differential
  spatial modulation},'' \emph{IEEE Transactions on Vehicular Technology},
  vol.~64, no.~7, pp. 3262--3268, 2015.

\bibitem{rajashekar2017dsm}
R.~Rajashekar, N.~Ishikawa, S.~Sugiura, K.~V.~S. Hari, and L.~Hanzo,
  ``{Full-diversity dispersion matrices from algebraic field extensions for
  differential spatial modulation},'' \emph{IEEE Transactions on Vehicular
  Technology}, vol.~66, no.~1, pp. 385--394, 2017.

\bibitem{ishikawa2017mmgsm}
N.~Ishikawa, R.~Rajashekar, S.~Sugiura, and L.~Hanzo,
  ``{Generalized-spatial-modulation-based reduced-RF-chain millimeter-wave
  communications},'' \emph{IEEE Transactions on Vehicular Technology}, vol.~66,
  no.~1, pp. 879--883, 2017.

\bibitem{liu2015ssk}
P.~Liu and A.~Springer, ``{Space shift keying for LOS communication at mmWave
  frequencies},'' \emph{IEEE Wireless Communications Letters}, vol.~4, no.~2,
  pp. 121--124, 2015.

\bibitem{liu2016mmsm}
P.~Liu, M.~{Di Renzo}, and A.~Springer, ``{Line-of-sight spatial modulation for
  indoor mmWave communication at 60 GHz},'' \emph{IEEE Transactions on Wireless
  Communications}, vol.~15, no.~11, pp. 7373--7389, 2016.

\bibitem{mesleh2011osm}
R.~Mesleh, H.~Elgala, and H.~Haas, ``{Optical spatial modulation},''
  \emph{IEEE/OSA Journal of Optical Communications and Networking}, vol.~3,
  no.~3, pp. 234--244, 2011.

\bibitem{fath2011osm}
T.~Fath, H.~Haas, M.~{Di Renzo}, and R.~Mesleh, ``{Spatial modulation applied
  to optical wireless communications in indoor LOS environments},'' in
  \emph{IEEE Global Telecommunications Conference}, Kathmandu, Nepal, 5-9 Dec.,
  2011.

\bibitem{fath2013comparison}
T.~Fath and H.~Haas, ``{Performance comparison of MIMO techniques for optical
  wireless communications in indoor environments},'' \emph{IEEE Transactions on
  Communications}, vol.~61, no.~2, pp. 733--742, 2013.

\bibitem{ishikawa2015osm}
N.~Ishikawa and S.~Sugiura, ``{Maximizing constrained capacity of
  power-imbalanced optical wireless MIMO communications using spatial
  modulation},'' \emph{Journal of Lightwave Technology}, vol.~33, no.~2, pp.
  519--527, 2015.

\bibitem{abu2009sim}
R.~Abu-Alhiga and H.~Haas, ``{Subcarrier-index modulation OFDM},'' in
  \emph{IEEE 20th International Symposium on Personal, Indoor and Mobile Radio
  Communications}, Tokyo, Japan, Sept. 13-16, 2009.

\bibitem{tsonev2011sim}
D.~Tsonev, S.~Sinanovic, and H.~Haas, ``{Enhanced subcarrier index modulation
  (SIM) OFDM},'' in \emph{IEEE GLOBECOM Workshops}, Texas, USA, Dec. 5-9, 2011.

\bibitem{basar2016im}
E.~Basar, ``{Index modulation techniques for 5G wireless networks},''
  \emph{IEEE Communications Magazine}, vol.~54, no.~7, pp. 168--175, 2016.

\bibitem{basar2016simmimo}
------, ``{On multiple-input multiple-output OFDM with index modulation for
  next generation wireless networks},'' \emph{IEEE Transactions on Signal
  Processing}, vol.~64, no.~1, pp. 3868--3878, 2016.

\bibitem{basar2013sim}
E.~Basar, U.~Aygolu, E.~Panayirci, and H.~V. Poor, ``{Orthogonal frequency
  division multiplexing with index modulation},'' \emph{IEEE Transactions of
  Signal Processing}, vol.~61, no.~22, pp. 5536--5549, 2013.

\bibitem{wen2015mar}
M.~Wen, X.~Cheng, M.~Ma, B.~Jiao, and H.~V. Poor, ``{On the achievable rate of
  OFDM with index modulation},'' \emph{IEEE Transactions on Signal Processing},
  vol.~64, no.~8, pp. 1919--1932, 2015.

\bibitem{ishikawa2016sim}
N.~Ishikawa, S.~Sugiura, and L.~Hanzo, ``{Subcarrier-index modulation aided
  OFDM -- will it work?}'' \emph{IEEE Access}, vol.~4, pp. 2580--2593, 2016.

\bibitem{chau2001sm}
Y.~A. Chau and S.-H. Yu, ``{Space modulation on wireless fading channels},'' in
  \emph{IEEE Vehicular Technology Conference}, Atlantic City, NJ, USA, Oct.
  7-11, 2001.

\bibitem{mesleh2006sm}
R.~Mesleh, H.~Haas, C.~W. Ahn, and S.~Yun, ``{Spatial modulation - A new low
  complexity spectral efficiency enhancing technique},'' in \emph{International
  Conference on Communications and Networking in China}, Beijing, China, Oct.
  25-27, 2006.

\bibitem{mesleh2008spatial}
R.~Y. Mesleh, H.~Haas, S.~Sinanovic, C.~Ahn, and S.~Yun, ``{Spatial
  modulation},'' \emph{IEEE Transactions on Vehicular Technology}, vol.~57,
  no.~4, pp. 2228--2241, 2008.

\bibitem{slepian1965pm}
D.~Slepian, ``{Permutation modulation},'' \emph{Proceedings of the IEEE},
  vol.~53, no.~3, pp. 228--236, 1965.

\bibitem{sasaki1991pcss}
S.~Sasaki, J.~Zhu, and G.~Marubayashi, ``{Performance of parallel combinatory
  spread spectrum multiple access communication systems},'' in \emph{IEEE
  International Symposium on Personal, Indoor and Mobile Radio Communications},
  London, UK, Sept. 23-25, 1991.

\bibitem{mittelholzer1999pm}
T.~Mittelholzer, ``{An information-theoretic approach to steganography and
  watermarking},'' in \emph{International Workshop on Information Hiding},
  Dresden, Germany, Sept. 29, 1999.

\bibitem{king2000sparse}
B.~M. King and M.~A. Neifeld, ``{Sparse modulation coding for increased
  capacity in volume holographic storage},'' \emph{Applied Optics}, vol.~39,
  no.~35, pp. 6681--6688, 2000.

\bibitem{jiang2009flash}
A.~Jiang, R.~Mateescu, M.~Schwarts, and J.~Bruck, ``{Rank modulation for flash
  memories},'' \emph{IEEE Transactions on Information Theory}, vol.~55, no.~6,
  pp. 2659--2673, 2009.

\bibitem{mittelholzer2013patent}
T.~Mittelholzer, N.~Papandreou, and C.~Pozidis, ``{Data encoding in solid-state
  storage devices},'' U.S. Patent 8\,578\,246, 2013.

\bibitem{nakao2017ftn}
M.~Nakao, T.~Ishihara, and S.~Sugiura, ``{Single-carrier frequency-domain
  equalization with index modulation},'' \emph{IEEE Communications Letters},
  vol.~21, no.~2, pp. 298--301, 2017.

\bibitem{sugiura2017access}
S.~Sugiura, T.~Ishihara, and M.~Nakao, ``{State-of-the-art design of index
  modulation in the space, time, and frequency domains: Benefits and
  fundamental limitations},'' \emph{IEEE Access}, vol.~5, pp. 21\,774--
  21\,790, 2017.

\bibitem{ishihara2017im}
T.~Ishihara and S.~Sugiura, ``{Faster-than-nyquist signaling with index
  modulation},'' \emph{IEEE Wireless Communications Letters}, vol.~6, no.~5,
  pp. 630--633, 2017.

\bibitem{nakao2017im}
M.~Nakao, T.~Ishihara, and S.~Sugiura, ``{Dual-mode time-domain index
  modulation for nyquist-criterion and faster-than-nyquist single-carrier
  transmissions},'' \emph{IEEE Access}, vol.~5, pp. 27\,659--27\,667, 2017.

\bibitem{wen2016underwater}
M.~Wen, X.~Cheng, L.~Yang, Y.~Li, X.~Cheng, and F.~Ji, ``{Index modulated OFDM
  for underwater acoustic communications},'' \emph{IEEE Communications
  Magazine}, vol.~54, no.~5, pp. 132--137, 2016.

\bibitem{wen2017imbook}
M.~Wen, X.~Cheng, and L.~Yang, \emph{{Index modulation for 5G wireless
  communications}}.\hskip 1em plus 0.5em minus 0.4em\relax Springer, 2017.

\bibitem{shamasundar2017im}
\BIBentryALTinterwordspacing
B.~Shamasundar, S.~Jacob, S.~Bhat, and A.~Chockalingam, ``{Multidimensional
  index modulation in wireless communications},'' \emph{arXiv preprint}, 2017.
  [Online]. Available: \url{http://arxiv.org/abs/1702.03250}
\BIBentrySTDinterwordspacing

\bibitem{slepian1965patent}
D.~Slepian, ``{Permutation code signaling},'' U.S. Patent 3\,196\,351, 1965.

\bibitem{li1996permutation}
W.~Li, \emph{{Study of hybrid permutation frequency phase modulation}}.\hskip
  1em plus 0.5em minus 0.4em\relax Master Thesis, University of Ottawa, 1996.

\bibitem{yongacoglu1997sim}
A.~Yongacoglu and W.~Li, ``{Hybrid permutation frequency phase modulation},''
  in \emph{IEEE Canadian Conference on Electrical and Computer Engineering},
  Saint Johns, Newfoundland, Canada, May 25-28, 1997.

\bibitem{jeganathan2008generalized}
J.~Jeganathan, A.~Ghrayeb, and L.~Szczecinski, ``{Generalized space shift
  keying modulation for MIMO channels},'' in \emph{IEEE International Symposium
  on Personal, Indoor and Mobile Radio Communications}, Cannes, France, Sept.
  15-18, 2008.

\bibitem{frenger1999pcofdm}
P.~K. Frenger and N.~A.~B. Svensson, ``{Parallel combinatory OFDM signaling},''
  \emph{IEEE Transactions on Communications}, vol.~47, no.~4, pp. 558--567,
  1999.

\bibitem{lehmer1960code}
D.~H. Lehmer, ``{Teaching combinatorial tricks to a computer},''
  \emph{Proceedings of Symposia in Applied Mathematics}, vol.~10, pp. 179--193,
  1960.

\bibitem{berger1972permutation}
T.~Berger, F.~Jelinek, and J.~K. Wolf, ``{Permutation codes for sources},''
  \emph{IEEE Transactions on Information Theory}, vol.~18, no.~1, pp. 160--169,
  1972.

\bibitem{atkin1989modulation}
G.~E. Atkin and H.~P. Corrales, ``{An efficient modulation/coding scheme for
  MFSK systems on bandwidth constrained channels},'' \emph{IEEE Journal on
  Selected Areas in Communications}, vol.~7, no.~9, pp. 1396--1401, 1989.

\bibitem{savage1997combinatorial}
C.~Savage, ``{A survey of combinatorial Gray codes},'' \emph{SIAM Review},
  vol.~39, no.~4, pp. 605--629, 1997.

\bibitem{king2001permutation}
B.~M. King and M.~A. Neifeld, ``{Low-complexity maximum-likelihood decoding of
  shortened enumerative permutation codes for holographic storage},''
  \emph{IEEE Journal on Selected Areas in Communications}, vol.~19, no.~4, pp.
  783--790, 2001.

\bibitem{silva2005permutation}
D.~Silva and W.~A. Finamore, ``{Vector permutation modulation},'' \emph{IEEE
  Communications Letters}, vol.~9, no.~8, pp. 673--675, 2005.

\bibitem{shi2010permutation}
M.~Shi, C.~D'Amours, and A.~Yongacoglu, ``{Design of spreading permutations for
  MIMO-CDMA based on space-time block codes},'' \emph{IEEE Communications
  Letters}, vol.~14, no.~1, pp. 36--38, 2010.

\bibitem{ishimura2015permutation}
S.~Ishimura and K.~Kikuchi, ``{Multi-dimensional permutation-modulation format
  for coherent optical communications},'' \emph{Optics Express}, vol.~23,
  no.~12, pp. 15\,587--15\,597, 2015.

\bibitem{tachikawa1992ss}
S.~Tachikawa, ``{Performance of M-ary/ spread spectrum multiple access
  communication systems using co-channel interference cancellation
  techniques},'' in \emph{IEEE International Symposium on Spread Spectrum
  Techniques and Applications}, Yokohama, Japan, Nov. 29 - Dec. 2, 1992.

\bibitem{dillard2003ccsk}
G.~M. Dillard, M.~Reuter, J.~Zeidler, and B.~Zeidler, ``{Cyclic code shift
  keying: A low probability of intercept communication technique},'' \emph{IEEE
  Transactions on Aerospace and Electronic Systems}, vol.~39, no.~3, pp.
  786--798, 2003.

\bibitem{hou2009papr}
Y.~Hou and T.~Hase, ``{New OFDM structure with parallel combinatory code},''
  \emph{IEEE Transactions on Consumer Electronics}, vol.~55, no.~4, pp.
  1854--1859, 2009.

\bibitem{xiaojie2015sec}
F.~Xiaojie, S.~Xuejun, and L.~Yong, ``{Secret communication using parallel
  combinatory spreading WFRFT},'' \emph{IEEE Communications Letters}, vol.~19,
  no.~1, pp. 62--65, 2015.

\bibitem{kaddoum2015cim}
G.~Kaddoum, M.~F.~A. Ahmed, and Y.~Nijsure, ``{Code index modulation: a high
  data rate and energy efficient communication system},'' \emph{IEEE
  Communications Letters}, vol.~19, no.~2, pp. 175--178, 2015.

\bibitem{kaddoum2016cim}
G.~Kaddoum, Y.~Nijsure, and H.~Tran, ``{Generalized code index modulation
  technique for high-data-rate communication systems},'' \emph{IEEE
  Transactions on Vehicular Technology}, vol.~65, no.~9, pp. 7000--7009, 2016.

\bibitem{mesleh2017imp}
R.~Mesleh, O.~Hiari, A.~Younis, and S.~Alouneh, ``{Transmitter design and
  hardware considerations for different space modulation techniques},''
  \emph{IEEE Transactions on Wireless Communications}, vol.~16, no.~11, pp.
  7512--7522, 2017.

\bibitem{ishibashi2014sm}
K.~Ishibashi and S.~Sugiura, ``{Effects of antenna switching on band-limited
  spatial modulation},'' \emph{IEEE Wireless Communications Letters}, vol.~3,
  no.~4, pp. 345--348, 2014.

\bibitem{xu2013reduced}
C.~Xu, S.~Sugiura, S.~X. Ng, and L.~Hanzo, ``{Spatial modulation and space-time
  shift keying: Optimal performance at a reduced detection complexity},''
  \emph{IEEE Transactions on Communications}, vol.~61, no.~1, pp. 206--216,
  2013.

\bibitem{younis2013spsm}
A.~Younis, S.~Sinanovic, M.~{Di Renzo}, R.~Mesleh, and H.~Haas, ``{Generalised
  sphere decoding for spatial modulation},'' \emph{IEEE Transactions on
  Communications}, vol.~61, no.~7, pp. 2805--2815, 2013.

\bibitem{basnayaka2015massive}
D.~{A. Basnayaka}, M.~{Di Renzo}, and H.~Haas, ``{Massive but few active
  MIMO},'' \emph{IEEE Transactions on Vehicular Technology}, vol.~65, no.~9,
  pp. 6861--6877, 2016.

\bibitem{mesleh2010osm}
R.~Mesleh, R.~Mehmood, H.~Elgala, and H.~Haas, ``{Indoor MIMO optical wireless
  communication using spatial modulation},'' in \emph{IEEE International
  Conference on Communications}, Cape Town, South Africa, May 23-27, 2010.

\bibitem{peterson1942patent}
H.~{O. Peterson}, ``{Diversity receiving system},'' U.S. Patent 2\,290\,992,
  1942.

\bibitem{clark1966patent}
R.~{G. Clark}, ``{Communications system for simultaneous communications on a
  single channel},'' U.S. Patent 3\,384\,894, 1968.

\bibitem{schmidt1973patent}
W.~Schmidt and N.~Shimasaki, ``{Satellite on-board switching utilizing
  space-division and spot beam antennas},'' U.S. Patent 3\,711\,855, 1973.

\bibitem{winters1987capacity}
J.~H. Winters, ``{On the capacity of radio communication systems with diversity
  in a Rayleigh fading environment},'' \emph{IEEE Journal on Selected Areas in
  Communications}, vol. SAC-5, no.~5, pp. 871--878, 1987.

\bibitem{amitay1984channel}
N.~Amitay and J.~Salz, ``{Linear equalization theory in digital data
  transmission over dually polarized fading radio channels},'' \emph{AT{\&}T
  Bell Laboratories Technical Journal}, vol.~63, no.~10, pp. 2215--2259, 1984.

\bibitem{foschini1996layered}
G.~Foschini, ``{Layered space-time architecture for wireless communication in a
  fading environment when using multi-element antennas},'' \emph{Bell labs
  technical journal}, vol.~1, no.~2, pp. 41--59, 1996.

\bibitem{foschini1998limits}
G.~Foschini and M.~Gans, ``{On limits of wireless communications in a fading
  environment when using multiple antennas},'' \emph{Wireless personal
  communications}, vol.~6, no.~3, pp. 311--335, 1998.

\bibitem{wolniansky1998v}
P.~Wolniansky, G.~Foschini, G.~Golden, and R.~Valenzuela, ``{V-BLAST: an
  architecture for realizing very high data rates over the rich-scattering
  wireless channel},'' in \emph{Proceedings of the International Symposium on
  Signals, Systems, and Electronics}, Pisa, Italy, Oct. 2, 1998.

\bibitem{alamouti1998std}
S.~Alamouti, ``{A simple transmit diversity technique for wireless
  communications},'' \emph{IEEE Journal on Selected Areas in Communications},
  vol.~16, no.~8, pp. 1451--1458, 1998.

\bibitem{zheng2003dam}
L.~Zheng and D.~Tse, ``{Diversity and multiplexing: a fundamental tradeoff in
  multiple-antenna channels},'' \emph{IEEE Transactions on Information Theory},
  vol.~49, no.~5, pp. 1073--1096, 2003.

\bibitem{hassibi2002hrc}
B.~Hassibi and B.~Hochwald, ``{High-rate codes that are linear in space and
  time},'' \emph{IEEE Transactions on Information Theory}, vol.~48, no.~7, pp.
  1804--1824, 2002.

\bibitem{spencer2004introduction}
Q.~H. Spencer, C.~B. Peel, A.~L. Swindlehurst, and M.~Haardt, ``{An
  introduction to the multi-user MIMO downlink},'' \emph{IEEE Communications
  Magazine}, vol.~42, no.~10, pp. 60--67, 2004.

\bibitem{xu2017bf}
Q.~Xu, C.~Jiang, Y.~Han, B.~Wang, and K.~J.~R. Liu, ``{Waveforming: An Overview
  with Beamforming},'' \emph{IEEE Communications Surveys {\&} Tutorials, in
  press}.

\bibitem{marzetta2010noncooperative}
T.~L. Marzetta, ``{Noncooperative cellular wireless with unlimited numbers of
  base station antennas},'' \emph{IEEE Transactions on Wireless
  Communications}, vol.~9, no.~11, pp. 3590--3600, 2010.

\bibitem{goldsmith2005wireless}
A.~Goldsmith, \emph{{Wireless Communications}}.\hskip 1em plus 0.5em minus
  0.4em\relax Cambridge University Press, 2005.

\bibitem{proakis2008}
J.~G. Proakis and M.~Salehi, \emph{{Digital Communications}}, 5th~ed.\hskip 1em
  plus 0.5em minus 0.4em\relax McGraw-Hill, 2008.

\bibitem{yang2015detect}
S.~Yang and L.~Hanzo, ``{Fifty years of MIMO detection: The road to large-scale
  MIMOs},'' \emph{IEEE Communications Surveys {\&} Tutorials}, vol.~17, no.~4,
  pp. 1941--1988, 2015.

\bibitem{bayes1763}
T.~Bayes and R.~Price, ``{An essay towards solving a problem in the doctrine of
  chances. By the late rev. Mr. Bayes, communicated by Mr. Price, in a letter
  to John Canton, M. A. and F. R. S.}'' \emph{Philosophical Transactions of the
  Royal Society of London}, vol.~53, pp. 370--418, 1763.

\bibitem{sugiura2010cad}
S.~Sugiura, S.~Chen, and L.~Hanzo, ``{Coherent and differential space-time
  shift keying: a dispersion matrix approach},'' \emph{IEEE Transactions on
  Communications}, vol.~58, no.~11, pp. 3219--3230, 2010.

\bibitem{hochwald2000ray}
B.~M. Hochwald and T.~L. Marzetta, ``{Unitary space-time modulation for
  multiple-antenna communications in Rayleigh flat fading},'' \emph{IEEE
  Transactions on Information Theory}, vol.~46, no.~2, pp. 543--564, 2000.

\bibitem{tarokh2000dostbc}
V.~Tarokh and H.~Jafarkhani, ``{A differential detection scheme for transmit
  diversity},'' \emph{IEEE Journal on Selected Areas in Communications},
  vol.~18, no.~7, pp. 1169--1174, 2000.

\bibitem{hughes2000dstm}
B.~L. Hughes, ``{Differential space-time modulation},'' \emph{IEEE Transactions
  on Information Theory}, vol.~46, no.~7, pp. 2567--2578, 2000.

\bibitem{wang2014diff}
L.~Wang, L.~Li, C.~Xu, D.~Liang, S.~X. Ng, and L.~Hanzo, ``{Multiple-symbol
  joint signal processing for differentially encoded single- and multi-carrier
  communications: Principles, designs and applications},'' \emph{IEEE
  Communications Surveys {\&} Tutorials}, vol.~16, no.~2, pp. 689--712, 2014.

\bibitem{hochwald2000unitary}
B.~M. Hochwald, T.~L. Marzetta, T.~J. Richardson, W.~Sweldens, and R.~Urbanke,
  ``{Systematic design of unitary space-time constellations},'' \emph{IEEE
  Transactions on Information Theory}, vol.~46, no.~6, pp. 1962--1973, 2000.

\bibitem{hochwald2000dstm}
B.~M. Hochwald and W.~Sweldens, ``{Differential unitary space-time
  modulation},'' \emph{IEEE Transactions on Communications}, vol.~48, no.~12,
  pp. 2041--2052, 2000.

\bibitem{hassibi2002cdu}
B.~Hassibi and B.~Hochwald, ``{Cayley differential unitary space-time codes},''
  \emph{IEEE Transactions on Information Theory}, vol.~48, no.~6, pp.
  1485--1503, 2002.

\bibitem{xia2002differential}
X.~G. Xia, ``{Differentially en/decoded orthogonal space-time block codes with
  APSK signals},'' \emph{IEEE Communications Letters}, vol.~6, no.~4, pp.
  150--152, 2002.

\bibitem{zhu2005differential}
Y.~Zhu and H.~Jafarkhani, ``{Differential modulation based on quasi-orthogonal
  codes},'' \emph{IEEE Transactions on Wireless Communications}, vol.~4, no.~6,
  pp. 3018--3030, 2005.

\bibitem{bhatnagar2009differential}
M.~Bhatnagar, A.~Hjorungnes, and L.~S.~L. Song, ``{Differential coding for
  non-orthogonal space-time block codes with non-unitary constellations over
  arbitrarily correlated rayleigh channels},'' \emph{IEEE Transactions on
  Wireless Communications}, vol.~8, no.~8, pp. 3985--3995, 2009.

\bibitem{ishikawa2017rdsm}
N.~Ishikawa and S.~Sugiura, ``{Rectangular differential spatial modulation for
  open-loop noncoherent massive-MIMO downlink},'' \emph{IEEE Transactions on
  Wireless Communications}, vol.~16, no.~3, pp. 1908--1920, 2017.

\bibitem{shannon1948mathematical}
C.~E. Shannon, ``{A mathematical theory of communication},'' \emph{Bell System
  Technical Journal}, vol.~27, pp. 379--423, 623--656, 1948.

\bibitem{rappaport2013mm}
T.~S. Rappaport, R.~Mayzus, Y.~Azar, K.~Wang, G.~N. Wong, J.~K. Schulz,
  M.~Samimi, and F.~Gutierrez, ``{Millimeter wave mobile communications for 5G
  cellular: it will work!}'' \emph{IEEE Access}, vol.~1, pp. 335--349, 2013.

\bibitem{busari2017massive}
S.~A. Busari, K.~M.~S. Huq, S.~Mumtaz, L.~Dai, and J.~Rodriguez,
  ``{Millimeter-wave massive MIMO communication for future wireless systems: A
  survey},'' \emph{IEEE Communications Surveys {\&} Tutorials, in press}, 2017.

\bibitem{hemadeh2018mm}
I.~A. Hemadeh, K.~Satyanarayana, M.~El-Hajjar, and L.~Hanzo, ``{Millimeter-wave
  communications: Physical channel models, design considerations, antenna
  constructions and link-budget},'' \emph{IEEE Communications Surveys {\&}
  Tutorials, in press}.

\bibitem{rangan2014mm}
S.~Rangan, T.~S. Rappaport, and E.~Erkip, ``{Millimeter-wave cellular wireless
  networks: Potentials and challenges},'' \emph{Proceedings of the IEEE}, vol.
  102, no.~3, pp. 366--385, 2014.

\bibitem{atta1959array}
A.~Van, ``{Electromagnetic reflector},'' 1959.

\bibitem{celik2008exp}
N.~Celik, M.~F. Iskander, R.~Emrick, S.~J. Franson, and J.~Holmes,
  ``{Implementation and experimental verification of a smart antenna system
  operating at 60 GHz band},'' \emph{IEEE Transactions on Antennas and
  Propagation}, vol.~56, no.~9, pp. 2790--2800, 2008.

\bibitem{zhang2003hybrid}
Z.~Zhang, M.~F. Iskander, Z.~Yun, and A.~H{\o}st-Madsen, ``{Hybrid smart
  antenna system using directional elements — performance analysis in flat
  Rayleigh fading},'' \emph{IEEE Transactions on Antennas and Propagation},
  vol.~51, no.~10, pp. 2926--2935, 2003.

\bibitem{sugiura2008iet}
S.~Sugiura, N.~Kikuma, and H.~Iizuka, ``{Eigenspace-based blind pattern
  optimisations of steerable antenna array for interference cancellation},''
  \emph{IET Microwaves, Antennas {\&} Propagation}, vol.~2, no.~4, pp.
  358--366, 2008.

\bibitem{torkildson2011mmmimo}
E.~Torkildson, U.~Madhow, and M.~Rodwell, ``{Indoor millimeter wave MIMO:
  Feasibility and performance},'' \emph{IEEE Transactions on Wireless
  Communications}, vol.~10, no.~12, pp. 4150--4160, 2011.

\bibitem{guo2012hybrid}
Y.~J. Guo, X.~Huang, and V.~Dyadyuk, ``{A hybrid adaptive antenna array for
  long-range mm-wave communications},'' \emph{IEEE Antennas and Propagation
  Magazine}, vol.~54, no.~2, pp. 271--282, 2012.

\bibitem{alkhateeb2013hybrid}
A.~Alkhateeb, O.~{El Ayach}, G.~Leus, and R.~W. Heath, ``{Hybrid precoding for
  millimeter wave cellular systems with partial channel knowledge},'' in
  \emph{Information Theory and Applications Workshop}, San Diego, CA, USA, Feb.
  10-15, 2013.

\bibitem{ayach2014sparse}
O.~E. Ayach, S.~Rajagopal, S.~Abu-Surra, Z.~Pi, and R.~W. Heath, ``{Spatially
  sparse precoding in millimeter wave MIMO systems},'' \emph{IEEE Transactions
  on Wireless Communications}, vol.~13, no.~3, pp. 1499--1513, 2014.

\bibitem{han2015mm5g}
S.~Han, I.~Chih-Lin, Z.~Xu, and C.~Rowell, ``{Large-scale antenna systems with
  hybrid analog and digital beamforming for millimeter wave 5G},'' \emph{IEEE
  Communications Magazine}, vol.~53, no.~1, pp. 186--194, 2015.

\bibitem{shoji2009sv}
Y.~Shoji, H.~Sawada, C.~S. Choi, and H.~Ogawa, ``{A modified SV-model suitable
  for line-of-sight desktop usage of millimeter-wave WPAN systems},''
  \emph{IEEE Transactions on Antennas and Propagation}, vol.~57, no.~10, pp.
  2940--2948, 2009.

\bibitem{saleh1987sv}
A.~Saleh and R.~Valenzuela, ``{A Statistical model for indoor multipath
  propagation},'' \emph{IEEE Journal on Selected Areas in Communications},
  vol.~5, no.~2, pp. 128--137, 1987.

\bibitem{bohagen2007los}
F.~B{\o}hagen, P.~Orten, and G.~E. {\O}ien, ``{Design of optimal high-rank
  line-of-sight MIMO channels},'' \emph{IEEE Transactions on Wireless
  Communications}, vol.~6, no.~4, pp. 1420--1424, 2007.

\bibitem{smulders1997prop}
P.~Smulders and L.~Correia, ``{Characterisation of propagation in 60 GHz radio
  channels},'' \emph{Electronics {\&} Communications Engineering Journal},
  vol.~9, no.~2, pp. 73--80, 1997.

\bibitem{xu2002mmch}
H.~Xu, V.~Kukshya, and T.~S. Rappaport, ``{Spatial and temporal characteristics
  of 60-GHz indoor channels},'' \emph{IEEE Journal on Selected Areas in
  Communications}, vol.~20, no.~3, pp. 620--630, 2002.

\bibitem{nix2007mmk}
I.~Sarris and A.~R. Nix, ``{Ricean K-factor measurements in a home and an
  office environment in the 60 GHz band},'' in \emph{IST Mobile and Wireless
  Communications Summit}, Budapest, Hungary, July 1-5, 2007.

\bibitem{sridhar2016bf5g}
V.~Sridhar, T.~Gabillard, and A.~Manikas, ``{Spatiotemporal-MIMO channel
  estimator and beamformer for 5G},'' \emph{IEEE Transactions on Wireless
  Communications}, vol.~15, no.~12, pp. 8025--8038, 2016.

\bibitem{zhou2015fast}
L.~Zhou and Y.~Ohashi, ``{Fast codebook-based beamforming training for mmWave
  MIMO systems with subarray structures},'' in \emph{IEEE Vehicular Technology
  Conference Fall}, Boston, MA, USA, Sept. 6-9, 2015.

\bibitem{bell1880photophone}
A.~G. Bell, ``{The photophone},'' \emph{Science}, vol.~1, no.~11, pp. 130--134,
  1880.

\bibitem{kamiya2002patent}
K.~Kamiya, ``{Optical data transmission system},'' J.P. Patent
  2\,000\,387\,660, 2002.

\bibitem{tanaka2003vlc}
Y.~Tanaka, T.~Komine, S.~Haruyama, and M.~Nakagawa, ``{Indoor visible light
  data transmission system utilizing white LED lights},'' \emph{IEICE
  Transactions on Communications}, vol. E86-B, no.~8, pp. 2440--2454, 2003.

\bibitem{komine2004vlc}
T.~Komine and M.~Nakagawa, ``{Fundamental analysis for visible-light
  communication system using LED lights},'' \emph{IEEE Transactions on Consumer
  Electronics}, vol.~50, no.~1, pp. 100--107, 2004.

\bibitem{komine2005shadowing}
T.~Komine, S.~Haruyama, and M.~Nakagawa, ``{A study of shadowing on indoor
  visible-light wireless communication utilizing plural white LED lightings},''
  \emph{Wireless Personal Communications}, vol.~34, no.~1, pp. 211--225, 2005.

\bibitem{minh2008vlc}
H.~L. Minh, D.~O. Brien, G.~Faulkner, L.~Zeng, K.~Lee, D.~Jung, and Y.~Oh,
  ``{High-speed visible light communications using multiple-resonant
  equalization},'' \emph{IEEE Photonics Technology Letters}, vol.~20, no.~14,
  pp. 1243--1245, 2008.

\bibitem{karunatilaka2015vlc}
D.~Karunatilaka, F.~Zafar, V.~Kalavally, and R.~Parthiban, ``{LED based indoor
  visible light communications: state of the art},'' \emph{IEEE Communications
  Surveys {\&} Tutorials}, vol.~17, no.~3, pp. 1649--1678, 2015.

\bibitem{wang2015vlc}
Q.~Wang, Z.~Wang, and L.~Dai, ``{Multiuser MIMO-OFDM for visible light
  communications},'' \emph{IEEE Photonics Journal}, vol.~7, no.~6, 2015.

\bibitem{elgala2011vlc}
H.~Elgala, R.~Mesleh, and H.~Haas, ``{Indoor optical wireless communication:
  potential and state-of-the-art},'' \emph{IEEE Communications Magazine},
  vol.~49, no.~9, pp. 56--62, 2011.

\bibitem{mostafa2015vlcsec}
A.~Mostafa and L.~Lampe, ``{Physical-layer security for MISO visible light
  communication channels},'' \emph{IEEE Journal on Selected Areas in
  Communications}, vol.~33, no.~9, pp. 1806--1818, 2015.

\bibitem{hamme1955infrared}
R.~{N. Hamme} and E.~{A. Boettner}, ``{Infrared communication and test devices:
  final report},'' \emph{Department of the Navy}, 1955.

\bibitem{gfeller1980infrared}
F.~Gfeller and U.~Bapst, ``{Wireless in-house data communication via diffuse
  infrared radiation},'' \emph{Proceedings of the IEEE}, vol.~67, no.~22, pp.
  1474--1486, 1979.

\bibitem{kahn1997ow}
J.~Kahn and J.~Barry, ``{Wireless infrared communications},'' \emph{Proceedings
  of the IEEE}, vol.~85, no.~2, pp. 265--298, 1997.

\bibitem{chan2006fso}
V.~{W. S. Chan}, ``{Free-space optical communications},'' \emph{Journal of
  Lightwave Technology}, vol.~24, no.~12, pp. 4750--4762, 2006.

\bibitem{zeng2009highrate}
L.~Zeng, D.~C. O'Brien, H.~{Le Minh}, G.~E. Faulkner, K.~Lee, D.~Jung, Y.~Oh,
  and E.~T. Won, ``{High data rate multiple input multiple output (MIMO)
  optical wireless communications using white LED lighting},'' \emph{IEEE
  Journal on Selected Areas in Communications}, vol.~27, no.~9, pp. 1654--1662,
  2009.

\bibitem{hanzo2003ofdm}
L.~Hanzo, M.~M{\"{u}}nster, B.~{J. Choi}, and T.~Keller, \emph{{OFDM and
  MC-CDMA for broadband multi-user communications, WLANs and
  broadcasting}}.\hskip 1em plus 0.5em minus 0.4em\relax Wiley-IEEE Press,
  2003.

\bibitem{hanzo2010mimoofdm}
L.~Hanzo, Y.~Akhtman, L.~Wang, and M.~Jiang, \emph{{MIMO-OFDM
  turbo-transceivers for LTE, WIFI and WIMAX}}.\hskip 1em plus 0.5em minus
  0.4em\relax Wiley-IEEE Press, 2010.

\bibitem{mosier1958ofdm}
R.~{R. Mosier} and R.~{G. Clabaugh}, ``{Kineplex, a bandwidth-efficient binary
  transmission system},'' \emph{Transactions of the American Institute of
  Electrical Engineers}, vol.~76, no.~6, pp. 723--728, 1958.

\bibitem{chang1966ofdm}
R.~W. Chang, ``{Synthesis of band‐limited orthogonal signals for multichannel
  data transmission},'' \emph{Bell System Technical Journal}, vol.~45, no.~10,
  pp. 1775--1796, 1966.

\bibitem{chang1970patent}
C.~Chang, ``{Orthogonal frequency multplex data transmission system},'' U.S.
  Patent 3\,488\,445, 1970.

\bibitem{weinstein1971ofdm}
S.~{B. Weinstein} and P.~{M. Ebert}, ``{Data transmission by frequency-division
  multiplexing using the discrete Fourier transform},'' \emph{IEEE Transactions
  on Communication Technology}, vol.~19, no.~5, pp. 628--634, 1971.

\bibitem{cimini1985ofdm}
L.~J. Cimini, ``{Analysis and simulation of a digital mobile channel using
  orthogonal frequency division multiplexing},'' \emph{IEEE Transactions on
  Communications}, vol.~33, no.~7, pp. 665--675, 1985.

\bibitem{ochiai2001papr}
H.~Ochiai and H.~Imai, ``{On the distribution of the peak-to-average power
  ratio in OFDM signals},'' \emph{IEEE Transactions on Communications},
  vol.~49, no.~2, pp. 282--289, 2001.

\bibitem{ochiai2015moment}
H.~Ochiai, ``{High-order moments and Gaussianity of single-carrier and OFDM
  signals},'' \emph{IEEE Transactions on Communications}, vol.~63, no.~12, pp.
  4964--4976, 2015.

\bibitem{sari1995ofdm}
H.~Sari, G.~Karam, and I.~Jeanclaude, ``{Transmission techniques for digital
  terrestrial TV broadcasting},'' \emph{IEEE Communications Magazine}, vol.~33,
  no.~2, pp. 100--109, 1995.

\bibitem{pancaldi2008fde}
F.~Pancaldi, G.~Vitetta, R.~Kalbasi, N.~Al-Dhahir, M.~Uysal, and H.~Mheidat,
  ``{Single-carrier frequency domain equalization},'' \emph{IEEE Signal
  Processing Magazine}, vol.~25, no.~5, pp. 37--56, 2008.

\bibitem{banelli2014ofdm}
P.~Banelli, S.~Buzzi, G.~Colavolpe, A.~Modenini, F.~Rusek, and A.~Ugolini,
  ``{Modulation formats and waveforms for 5G networks: Who will be the heir of
  OFDM?: An overview of alternative modulation schemes for improved spectral
  efficiency},'' \emph{IEEE Signal Processing Magazine}, vol.~31, no.~6, pp.
  80--93, 2014.

\bibitem{baghdady1990directional}
E.~J. Baghdady, ``{Directional signal modulation by means of switched spaced
  antennas},'' \emph{IEEE Transactions on Communications}, vol.~38, no.~4, pp.
  399--403, 1990.

\bibitem{yang2008capacity}
Y.~Yang and B.~Jiao, ``{Information-guided channel-hopping for high data rate
  wireless communication},'' \emph{IEEE Communications Letters}, vol.~12,
  no.~4, pp. 225--227, 2008.

\bibitem{jeganathan2008spatial}
J.~Jeganathan, A.~Ghrayeb, and L.~Szczecinski, ``{Spatial modulation: Optimal
  detection and performance analysis},'' \emph{IEEE Communications Letters},
  vol.~12, no.~8, pp. 545--547, 2008.

\bibitem{ngo2011stfsk}
H.~Ngo, C.~Xu, S.~Sugiura, and L.~Hanzo, ``{Space-time-frequency shift keying
  for dispersive channels},'' \emph{IEEE Signal Processing Letters}, vol.~18,
  no.~3, pp. 177--180, 2011.

\bibitem{sugiura2011gstsk}
S.~Sugiura, S.~Chen, and L.~Hanzo, ``{Generalized space-time shift keying
  designed for flexible diversity-, multiplexing-and complexity-tradeoffs},''
  \emph{IEEE Transactions on Wireless Communications}, vol.~10, no.~4, pp.
  1144--1153, 2011.

\bibitem{yang2012smcapacity}
Y.~Yang and S.~A{\'{i}}ssa, ``{Information guided channel hopping with an
  arbitrary number of transmit antennas},'' \emph{IEEE Communications Letters},
  vol.~16, no.~10, pp. 1552--1555, 2012.

\bibitem{rajashekar2013select}
R.~Rajashekar, K.~V.~S. Hari, and L.~Hanzo, ``{Antenna selection in spatial
  modulation systems},'' \emph{IEEE Communications Letters}, vol.~17, no.~3,
  pp. 521--524, 2013.

\bibitem{wu2016sec}
F.~Wu, R.~Zhang, L.~L. Yang, and W.~Wang, ``{Transmitter precoding-aided
  spatial modulation for secrecy communications},'' \emph{IEEE Transactions on
  Vehicular Technology}, vol.~65, no.~1, pp. 467--471, 2016.

\bibitem{wang2017huff}
W.~Wang and W.~Zhang, ``{Huffman coding-based adaptive spatial modulation},''
  \emph{IEEE Transactions on Wireless Communications}, vol.~16, no.~8, pp.
  5090--5101, 2017.

\bibitem{rajashekar2013structured}
R.~Rajashekar, K.~V.~S. Hari, and L.~Hanzo, ``{Structured dispersion matrices
  from division algebra codes for space-time shift keying},'' \emph{IEEE Signal
  Processing Letters}, vol.~20, no.~4, pp. 371--374, 2013.

\bibitem{wen2016equiprobable}
M.~Wen, Y.~Zhang, J.~Li, E.~Basar, and F.~Chen, ``{Equiprobable subcarrier
  activation method for OFDM with index modulation},'' \emph{IEEE
  Communications Letters}, vol.~20, no.~12, pp. 2386--2389, 2016.

\bibitem{younis2010gsm}
A.~Younis, N.~Serafimovski, R.~Mesleh, and H.~Haas, ``{Generalised spatial
  modulation},'' in \emph{Conference Record of the Forty Fourth Asilomar
  Conference on Signals, Systems and Computers}, Pacific Grove, CA, USA, Nov.
  7-10, 2010.

\bibitem{khandani2013mbm}
A.~K. Khandani, ``{Media-based modulation: A new approach to wireless
  transmission},'' in \emph{IEEE International Symposium on Information
  Theory}, Istanbul, Turkey, Jul. 7-12, 2013.

\bibitem{khandani2014mbm}
------, ``{Media-based modulation: Converting static Rayleigh fading to
  AWGN},'' in \emph{IEEE International Symposium on Information Theory},
  Honolulu, HI, USA, Jun. 29-Jul. 4, 2014.

\bibitem{naresh2017mbm}
Y.~Naresh and A.~Chockalingam, ``{On media-based modulation using RF
  mirrors},'' \emph{IEEE Transactions on Vehicular Technology}, vol.~66, no.~6,
  pp. 4967--4983, 2017.

\bibitem{seifi2016mbm}
E.~Seifi, M.~Atamanesh, and A.~K. Khandani, ``{Media-based MIMO: Outperforming
  known limits in wireless},'' in \emph{IEEE International Conference on
  Communications}, Kuala Lumpur, Malaysia, May 22-27, 2016.

\bibitem{yapeng2016mbm}
L.~Yapeng, T.~Cheng, L.~Liu, and L.~Yongzhi, ``{Novel reduced-complexity
  channel state selection algorithms for media-based modulation},'' in
  \emph{IEEE International Conference on Signal Processing}, Chengdu, China,
  Nov. 6-10, 2016.

\bibitem{bandi2017mbm}
A.~Bandi and C.~R. Murthy, ``{Structured sparse recovery algorithms for data
  decoding in media based modulation},'' in \emph{IEEE International Conference
  on Communications}, Paris, France, May 21-25, 2017.

\bibitem{yildirim2017mbm}
I.~Yildirim, E.~Basar, and I.~Altunbas, ``{Quadrature channel modulation},''
  \emph{IEEE Wireless Communications Letters, in press}.

\bibitem{basar2017mbm}
E.~Basar and I.~Altunbas, ``{Space-time channel modulation},'' \emph{IEEE
  Transactions on Vehicular Technology}, vol.~66, no.~8, pp. 7609--7614, 2017.

\bibitem{oggier2007cyclic}
F.~Oggier, ``{Cyclic algebras for noncoherent differential space-time
  coding},'' \emph{IEEE Transactions on Information Theory}, vol.~53, no.~9,
  pp. 3053--3065, 2007.

\bibitem{li2016dsm}
J.~Li, M.~Wen, X.~Cheng, Y.~Yan, S.~Song, and M.~H. Lee, ``{Differential
  spatial modulation with gray coded antenna activation order},'' \emph{IEEE
  Communications Letters}, vol.~20, no.~6, pp. 1100--1103, 2016.

\bibitem{xiao2017stdsm}
L.~Xiao, Y.~Xiao, P.~Yang, J.~Liu, S.~Li, and W.~Xiang, ``{Space-time block
  coded differential spatial modulation},'' \emph{IEEE Transactions on
  Vehicular Technology}, vol.~66, no.~10, pp. 8821--8834, 2017.

\bibitem{rajashekar2017afedsm}
R.~Rajashekar, C.~Xu, N.~Ishikawa, S.~Sugiura, K.~V.~S. Hari, and L.~Hanzo,
  ``{Algebraic differential spatial modulation is capable of approaching the
  performance of its coherent counterpart},'' \emph{IEEE Transactions on
  Communications, in press}, vol.~65, no.~10, pp. 4260--4273, 2017.

\bibitem{xu2018isk}
C.~Xu, R.~Rajashekar, N.~Ishikawa, S.~Sugiura, and L.~Hanzo, ``{Single-RF index
  shift keying aided differential space-time block coding},'' \emph{IEEE
  Transactions on Signal Processing}, vol.~66, no.~3, pp. 773--788, 2018.

\bibitem{sugiura2012smchannel}
S.~Sugiura and L.~Hanzo, ``{Effects of channel estimation on spatial
  modulation},'' \emph{IEEE Signal Processing Letters}, vol.~19, no.~12, pp.
  805--808, 2012.

\bibitem{xu2018tast}
C.~Xu, R.~Rajashekar, N.~Ishikawa, S.~Sugiura, and L.~Hanzo,
  ``Finite-cardinality single-rf differential space-time modulation: Improved
  performance at a reduced complexity,'' \emph{IEEE Transactions on
  Communications}, submitted.

\bibitem{ishikawa2014ncgsm}
N.~Ishikawa and S.~Sugiura, ``{Single- and multiple-RF aided non-coherent
  generalized spatial modulation},'' in \emph{IEEE Vehicular Technology
  Conference Spring}, Seoul, Korea, May 18-21, 2014.

\bibitem{babakhani2008direct}
A.~Babakhani, D.~B. Rutledge, and A.~Hajimiri, ``{Transmitter architectures
  based on near-field direct antenna modulation},'' \emph{IEEE Journal of
  Solid-State Circuits}, vol.~43, no.~12, pp. 2674--2692, 2008.

\bibitem{valliappan2013asm}
N.~Valliappan, A.~Lozano, and R.~W. Heath, ``{Antenna subset modulation for
  secure millimeter-wave wireless communication},'' \emph{IEEE Transactions on
  Communications}, vol.~61, no.~8, pp. 3231--3245, 2013.

\bibitem{mesleh2017qsm}
R.~Mesleh and A.~Younis, ``{Capacity analysis for LOS millimeter-wave
  quadrature spatial modulation},'' \emph{Wireless Networks}, pp. 1--10, 2017.

\bibitem{perovic2017mmgsm}
N.~S. Perovic, P.~Liu, and A.~Springer, ``{Design of a simple phase precoder
  for generalized spatial modulation in LOS millimeter wave channels},'' in
  \emph{International ITG Conference on Systems, Communications and Coding},
  Hamburg, Germany, Feb. 6-9, 2017.

\bibitem{perovic2017rsm}
N.~S. Perovic, P.~Liu, M.~D. Renzo, and A.~Springer, ``{Receive spatial
  modulation for LOS mmWave communications based on TX beamforming},''
  \emph{IEEE Communications Letters}, vol.~21, no.~4, pp. 1089--7798, 2017.

\bibitem{ding2017mmsm}
\BIBentryALTinterwordspacing
Y.~Ding, K.~J. Kim, T.~Koike-Akino, M.~Pajovic, P.~Wang, and P.~Orlik,
  ``{Spatial Scattering Modulation for Uplink Millimeter-Wave Systems},''
  \emph{IEEE Communications Letters}, vol.~21, no.~7, pp. 1493--1496, jul 2017.
  [Online]. Available: \url{http://ieeexplore.ieee.org/document/7880654/}
\BIBentrySTDinterwordspacing

\bibitem{hemadeh2016mmstsk}
I.~A. Hemadeh, M.~El-Hajjar, S.~Won, and L.~Hanzo, ``{Layered multi-group
  steered space-time shift-keying for millimeter-wave communications},''
  \emph{IEEE Access}, vol.~4, pp. 3708--3718, 2016.

\bibitem{hemadeh2016reduced}
------, ``{Multi-set space-time shift-keying with reduced detection
  complexity},'' \emph{IEEE Access}, vol.~4, pp. 4234--4246, 2016.

\bibitem{hemadeh2017multi}
------, ``{Multiuser steered multiset space-time shift keying for
  millimeter-wave communications},'' \emph{IEEE Transactions on Vehicular
  Technology}, vol.~66, no.~6, pp. 5494--5498, 2017.

\bibitem{hemadeh2017reduced}
I.~A. Hemadeh, P.~Botsinis, M.~El-Hajjar, S.~Won, and L.~Hanzo,
  ``{Reduced-RF-chain aided soft-decision multi-set steered space-time
  shift-keying for millimeter-wave communications},'' \emph{IEEE Access},
  vol.~5, pp. 7223--7243, 2017.

\bibitem{hemadeh2017sfstsk}
I.~A. Hemadeh, M.~El-Hajjar, S.~Won, and L.~Hanzo, ``{Multi-set space-time
  shift keying and space-frequency space-time shift keying for millimeter-wave
  communications},'' \emph{IEEE Access}, vol.~5, pp. 8324--8342, 2017.

\bibitem{yang2011rsm}
L.~L. Yang, ``{Transmitter preprocessing aided spatial modulation for
  multiple-input multiple-output systems},'' in \emph{IEEE Vehicular Technology
  Conference}, Budapest, Hungary, May 15-18, 2011.

\bibitem{sacchi2017mmstsk}
C.~Sacchi, T.~Rahman, I.~A. Hemadeh, and M.~El-Hajjar, ``{Millimeter-wave
  transmission for small-cell backhaul in dense urban environment: a solution
  based on MIMO-OFDM and space-time shift keying (STSK)},'' \emph{IEEE Access},
  vol.~5, pp. 4000 -- 4017, 2017.

\bibitem{he2017mmgsm}
L.~He, J.~Wang, and J.~Song, ``{Spectral-efficient analog precoding for
  generalized spatial modulation aided mmWave MIMO},'' \emph{IEEE Transactions
  on Vehicular Technology}, vol.~66, no.~10, pp. 9598--9602, 2017.

\bibitem{botsinis2017mm}
P.~Botsinis, I.~Hemadeh, D.~Alanis, Z.~Babar, H.~Nguyen, D.~Chandra, S.~X. Ng,
  M.~El-Hajjar, and L.~Hanzo, ``{Joint-alphabet space time shift keying in
  mm-wave non-orthogonal multiple access},'' \emph{IEEE Access, in press},
  2017.

\bibitem{fath2010ssk}
T.~Fath, M.~D. Renzo, and H.~Haas, ``{On the performance of space shift keying
  for optical wireless communications},'' in \emph{IEEE GLOBECOM Workshops},
  Miami, FL, USA, Dec. 6-10, 2010.

\bibitem{popoola2012ppm}
W.~Popoola, E.~Poves, and H.~Haas, ``{Spatial pulse position modulation for
  optical communications},'' \emph{Journal of Lightwave Technology}, vol.~30,
  no.~18, pp. 2948--2954, 2012.

\bibitem{Popoola2013a}
W.~Popoola, ``{Merits and limitations of spatial modulation for optical
  wireless communications},'' in \emph{International Workshop on Optical
  Wireless Communications}, Newcastle upon Tyne, UK, Oct. 21, 2013.

\bibitem{popoola2014merit}
W.~Popoola and H.~Haas, ``{Demonstration of the merit and limitation of
  generalised space shift keying for indoor visible light communications},''
  \emph{Journal of Lightwave Technology}, vol.~32, no.~10, pp. 1960--1965,
  2014.

\bibitem{popoola2013error}
W.~Popoola, E.~Poves, and H.~Haas, ``{Error performance of generalised space
  shift keying for indoor visible light communications},'' \emph{IEEE
  Transactions on Communications}, vol.~61, no.~5, pp. 1968--1976, 2013.

\bibitem{poves2012exp}
E.~Poves, W.~Popoola, H.~Haas, J.~Thompson, and D.~C{\'{a}}rdenas,
  ``{Experimental results on the performance of optical spatial modulation
  systems},'' in \emph{IEEE Vehicular Technology Conference Fall}, Quebec City,
  QC, Canada, Sept. 3-6, 2012.

\bibitem{fath2013color}
T.~Fath and H.~Haas, ``{Optical spatial modulation using colour LEDs},'' in
  \emph{IEEE International Conference on Communications}, Budapest, Hungary,
  June 9-13, 2013.

\bibitem{ozbilgin2015osm}
T.~{\"{O}}zbilgin and M.~Koca, ``{Optical spatial modulation over atmospheric
  turbulence channels},'' \emph{Journal of Lightwave Technology}, vol.~33,
  no.~11, pp. 2313--2323, 2015.

\bibitem{peppas2015osm}
K.~P. Peppas and P.~T. Mathiopoulos, ``{Free-space optical communication with
  spatial modulation and coherent detection over H-K atmospheric turbulence
  channels},'' \emph{Journal of Lightwave Technology}, vol.~33, no.~20, pp.
  4221--4232, 2015.

\bibitem{cai2016osm}
K.~Cai and M.~Jiang, ``{SM/SPPM aided multiuser precoded visible light
  communication systems},'' \emph{IEEE Photonics Journal}, vol.~8, no.~2, 2016.

\bibitem{wang2016osm}
J.~{Y. Wang}, Z.~Yang, Y.~Wang, and M.~Chen, ``{On the performance of spatial
  modulation-based optical wireless communications},'' \emph{IEEE Photonics
  Technology Letters}, vol.~28, no.~19, pp. 2094--2097, 2016.

\bibitem{ishikawa2015exitpiosm}
N.~Ishikawa and S.~Sugiura, ``{EXIT-chart-based design of irregular-precoded
  power-imbalanced optical spatial modulation},'' in \emph{IEEE Vehicular
  Technology Conference Fall}, Boston, USA, Sept. 6-9, 2015.

\bibitem{he2016osm}
C.~He, T.~Q. Wang, and J.~Armstrong, ``{Performance comparison between spatial
  multiplexing and spatial modulation in indoor MIMO visible light
  communication systems},'' in \emph{IEEE International Conference on
  Communications}, Kuala Lumpur, Malaysia, May 22-27, 2016.

\bibitem{safari2008needostbc}
M.~Safari and M.~Uysal, ``{Do we really need OSTBCs for free-space optical
  communication with direct detection?}'' \emph{IEEE Transactions on Wireless
  Communications}, vol.~7, no.~11, pp. 4445--4448, 2008.

\bibitem{schneider1968permutation}
H.~{L. Schneider}, ``{Data transmission with FSK permutation modulation},''
  \emph{Bell Labs Technical Journal}, vol.~47, pp. 1131--1138, 1968.

\bibitem{padovani1986mod}
R.~Padovani and J.~K. Wolf, ``{Coded phase/frequency modulation},'' \emph{IEEE
  Transactions on Communications}, vol.~34, no.~5, pp. 446--453, 1986.

\bibitem{sasaki1994performance}
S.~Sasaki, H.~Kikuchi, Z.~Jinkang, and G.~Marubayashi, ``{Performance of
  parallel combinatory SS communication systems in Rayleigh fading channel},''
  \emph{IEICE Transactions on Fundamentals of Electronics, Communications and
  Computer Sciences}, vol.~77, no.~6, pp. 1028--1032, 1994.

\bibitem{sasaki1995mapc}
S.~Sasaki, H.~Kikuchi, J.~Zhu, and G.~Marubayashi, ``{Multiple access
  performance of parallel combinatory spread spectrum communication systems in
  nonfading and Rayleigh fading channels},'' \emph{IEICE Transactions on
  Communications}, vol. E78-B, no.~8, pp. 1152--1161, 1995.

\bibitem{zhang2016cs}
H.~Zhang, L.~Yang, and L.~Hanzo, ``{Compressed sensing improves the performance
  of subcarrier index-modulation assisted OFDM},'' \emph{IEEE Access}, vol.~4,
  pp. 7859--7873, 2016.

\bibitem{hong2014fqam}
S.~Hong, M.~Sagong, C.~Lim, S.~Cho, K.~Cheun, and K.~Yang, ``{Frequency and
  quadrature-amplitude modulation for downlink cellular OFDMA networks},''
  \emph{IEEE Journal on Selected Areas in Communications}, vol.~32, no.~6, pp.
  1256--1267, 2014.

\bibitem{hori2015hop}
Y.~Hori and H.~Ochiai, ``{A low PAPR subcarrier hopping multiple access with
  coded OFDM for low latency wireless networks},'' in \emph{IEEE Global
  Communications Conference}, San Diego, CA, Dec 6-10, 2015.

\bibitem{hori2016ma}
------, ``{A design of multiuser detection and decoding for subcarrier hopping
  multiple access based on coded OFDM},'' in \emph{IEEE International
  Conference on Communications}, Kuala Lumpur, Malaysia, May 22-27, 2016.

\bibitem{kitamoto2005pcoptical}
N.~Kitamoto and T.~Ohtsuki, ``{Parallel combinatory multiple-subcarrier optical
  wireless communication systems},'' \emph{International Journal of
  Communication Systems}, vol.~18, no.~3, pp. 195--203, 2005.

\bibitem{hou2007hc}
Y.~Hou and M.~Hamamura, ``{A novel modulation with parallel combinatory and
  high compaction multi-carrier modulation},'' \emph{IEICE Transactions on
  Fundamentals}, vol. E90-A, no.~11, pp. 2556--2567, 2007.

\bibitem{xiao2014sim}
Y.~Xiao, S.~Wang, L.~Dan, X.~Lei, P.~Yang, and W.~Xiang, ``{OFDM with
  interleaved subcarrier-index modulation},'' \emph{IEEE Communications
  Letters}, vol.~18, no.~8, pp. 1447--1450, 2014.

\bibitem{basar2015interleave}
E.~Basar, ``{OFDM with index modulation using coordinate interleaving},''
  \emph{IEEE Wireless Communications Letters}, vol.~4, no.~4, pp. 381--384,
  2015.

\bibitem{fan2014generalizedsim}
R.~Fan, Y.~J. Yu, and Y.~L. Guan, ``{Generalization of orthogonal frequency
  division multiplexing with index modulation},'' \emph{IEEE Transactions on
  Wireless Communications}, vol.~14, no.~10, pp. 5350--5359, 2015.

\bibitem{zheng2015sim}
B.~Zheng, F.~Chen, M.~Wen, F.~Ji, H.~Yu, and Y.~Liu, ``{Low-complexity ML
  detector and performance analysis for OFDM with in-phase/quadrature index
  modulation},'' \emph{IEEE Communications Letters}, vol.~19, no.~11, pp.
  1893--1896, 2015.

\bibitem{basar2015mimosim}
E.~Basar, ``{Multiple-input multiple-output OFDM with index modulation},''
  \emph{IEEE Signal Processing Letters}, vol.~22, no.~12, pp. 2259--2263, 2015.

\bibitem{datta2015spacefreqindexmod}
T.~Datta, H.~Eshwaraiah, and A.~Chockalingam, ``{Generalized
  space-and-frequency index modulation},'' \emph{IEEE Transactions on Vehicular
  Technology}, vol.~65, no.~7, pp. 4911--4924, 2015.

\bibitem{ma2016sim}
Q.~Ma, Y.~Xiao, L.~Dan, P.~Yang, L.~Peng, and S.~Li, ``{Subcarrier allocation
  for OFDM with index modulation},'' \emph{IEEE Communications Letters},
  vol.~20, no.~7, pp. 1469--1472, 2016.

\bibitem{wang2016sim}
L.~Wang, Z.~Chen, Z.~Gong, and M.~Wu, ``{Space-frequency coded index modulation
  with linear-complexity maximum likelihood receiver in the MIMO-OFDM
  system},'' \emph{IEEE Signal Processing Letters}, vol.~23, no.~10, pp.
  1439--1443, 2016.

\bibitem{mao2016sim}
T.~Mao, Z.~Wang, Q.~Wang, S.~Chen, and L.~Hanzo, ``{Dual-mode index modulation
  aided OFDM},'' \emph{IEEE Access}, vol.~5, pp. 50--60, 2016.

\bibitem{mao2017gen}
T.~Mao, Q.~Wang, and Z.~Wang, ``{Generalized dual-mode index modulation aided
  OFDM},'' \emph{IEEE Communications Letters}, vol.~21, no.~4, pp. 761--764,
  2017.

\bibitem{ng2006mimo}
S.~X. Ng and L.~Hanzo, ``{On the MIMO channel capacity of multi-dimensional
  signal sets},'' \emph{IEEE Transactions on Vehicular Technology}, vol.~55,
  no.~2, pp. 528--536, 2006.

\bibitem{hanzo2009nearcapacity}
L.~Hanzo, O.~Alamri, M.~El-Hajjar, and N.~Wu, \emph{{Near-capacity
  multi-functional MIMO systems}}.\hskip 1em plus 0.5em minus 0.4em\relax John
  Wiley {\&} Sons, Ltd, 2009.

\bibitem{nguyen2015capacity}
H.~V. Nguyen, C.~Xu, S.~X. Ng, and L.~Hanzo, ``{Near-capacity wireless system
  design principles},'' \emph{IEEE Communications Surveys {\&} Tutorials},
  vol.~17, no.~4, pp. 1806--1833, 2015.

\bibitem{sugiura2013jointopt}
S.~Sugiura and L.~Hanzo, ``{On the joint optimization of dispersion matrices
  and constellations for near-capacity irregular precoded space-time shift
  keying},'' \emph{IEEE Transactions on Wireless Communications}, vol.~12,
  no.~1, pp. 380--387, 2013.

\bibitem{tarokh1998space}
V.~Tarokh, N.~Seshadri, and A.~Calderbank, ``{Space-time codes for high data
  rate wireless communication: performance criterion and code construction},''
  \emph{IEEE Transactions on Information Theory}, vol.~44, no.~2, pp. 744--765,
  1998.

\bibitem{heathjr2002ldc}
R.~Heath and A.~Paulraj, ``{Linear dispersion codes for MIMO systems based on
  frame theory},'' \emph{IEEE Transactions on Signal Processing}, vol.~50,
  no.~10, pp. 2429--2441, 2002.

\bibitem{brent2010arith}
R.~P. Brent and P.~Zimmermann, \emph{{Modern Computer Arithmetic}}.\hskip 1em
  plus 0.5em minus 0.4em\relax Cambridge University Press, 2010.

\bibitem{cavus2006low}
E.~Cavus and B.~Daneshrad, ``{A very low-complexity space-time block decoder
  (STBD) ASIC for wireless systems},'' \emph{IEEE Transactions on Circuits and
  Systems I: Regular Papers}, vol.~53, no.~1, pp. 60--69, 2006.

\bibitem{knuth1976bigomega}
D.~E. Knuth, ``{Big Omicron and big Omega and big Theta},'' \emph{ACM SIGACT
  News}, vol.~8, no.~2, pp. 18--24, 1976.

\bibitem{schober1999dpsk}
R.~Schober, S.~Member, W.~H. Gerstacker, and J.~B. Huber, ``{Decision-feedback
  differential detection of MDPSK for flat Rayleigh fading channels},''
  \emph{IEEE Transactions on Communications}, vol.~47, no.~7, pp. 1025--1035,
  1999.

\bibitem{hwang2005differential}
C.~S. Hwang, S.~H. Nam, J.~Chung, and V.~Tarokh, ``{Differential space time
  block codes using nonconstant modulus constellations},'' \emph{IEEE
  Transactions on Signal Processing}, vol.~51, no.~11, pp. 2955--2964, 2003.

\bibitem{wen2017mode}
M.~Wen, E.~Basar, Q.~Li, B.~Zheng, and M.~Zhang, ``{Multiple-mode orthogonal
  frequency division multiplexing with index modulation},'' \emph{IEEE
  Transactions on Communications}, vol.~65, no.~9, pp. 3892--3906, 2017.

\end{thebibliography}

%
%

\begin{IEEEbiography}[{\includegraphics[width=1in,height=1.25in,clip,keepaspectratio]{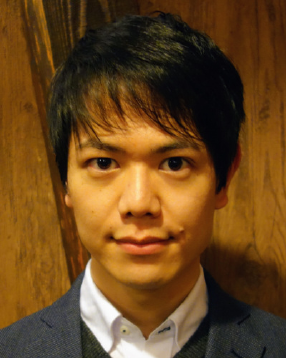}}]{Naoki~Ishikawa}
	(S'13--M'17) \url{https://ishikawa.cc} was born in Kanagawa, Japan, in 1991. He received the B.E., M.E., and Ph.D. degrees from the Tokyo University of Agriculture and Technology, Tokyo, Japan, in 2014, 2015, and 2017, respectively. From June 2015 to September 2015, he was an academic visitor with the School of Electronics and Computer Science, University of Southampton, UK. From April 2016 to March 2017, he was a research fellow of the Japan Society for the Promotion of Science. From April 2017, he has been an assistant professor in the Graduate School of Information Sciences, Hiroshima City University, Japan.
	
	He received eight domestic awards, including the Yasujiro Niwa Outstanding Paper Award from Tokyo Denki University in 2018, the Telecom System Technology student Award (honorable mention) from Telecommunications Advancement Foundation of Japan in 2014, the Outstanding Paper Award for Young C\&C Researchers from NEC C\&C Foundation in 2014, and the Young Researcher's Encouragement Award from the IEEE VTS Japan Chapter in 2014. 
\end{IEEEbiography}

\begin{IEEEbiography}[{\includegraphics[width=1in,height=1.25in,clip,keepaspectratio]{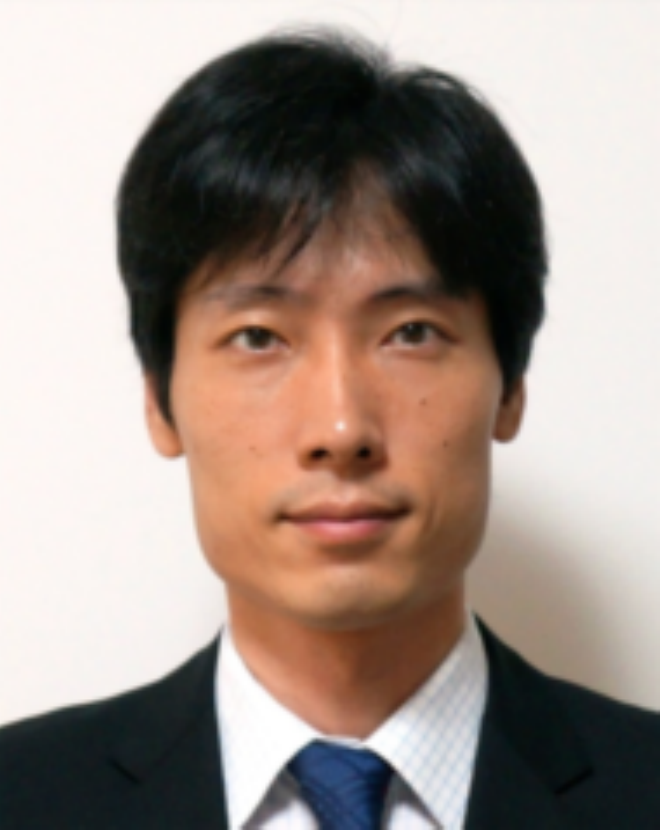}}]{Shinya~Sugiura}
	(M'06--SM'12) received the B.S.\ and M.S.\ degrees in aeronautics and astronautics from Kyoto University, Kyoto, Japan, in 2002 and 2004, respectively, and the Ph.D.\ degree in electronics and electrical engineering from the University of Southampton, Southampton, U.K., in 2010.
	
	From 2004 to 2012, he was a Research Scientist with Toyota Central Research and Development Laboratories, Inc., Aichi, Japan. Since 2013, he has been an Associate Professor with the Department of Computer and Information Sciences, Tokyo University of Agriculture and Technology, Tokyo, Japan, where he heads the Wireless Communications Research Group. His research has covered a range of areas in wireless communications, networking, signal processing, and antenna technology. He has authored or coauthored over 50 IEEE journal papers in these research fields.
	
	Dr.\ Sugiura has been a recipient of a number of awards, including the RIEC Award from the Foundation for the Promotion of Electrical Communication in 2016, the Young Scientists' Prize by the Minister of Education, Culture, Sports, Science and Technology of Japan in 2016, the 14th Funai Information Technology Award (First Prize) from the Funai Foundation in 2015, the 28th Telecom System Technology Award from the Telecommunications Advancement Foundation in 2013, the Sixth IEEE Communications Society Asia-Pacific Outstanding Young Researcher Award in 2011, the 13th Ericsson Young Scientist Award in 2011, and the 2008 IEEE Antennas and Propagation Society Japan Chapter Young Engineer Award. He was also certified as an Exemplary Reviewer of \textsc{IEEE Communications Letters} in 2013 and 2014.
\end{IEEEbiography}

\begin{IEEEbiography}[{\includegraphics[width=1in,height=1.25in,clip,keepaspectratio]{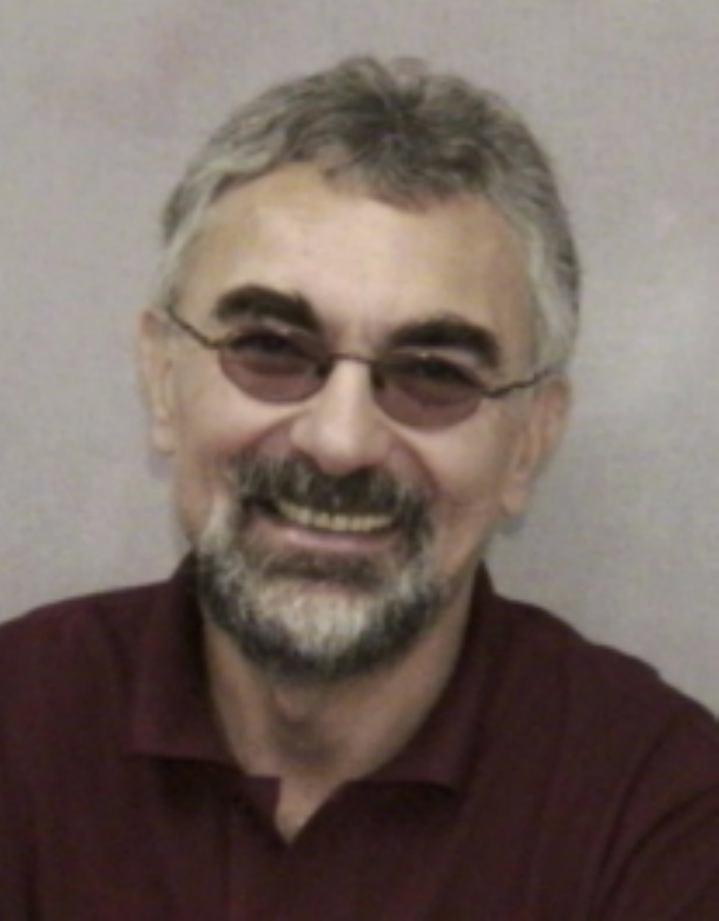}}]{Lajos~Hanzo}
	(\url{http://www-mobile.ecs.soton.ac.uk}) FREng, FIEEE,
	FIET, Fellow of EURASIP, DSc
	received his degree in electronics in
	1976 and his doctorate in 1983.  In 2009 he was awarded an honorary
	doctorate by the Technical University of Budapest and in 2015 by the
	University of Edinburgh.  In 2016 he was admitted to the Hungarian
	Academy of Science. During his 40-year career in telecommunications he
	has held various research and academic posts in Hungary, Germany and
	the UK. Since 1986 he has been with the School of Electronics and
	Computer Science, University of Southampton, UK, where he holds the
	chair in telecommunications.  He has successfully supervised 111
	PhD students, co-authored 18 John Wiley/IEEE Press books on mobile
	radio communications totalling in excess of 10 000 pages, published
	1700+ research contributions at IEEE Xplore, acted both as TPC and General
	Chair of IEEE conferences, presented keynote lectures and has been
	awarded a number of distinctions. Currently he is directing a
	60-strong academic research team, working on a range of research
	projects in the field of wireless multimedia communications sponsored
	by industry, the Engineering and Physical Sciences Research Council
	(EPSRC) UK, the European Research Council's Advanced Fellow Grant and
	the Royal Society's Wolfson Research Merit Award.  He is an
	enthusiastic supporter of industrial and academic liaison and he
	offers a range of industrial courses.  He is also a Governor of the
	IEEE ComSoc and VTS.  During 2008 - 2012 he was the Editor-in-Chief of the IEEE
	Press and a Chaired Professor also at Tsinghua University, Beijing.
	For further information on research in progress and associated
	publications please refer to \url{http://www-mobile.ecs.soton.ac.uk} Lajos
	has 34 000+ citations.
\end{IEEEbiography}

\end{document}